\documentclass[12pt,a4paper]{article}
\usepackage{fullpage}

\usepackage{setspace}

\usepackage{epic,eepic,graphicx} 
\usepackage{psfrag,epsfig}

\usepackage{subfigure} 
\usepackage{eepic,epic,rotating} 
\usepackage{ccaption}
         
\captiondelim{. }  
\captionnamefont{\small\bfseries} 
\captiontitlefont{\small}
\changecaptionwidth
\captionwidth{\linewidth}

\usepackage{appendix}

\usepackage{topcapt}

\def\re{\mathrm{Re\,}}
\def\im{\mathrm{Im\,}}

\usepackage{amsmath,amssymb,esint,bm,cases} 
      
\usepackage{titlesec}

\title{\bf \Large The role of a delay time on the spatial structure of chaotically advected reactive scalars}
\author{Alexandra Tzella\footnote{Corresponding author email: tzella@lmd.ens.fr} \\ {\small Laboratoire de M\'et\'eorologie Dynamique, ENS, 24 rue Lhomond, F-75231, Paris, France.} \medskip \\
Peter H. Haynes\\
{\small Department of Applied Mathematics and Theoretical Physics,}\\
{\small University of Cambridge, CB3 0WA, Cambridge, United Kingdom.}}
\date{\bigskip \normalsize January 2009}

\begin{document}

\maketitle
\thispagestyle{empty}

\newpage

\centerline{\Large\bf Abstract}
\vspace{1cm}

The stationary-state spatial structure of reacting scalar fields,
chaotically advected by a two-dimensional large-scale flow, is
examined for the case for which the reaction equations contain delay
terms. Previous theoretical investigations have shown that, in the
absence of delay terms and in a regime where diffusion can be
neglected (large P\'eclet number), the emergent
spatial structures are filamental and characterized by a single
scaling regime with a H\"older exponent that depends on the 
rate of convergence of the reactive processes and the
strength of the stirring measured by the average stretching rate.
In the presence of delay terms, we show that for
sufficiently small scales all interacting fields should share the same
spatial structure, as found in the absence of delay terms.
Depending on the strength of the stirring and the magnitude of
the delay time, two further scaling regimes that are unique to the delay
system may appear at intermediate length scales. An expression for the
transition length scale dividing small-scale and intermediate-scale
regimes is obtained and the scaling behavior of the scalar field
is explained. The theoretical results are illustrated by numerical
calculations for two types of reaction models, both based on delay
differential equations, coupled to a two-dimensional chaotic
advection flow. The first corresponds to a single reactive scalar
and the second to a nonlinear biological model that includes
nutrients, phytoplankton and zooplankton. As in the no-delay case, the
presence of asymmetrical couplings among the biological species
results in a non-generic scaling behavior.
  
\newpage                           


\maketitle


\section{Introduction}
The transport and stirring of reactive scalars 
is a problem that naturally arises in many environmental and geophysical situations as well as in engineering applications.
Important examples of reactive scalars may be found in oceanic ecosystems 
e.g. interacting nutrient and plankton populations,   
in atmospheric chemistry e.g. stratospheric ozone as well as in microfluidics and combustion.   
In all of these examples, fine-scale strongly-inhomogeneous structures, usually in the form of filaments, 
characterize the spatial structure of the corresponding reactive scalar fields \cite{Abraham_etal2000, Neufeld_etal2002, Nieves_etal2007, TuckHovde1999, Stremler_etal2004, Kiss_etal2003}.  
Understanding 
the main mechanisms controlling the nature of these small-scale structures is important
as they can have a large-scale impact  
for instance on the global ozone depletion \cite{Edouard_etal1996} or on the total plankton production \cite{MahadevanArcher2000}.

It is now well known that small-scale filamentary structures arise
naturally through chaotic advection in spatially smooth
(differentiable) and time-dependent velocity fields \cite{Aref1984,Crisanti_etal1999,Cartwright_etal1999}, 
relevant to a
broad set of applications ranging from stably stratified flows in
the atmosphere and the ocean \cite{Haynes1999} to microfluidic devices \cite{Aref2002}. 
Scalar mixing is
induced through the continual stretching and folding of fluid elements
by which large-scale scalar variability is transferred into small scales
until it is dissipated by molecular diffusion. The rate at which the
scalar is mixed is insensitive to the details of the diffusion and
depends primarily on the stirring strength of the flow. 
A measure for the latter is given by the exponential rate at which neighboring fluid parcel trajectories separate in backward time. 
Following previous work \cite{Ottino1989,Ott1993} on dynamical systems theory applied to chaotic advection, we call this rate the flow Lyapunov exponent. More precisely, it is the most positive Lyapunov exponent associated with the backward dynamics.

A non-trivial stationary-state spatial distribution is obtained in the
presence of a large- scale space-dependent forcing \cite{Batchelor1959}. In the presence
of reactions whose dynamics are stable and for a spatially smooth
force, the distribution is filamental or smooth depending on whether
the stirring of the flow is stronger or weaker than the rate of
convergence of the reaction dynamics. The latter is measured by the
set of Lyapunov exponents associated with the reaction dynamics,
better known as the chemical Lyapunov exponents \cite{Neufeld_etal1999}, whose values
depend on the reaction system and, to a lesser extent, on the driving
induced by  chaotic advection. A useful way to characterize the
scaling behavior of the spatial distribution is by
investigating the 
 scaling exponents of statistical quantities such as structure functions. 
For closed chaotic flows (bounded flow domain) 
and at scales for which diffusion can be neglected, the small-scale structure of all the reactive scalar fields is shared 
and characterized by a single scaling regime (special conditions that give rise to exceptions will be discussed later). The theoretical prediction for the H\"older exponent, the scaling exponent associated with the field's first-order structure function,  was found by \cite{Neufeld_etal1999} 
to be determined by the ratio of the 
least negative chemical Lyapunov exponent to the flow Lyapunov exponent (as defined previously) (see also \cite{Hernandez-Garcia_etal2002} for an extension to a multi-species reaction model). 

This theoretical prediction, deduced for reaction systems that are based on ordinary differential equations, was found to be in contradiction with the numerical results 
that \cite{Abraham1998} 
obtained for a reaction model that is based on delay  differential equations.  
The latter is a model that describes the biological interactions among nutrients, phytoplankton and zooplankton 
and is in this paper referred to as the {\it delay plankton model}. 
The numerical results of \cite{Abraham1998} appeared to show that
introducing a delay time into the reactions led to the decoupling
among the phytoplankton and zooplankton distributions at all length
scales.
Moreover, as the value of the delay time was increased, 
the zooplankton distribution was found to become increasingly filamental, 
ultimately behaving like a passive, non-reactive scalar, in agreement with most oceanic observations at the mesoscale \cite{MackasBoyd1979,Mackas_etal1985, Tsuda1995}.

The relation between the numerical work of \cite{Abraham1998} for the system with delay  
and the theoretical and numerical work of \cite{Neufeld_etal1999} and \cite{Hernandez-Garcia_etal2002} 
for the system without delay 
has recently been addressed in \cite{TzellaHaynes2007}. 
Based on an alternative numerical method that permits the study of smaller length scales, 
a new set of carefully performed numerical simulations  revealed that
for sufficiently small length scales, 
the phytoplankton and zooplankton distributions share the same small-scale structure, 
as would be expected in the absence of delay. 
However, at scales larger than a  transition length scale, a second scaling regime appeared  
in which the scaling behavior that \cite{Abraham1998} observed was reproduced.

The main focus of this paper is to present a theory for the spatial
properties of reactive scalar fields whose reactions explicitly
contain a delay time and which are stirred by a chaotic advection
flow. One motivation is better understanding of the delay plankton
model discussed above, but broader motivation comes from the wide 
application of delay equations to model chemical \cite{Roussel1996}
and biological \cite{Murray1993} systems. By varying the delay time as well as the stirring strength of the flow and the reactions, 
two main issues are here investigated: firstly, the origin of the second scaling regime 
and secondly the parameters that control the transition length scale and scaling behavior in each of these two  regimes. In order to obtain a theoretical understanding of such a system, models of increasing complexity will be considered starting with a single linear delay reactive scalar field and moving on to a system of nonlinearly interacting scalar fields. Scalar fields evolving according to reaction equations containing a delay time are in the following referred to as {\it delay reactive scalar fields}. The theoretical development is accompanied by  a set of numerical results obtained for (i) a single linear delay reactive scalar and (ii) the delay plankton model, both coupled to a two-dimensional, unsteady and incompressible flow via a large scale spatially smooth source.

This paper is organized into two parts. The first part, Sec. \ref{sec:DelayTheory}, is solely devoted to the theoretical development of a single delay reactive scalar, complemented in the Appendix  for a system of such fields. A set of scaling laws are deduced describing the 
H\"older exponents associated within three scaling regimes. 
The transition length scale dividing small-scale and  intermediate-scale regimes
is found to depend on the product of the delay time and the stirring strength of the flow.  
The second part of the paper, Sec. \ref{sec:DelayNumericalResults}, consists of the numerical simulations to verify the theoretical results obtained in Sec. \ref{sec:DelayTheory}. 
The paper concludes with a summary and conclusion.

\section{Theoretical Development}\label{sec:DelayTheory} 
\subsection{Reactive Scalar Evolution Models}   
The spatial and temporal evolution of passively advected reactive tracers  
is described by the Advection Diffusion Reaction (ADR) equations.
For the case of an incompressible velocity field, $\bm{v}(\bm{x},t)$, and for $t>0$, the typical form of these equations is
\begin{equation}\label{eqn:ADR}
\frac{\partial}{\partial t}  \bm{c}(\bm{x},t)+ \bm{v}(\bm{x},t)\cdot \nabla \bm{c}(\bm{x},t) = 
\bm{\mathcal{F}}_{-\tau}+D\nabla^{2}\bm{c}(\bm{x},t),
\end{equation}  
where the fields $\bm{c}(\bm{x},t)=(c_1(\bm{x},t), c_2(\bm{x},t),\ldots,c_n(\bm{x},t))$,
$n$ being the number of chemical species,
are assumed to diffuse independently from one another with the same constant diffusivity $D$.

The interactions among these scalar fields e.g. chemical reactions or predator-prey interactions,  
are described by the forcing term $\bm{\mathcal{F}}_{-\tau}\equiv\bm{\mathcal{F}}(\bm{c}(\bm{x},t), \bm{c}(\bm{x},t-\tau), \bm{x})$ 
in which  the effects of sources and sinks are also included.   
The main feature of the forcing term is its dependence on a delay time $\tau$ 
associated with, e.g.  the time it takes for a biological species to mature.     
Note that for Eq. (\ref{eqn:ADR}) to be well-defined, 
$\bm{c}(\bm{x},t)$ needs to be initialized for $t\in[-\tau,0]$. 
 
The explicit dependence of the forcing term on the spatial coordinate $\bm{x}$ 
accounts for  the inhomogeneous distributions of these sources and sinks 
e.g. due to a spatially varying nutrient field,  
or for the spatial dependence of the reproduction and predation rates  of biological species e.g. due to a temperature dependence.                                 
If the forcing term does not depend on the spatial coordinate, the reactions are not coupled to the flow 
and any initial inhomogeneity in the concentration fields is stirred down by advection and eventually smoothed out by diffusion.                                                  

We  will here concentrate on a forcing term  that in the absence of advection has 
a single, stable, fixed point of equilibrium.
In this case, as it will be clear later, for a time $t$ that is large enough, $\bm{c}(\bm{x},t)$ is  assumed \cite{Neufeld_etal1999} to reach a statistical equilibrium.   

To tackle Eq. (\ref{eqn:ADR}) one can either consider the fields in the space domain the fluid is defined \cite{Corrsin1961} - the Eulerian approach - or instead consider their evolution along the trajectory traced by each fluid parcel that constitutes the fluid - the Lagrangian approach. The approach we will adopt is the Lagrangian one.

For cases for which advective transport dominates diffusion i.e. large P\'eclet number,       
a natural approach is to set $D=0$. 
The chemical evolution of a fluid parcel is then independent of all such parcels
and
Eq. (\ref{eqn:ADR}) is reduced
into a low-dimensional dynamical system given by        
\begin{subequations}
\begin{align}                                               
\frac{d\bm{X}(t)}{dt}&=\bm{v}(\bm{X}(t),t) \label{eqn:traj},\\
\frac{d\bm{C}_{\bm{X}(t)}(t)}{dt}&=\bm{\mathcal{F}}_{-\tau}(\bm{C}_{\bm{X}(t)}(t),\bm{X}(t)),\label{eqn:chem}     
\end{align}
\end{subequations}     
where $\bm{X}(t)$ denotes the fluid parcel's trajectory and
$\bm{C}_{\bm{X}(t)}(t)$  is a vector of its chemical concentration fields, satisfying 
$\bm{C}_{\bm{X}(t)}(t)=\bm{c}(\bm{x}=\bm{X}(t),t)$.
The implication of the neglect of diffusion is that any predictions concerning the spatial structure apply only above a certain spatial cut-off scale whose value  approaches zero for smaller and smaller diffusivities (see \cite{LopezHernandez-Garcia2002} where this argument is developed for a linearly decaying reactive scalar).

The principal aim here is to examine the small-scale structure of the scalar fields once statistical equilibrium has been attained and characterize this structure in terms of H\"older exponents. 
To do so, the concentration difference between neighboring points, given by  
$\delta\bm{c}(\delta\bm{x};\bm{x},t)\equiv\bm{c}(\bm{x}+\delta\bm{x},t)-\bm{c}(\bm{x},t)$, needs to be investigated as a function of $\delta \bm{x}$ from where the H\"older exponents 
$\bm{\gamma}=(\gamma_1, \gamma_2,\ldots,\gamma_n)$, defined by  
\begin{equation}\label{eqn:concentrationdiff2}
|\delta c_i(\delta\bm{x};\bm{x},t)|\sim |\delta \bm{x}|^{\gamma_i},  \quad 
|\delta \bm{x}|\rightarrow 0,   
\end{equation}
can de deduced. 
For a smooth field (i.e. differentiable) $\gamma_i=1$ at $\bm{x}$ while 
the range $0<\gamma_i<1$ corresponds to 
an irregular (e.g. filamental) field. This concentration difference can be estimated by considering the concentration difference between two neighboring fluid parcels
$\bm{X}(t)$, $\bm{X}+\delta \bm{X}(t)$ with
\begin{equation}\label{eqn:LagrangianEulerian2} 
\delta\bm{c}(\delta\bm{x};\bm{x},t)
=\delta\bm{C}_{\delta\bm{X}(t);\bm{X}(t)} 
\equiv\bm{C}_{\bm{X}+\delta\bm{X}(t)}-\bm{C}_{\bm{X}(t)}.	
\end{equation}

In order to simplify the analysis, in the following we will concentrate on the following simple example, 
\begin{equation}\label{eqn:1Ddelay}
\frac{d}{dt}C(t)=-aC(t)-bC(t-\tau)+C_0(\bm{x}(t)), 	   
\end{equation}                                 
where $a,b,\tau$ are constants with $a,\tau>0$ 
and $C_0(\bm{x})$ is a spatially smooth source.                                          
The more general case (\ref{eqn:chem}) is considered in the Appendix. 
We will only consider two-dimensional flows, however the theory presented is readily extendable to large-scale flows in higher dimensions.

\subsection{Key Properties of Forced Linear Delay Equations}\label{subsec:KeyProperties}   
To understand the role that a delay time plays on the fields' scaling behavior,  
the general properties of forced linear delay differential equations (DDEs) need to be considered.
An overview of those is now presented.
For more complete treatments see  \cite{HaleLunel1993}, \cite{BellmanCooke1963} and \cite{Diekmann_etal1995}.        

Take the one-dimensional  forced, linear DDE 
\begin{equation}\label{eqn:forced1D}
\dot{y}=-ay(t)-by(t-\tau)+f(t),
\end{equation}
where $a$, $b$ and $\tau$ are the same as before and $f$ is a real continuous function.
In order for $y(t)$ to be uniquely determined, it is necessary to prescribe an initial function
on the interval $[-\tau,0]$. Denoting this  function by $\phi(t)$, it follows that     
\begin{subequations}\label{eqn:general}
\begin{align}
y(t)&=\phi (t), \quad \text{for $t\in [-\tau,0]$,} \label{eqn:initialcd}\\  
y(t)&=
	e^{-at}\phi(0)+  \nonumber \\
\phantom{y(t)}&\phantom{=} 
\int_0^t e^{-a(t-t')}[-by(t'-\tau)+f(t')]dt',\quad \text{for $t>0$,}
	\label{eqn:variationconstants}               
\end{align}
\end{subequations} 
where Eq. (\ref{eqn:variationconstants}) is easily deduced using the well-known  variation of constants (or parameters) formula. 
Based on Eq. (\ref{eqn:general}), an expression for  $y(t)$ for $t\in[0,\tau]$ is readily determined.
Substituting this expression into (\ref{eqn:variationconstants}), $y(t)$ can be calculated for $t\in[\tau,2\tau]$       
and so on for larger time intervals. This method is called the {\it method of steps}.

In a similar way to  ordinary differential equations (ODEs), 
the {\it characteristic equation}  for the homogeneous part of Eq. (\ref{eqn:forced1D}) is obtained by looking for solutions of the form $ce^{\lambda t}$, where $c$ is a constant and $\lambda$ is complex.
The scalar equation 
\begin{equation}\label{eqn:1Dhom}
\dot{y}=-ay(t)-by(t-\tau)
\end{equation}
has a nontrivial solution, $ce^{\lambda t}$, if and only if
\begin{equation}\label{eqn:1Dchar}
h(\lambda)\equiv\lambda+a+be^{-\lambda \tau}=0.
\end{equation}
Eq. (\ref{eqn:1Dchar}) is transcendental and thus the number of roots is infinite.
At the same time, because $h(\lambda)$ is an entire function,  the number of roots is finite 
within any compact region in the complex plane.
Because $a$, $b$ and $\tau$ are real, the roots must come in complex conjugate pairs. 
It can be shown \cite{HaleLunel1993} that the real part of each root is bounded. Moreover, for $a>|b|$ and for all $\tau>0$, 
$\re\lambda<0$. 
The latter is the necessary condition for the solution to Eq. (\ref{eqn:1Dhom}) to be stable.

The solution to the forced delay equation (\ref{eqn:forced1D}) is closely dependent on a particularly initialized solution of the homogeneous delay equation (\ref{eqn:1Dhom}), called the {\it fundamental solution}. 
This function, denoted by $Y(t)$, is defined as the solution of (\ref{eqn:1Dhom}) which satisfies the following initial condition 
\begin{equation}\label{initialcds}
Y(t) = \left\{ \begin{array}{rl}
 0, &\mbox{ $t<0$,} \\
 1, &\mbox{ $t=0$.}
 \end{array} \right.
\end{equation}     
For $0\leq t\leq\tau$,  
an exact expression for $Y(t)$  may be obtained using the 
{\it method of steps}.  
Substituting into  Eq. (\ref{eqn:variationconstants})   the initial conditions  given by Eq. (\ref{initialcds}) and setting $f=0$ gives
\begin{equation}\label{eqn:relationa_eigenval}
Y(t)=e^{-at}.
\end{equation}
For $t>\tau$,  an expression for $Y(t)$, obtained using the {\it method of steps}, 
is no longer useful.    
This is because the expression involves terms in powers of $t$ and thus for large values of $t$  
it is difficult to extract any insight into the behavior of $Y(t)$.
Using Laplace transforms, it is possible to express $Y(t)$ in terms of an infinite sum of eigenfunctions. 
Taking the Laplace transform of Eq. (\ref{eqn:1Dhom}) with initial conditions given by Eq. (\ref{initialcds}) leads to    
\begin{equation}\label{eqn:fundamental_aux}
\mathcal{L}(Y)(\lambda)\equiv\int_0^{\infty}e^{-\lambda t}Y(t)\,dt =h^{-1}(\lambda),
\end{equation} 
where $\mathcal{L}$ stands for Laplace transform.                          
Employing the inversion theorem,      
\begin{equation}\label{eqn:fundamental_formal}
Y(t)=\int_{(\gamma)}e^{\lambda t}h^{-1}(\lambda)d\lambda, \quad t>0,
\end{equation}
where 
$\int_{(\gamma)}\equiv\lim_{T\rightarrow\infty} \,\frac{1}{2\pi i}\,\int_{\gamma-iT}^{\gamma+iT}$
with  $\gamma>\text{max}\{\re\lambda:h(\lambda)=0\}$.   
Using the Cauchy residue theorem to integrate $e^{\lambda t}h^{-1}(\lambda)$ along a suitably chosen contour, $Y(t)$ can be expressed as an infinite series of eigenfunctions  
\begin{equation}\label{eqn:Y_fundamental_alt}
Y(t)=\sum_{j=1}^{\infty}
\underset{\lambda=\lambda_j}{\text{Res}}\phantom{x} e^{\lambda t}h^{-1}(\lambda),   \quad t>0,   
\end{equation}
that is uniformly convergent in $t$ (see \cite{Lunel1995}). 
Since the roots are either real or come in complex conjugate pairs, Eq. (\ref{eqn:Y_fundamental_alt}) can be re-written as
\begin{subequations}\label{eqn:Y_fundamental_alt23}
\begin{equation}\label{eqn:Y_fundamental_alt2}  
Y(t)=\lim\limits_{N\rightarrow\infty}{Y_N(t)}, \quad t>0,
\end{equation}  
with $Y_N(t)$ defined by    
\begin{equation}\label{eqn:Y_fundamental_alt3} 
Y_N(t)\equiv 
\sum_{\substack{j=1\\ \{\lambda_j^+:\,\im\lambda_{j}\geq 0\}}}^N 
P_j(\lambda_j^+,t) \,e^{\re\lambda_j^+ t},\quad t>0
\end{equation}  
where  $\lambda_j^+$ represents a root of (\ref{eqn:1Dchar}) with a positive or zero imaginary part satisfying $\re\lambda_j^+>\re\lambda_{j+1}^+$ for all $j$ with   
\begin{equation}\label{eqn:Pjlambdat}  
P_j(\lambda_j^+,t)=
2^{\mathcal{H}(\im\lambda_j^+)}\cos(\im\lambda_j^+t-\phi_j^+)|h'(\lambda_j^+)|^{-1},
\end{equation}
and
\begin{equation}\label{eqn:angle1}
\phi_j^+=\tan^{-1}\left(\frac{\im h'(\lambda_j^+)}{\re h'(\lambda_j^+)}\right).
\end{equation} 
$\mathcal{H}(x)$ is defined as   
\begin{equation}\label{eqn:H(x)_back2}
\mathcal{H}(x)= 	
 \begin{cases}
 \phantom{x}1,&  \text{if $x> 0$},\\ 
 \phantom{x}0,&  \text{if $x\leq  0$}.   
 \end{cases}
 \end{equation}
\end{subequations}  
Note that by a suitable choice of parameters, all roots of (\ref{eqn:1Dchar}) are distinct and thus $e^{\lambda t}h^{-1}(\lambda)$
only has simple poles.

It follows that for sufficiently large 
values of $t$, 
$Y(t)$ is dominated by its slowest decaying eigenfunction and thus 
\begin{equation} 
Y(t)\sim Y_1(t).
\end{equation}

\begin{figure}[t] 
\begin{minipage}{\linewidth}   	
\begin{minipage}{0.48\linewidth}
\centerline{\includegraphics[width=7cm]{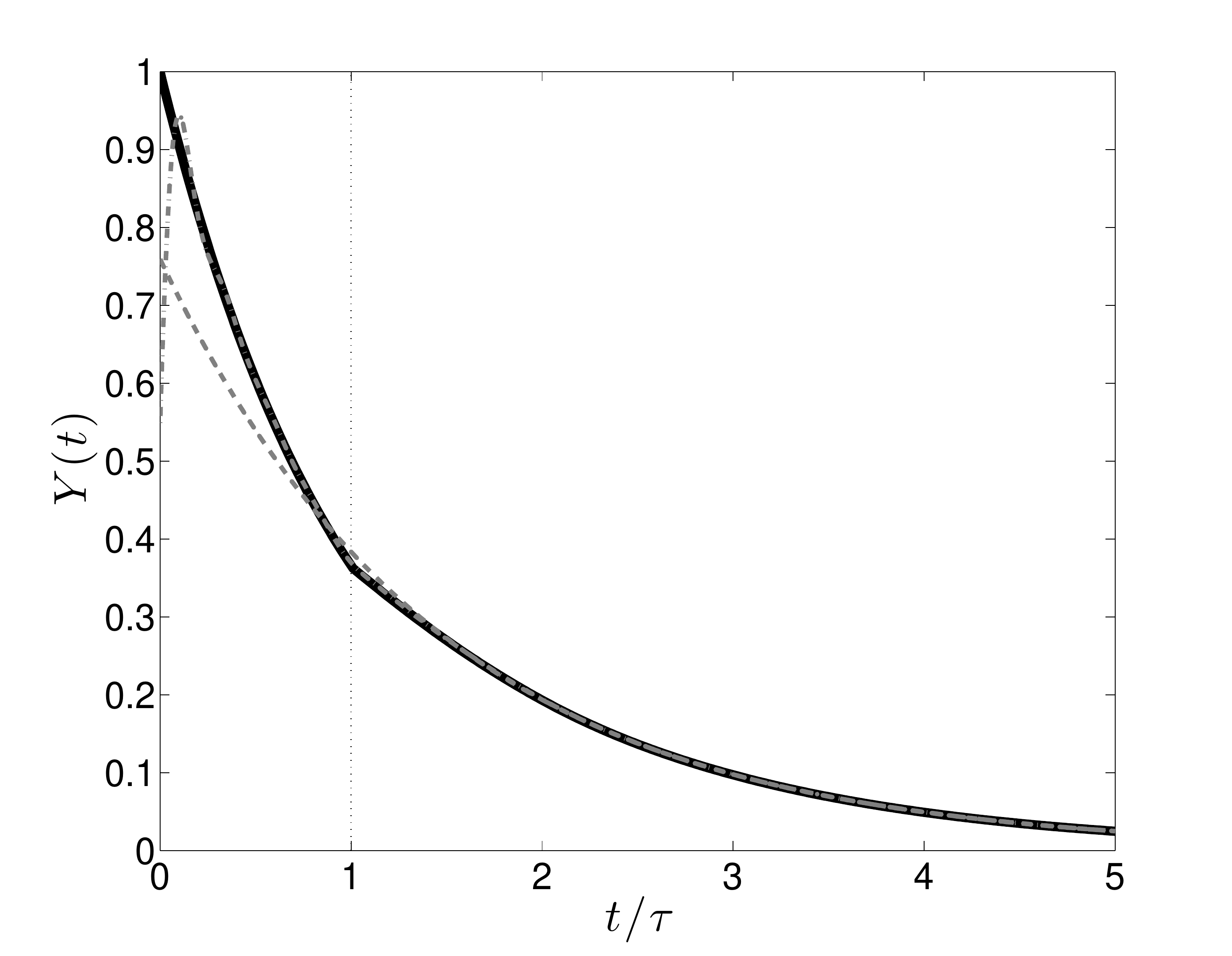}} 
\centerline{(a) $a=1$, $b=-0.16$, $\tau=1$} 
\end{minipage}
\hfill
\begin{minipage}{0.48\linewidth}
\centerline{\includegraphics[width=7cm]{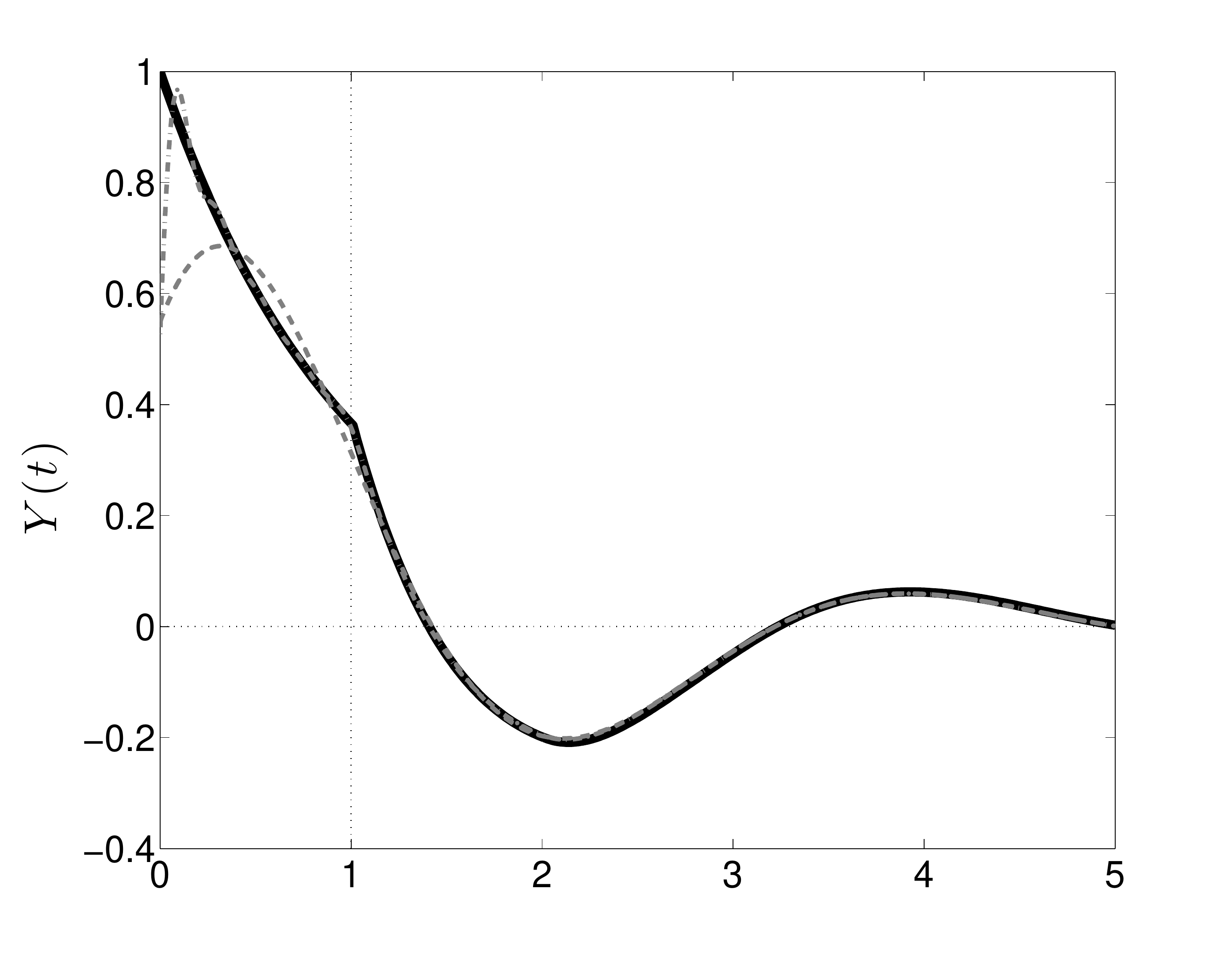}}
  \centerline{(b) $a=1$, $b=0.9$, $\tau=1$}
\end{minipage}
\end{minipage}        
\caption{The fundamental solution, $Y(t)$,  plotted as a function of $t/\tau$ (solid black).
Also plotted are  $Y_1(t)$ (dashed gray) and $Y_5(t)$ (dashed/dotted gray) (see Eq. (\ref{eqn:Y_fundamental_alt3})).
In both parameter sets $\re\lambda_1\simeq - 0.68$.}
\label{fig:SeriesFundamental1}
\end{figure}

$Y(t)$ is numerically determined and plotted for two sets of parameters $(a,b,\tau)$ in Fig. \ref{fig:SeriesFundamental1}.  
Both sets share the same $\re\lambda_1\simeq-0.68$; the difference is that 
$\lambda_1$ is real in Fig. \ref{fig:SeriesFundamental1}(a) and imaginary  in Fig. \ref{fig:SeriesFundamental1}(b).
Also plotted in Fig. \ref{fig:SeriesFundamental1} are the functions $Y_1(t)$ and $Y_5(t)$.
The roots of the  characteristic equation are determined using the DDE-BIFTOOL \cite{DDE-BIFTOOL}.
In both cases,  $Y_1(t)$, is found to be in good agreement with $Y(t)$ for $t\gtrsim\tau$.
This indicates that within this period, 
the remaining eigenfunctions have decayed sufficiently for $Y_1(t)$ to dominate the behavior of $Y(t)$. However for $0\leq t \leq \tau$,   
its dominant behavior depends on the contribution of many eigenfunctions, the number of which increases as $t\rightarrow 0$.  
Instead,  one needs to refer to Eq. (\ref{eqn:relationa_eigenval}).

\begin{figure}[!]
\begin{minipage}{\linewidth}
\begin{minipage}{0.48\linewidth}  
\centerline{\includegraphics[width=7cm]{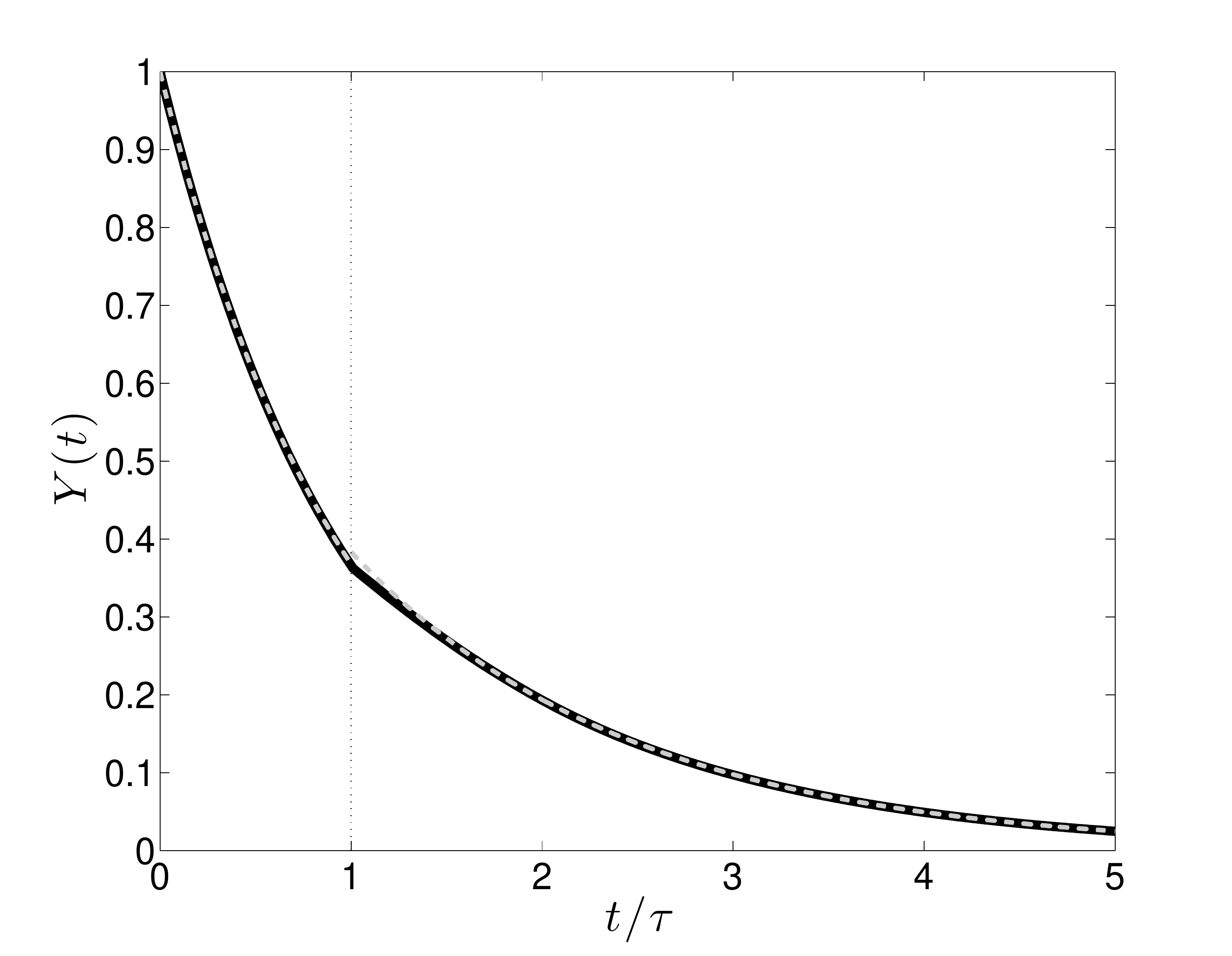}} 
\centerline{(a) $a=1$, $b=-0.16$, $\tau=1$} 
\end{minipage}
\hfill
\begin{minipage}{0.48\linewidth}
\centerline{\includegraphics[width=7cm]{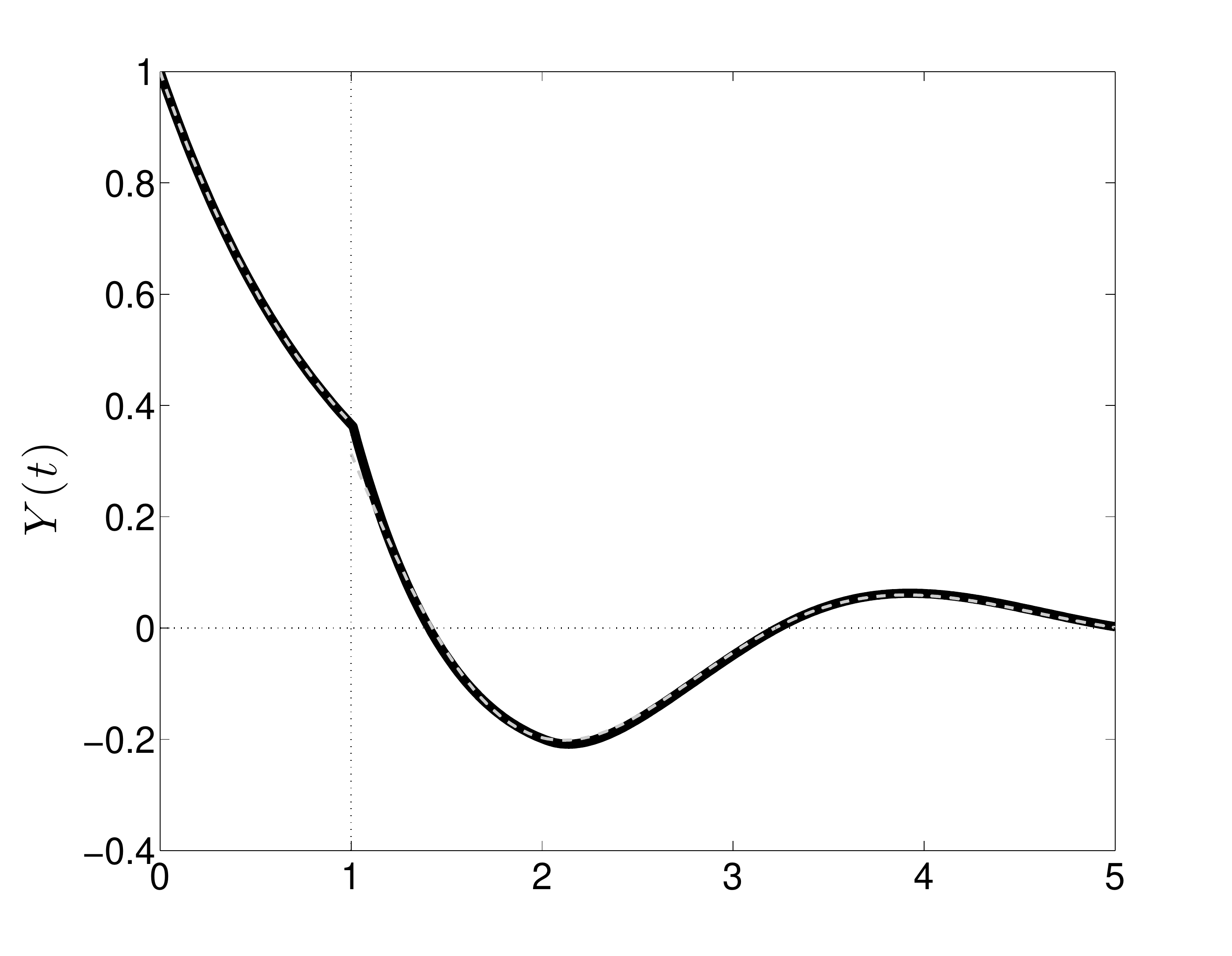}}
\centerline{(b) $a=1$, $b=0.9$, $\tau=1$}
\end{minipage}
\end{minipage}        
\caption{Same as Fig. \ref{fig:SeriesFundamental1} but this time the fundamental solution is compared to expression (\ref{eqn:YLongShort}).
$Y_1(t)$ is plotted (dashed gray) for $t>\tau$ and $e^{-at}$ for $0\leq t\leq \tau$ (dashed gray).
}
\label{fig:SeriesFundamental2}
\end{figure}

The above can be summarized into the following expression for the fundamental solution:
\begin{equation}\label{eqn:YLongShort}
Y(t) = \left\{ \begin{array}{rll}
 &e^{-at},  &\mbox{ $0\leq t\leq\tau$,} \\
\sim& Y_1(t) , &\mbox{ $t>\tau$.}
 \end{array} \right.
\end{equation}                                                                                      
The validity of expression (\ref{eqn:YLongShort}) is clearly depicted in Fig. \ref{fig:SeriesFundamental2} where it is plotted and compared to
$Y(t)$ for the two sets of parameters already shown in Fig. \ref{fig:SeriesFundamental1}.

Notice the central difference between the fundamental solution of an ODE 
and a DDE.
While in the former case, the behavior of the fundamental solution remains unaltered at all times,
in the latter case, a distinct transition takes place at $t=\tau$.            
At the same time,  for $t\leq \tau$, 
the fundamental solution of a DDE is identical to the fundamental solution of the ODE that is obtained by omitting from the DDE  the terms that contain a delay time i.e. equivalent to setting $b=0$ in Eq. (\ref{eqn:forced1D}).

The reason for which so much attention is given to the fundamental solution is  
that the general solution to the  forced delay equation (\ref{eqn:forced1D}) can be expressed in terms of it. 
To see this, consider the Laplace transform of Eq. (\ref{eqn:forced1D}) with initial conditions   given by (\ref{eqn:initialcd}).
Provided that the forcing $f(t)$  is exponentially bounded, 
\begin{equation}\label{eqn:variation_aux} 
 \begin{split}
 h(\lambda)\,\mathcal{L}(y)(\lambda)=\phi(0)
-be^{-\lambda\tau}&\int_{-\tau}^0 e^{-\lambda\theta}\phi(\theta)\,d\theta  \\
+&\int_0^\infty e^{-\lambda t}f(t)dt.  
\end{split}
\end{equation}	                    
Use of the convolution and inversion theorems leads to the following expression for the general solution 
\begin{subequations}\label{eqn:variation_tot} 
\begin{equation}\label{eqn:variation}
y(\phi,f)(t)=y(\phi,0)(t)+\int_{0}^{t}Y(t-t')f(t')\,dt',
\end{equation}
where $y(\phi,0)(t)$ represents the solution to the (unforced) homogeneous delay equation (\ref{eqn:1Dhom}) and is given by
\begin{equation}\label{eqn:variation_homogeneous}   
y(\phi,0)(t)=Y(t)\phi (0)-b\int_{-\tau}^{0}Y(t-\theta-\tau)\phi (\theta)\,d\theta.	
\end{equation}	  
\end{subequations}
Because of its similarity to ordinary differential equations, 
the representation of $y(\phi,f)(t)$ in this form is often referred \cite{HaleLunel1993} to as the variation of constants formula.
Using this representation, it is easily deduced that the solution of any, either homogeneous or forced, linear delay equation is governed by its fundamental solution with the roots of the characteristic equation controlling its asymptotic behavior.

\subsection{Scaling Behavior}
Having presented some basic properties concerning linear DDEs, the next objective is to consider their coupling to a chaotic advection flow.
For a chemical system satisfying Eq. (\ref{eqn:1Ddelay}), 
the evolution of the  chemical difference between a pair of fluid parcels can be obtained by
simultaneously linearizing  
the chemical (\ref{eqn:1Ddelay}) and trajectory (\ref{eqn:traj}) evolution Eqs. around a fluid parcel.                                     
Using the variation of constants formula (\ref{eqn:variation_tot}),  
\begin{equation}\label{eqn:chemicaldifference_delay}  
\begin{split}
\delta C(t)= 
Y(t)\delta C(0)
&-b \int_{-\tau}^0 Y(t-\theta-\tau)\, \delta C(\theta)\,d\theta \\
&+\phantom{b}\int_0^t Y(t-t') \, \left(\frac{\partial C_0}{\partial \bm{X}}\cdot \delta\bm{X}(t')\right) \,dt',   
\end{split}                                                                                                    
\end{equation}                                                                                                 
where   $\{\delta\bm{X}(t);\bm{X}(t)\}$, the label on the fluid parcel difference, has been suppressed for brevity. 
For $t\in [-\tau,0]$,  
$\delta C(t)=\phi(t)$ where $\phi(t)$ is a prescribed initial function.

To analyze the scaling behavior of the delay scalar field  
at statistical equilibrium,
the long-time limit of Eq. (\ref{eqn:chemicaldifference_delay}) needs to be considered.
A useful property for $Y(t)$ is that it is bounded with $|Y(t)|<K\exp[\re\lambda_1t]$ where $K>0$ 
(see \cite{HaleLunel1993}).  
We impose that 
$a>|b|$, thus ensuring that $\re\lambda_1<0$ for all $\tau>0$ (see \S\ref{subsec:KeyProperties}). 
It follows that in the long-time limit,
the first two terms that describe the evolution of the initial conditions vanish.  
Note that this is not the case for either marginally stable ($\re\lambda_1=0$) or unstable 
($\re\lambda_1>0$) chemical dynamics.    

At the same time, since the source depends smoothly on space, its spatial derivatives do not increase or decrease in a systematic way.
Thus,  the evolution of $|\delta_{\bm{X}}C_0(t')|$ is closely related to the evolution of the
separation between the pair of  fluid parcels  i.e. 
$\delta_{\bm{X}} C_0(t')=\frac{\partial C_0}{\partial\bm{X}}\cdot\delta\bm{X}(t')\sim  |\delta \bm{X}(t')|$. 
To obtain an expression for $|\delta\bm{X}(t')|$ in terms of $|\delta\bm{X}(t)|$, 
Eq. (\ref{eqn:traj}) is linearized around $\bm{X}(t')$ from where it can be deduced that  for $t>t'$,
\begin{subequations}
\begin{equation}
\delta\bm{X}(t')=\bm{N}(t',t) \delta \bm{X}(t),
\end{equation} 
with 
\begin{equation}
\bm{N}(t',t)=\exp\left[\int_{t}^{t'}\frac{\partial \bm{v}}{\partial\bm{X}}ds\right],
\end{equation}                                        
\end{subequations}
where $|\delta\bm{X}(t)|$ is considered to be much less than the characteristic length scale of the 
velocity field, $L$, where here $L=1$.  Consequently, the evolution of $|\delta \bm{X}(t')|$ is dictated by $\bm{N}^T\bm{N}(t',t)$ once calculated  along the fluid parcel trajectory in backward time. 
Because  $\bm{N}^T\bm{N}$ is a real, non-negative symmetric matrix, its eigenvalues are positive. 
Therefore, depending on its orientation at time $t$, 
as time $t'$ decreases $|\delta \bm{X}(t')|$ increases  or decreases exponentially  
according to a set of rates whose number equals the dimension of the flow and whose values depend on the eigenvalues of $\bm{N}^T\bm{N}$.    
                      
In the limit of $t-t'\rightarrow\infty$, these rates are defined as the Lyapunov exponents \cite{Ottino1989,Ott1993}.
For a two-dimensional, incompressible flow 
that is both ergodic and hyperbolic, all trajectories share the same set of Lyapunov exponents 
$\{h_0,-h_0\}$ with
$h_0>0$.
It follows that for almost all orientations at time $t$, 
the typical separation between a pair of neighboring fluid parcels increases exponentially in backward time at a rate given by the flow Lyapunov exponent $h_0$ with 
$|\delta \bm{X}(t')|\sim |\delta\bm{X}(t)|\exp[h_0(t-t')]$.

The exponential increase of $|\delta \bm{X} (t)|$ 
can only be valid for the time period for which its length remains considerably less than the characteristic length scale of the velocity field.
This is because for larger length scales ($\gtrsim 0.1$), 
linearizing the trajectory Eq. (\ref{eqn:traj}) 
is no longer valid.
For these larger length scales, 
finite-size effects become important and the value of $|\delta\bm{X}(t)|$ saturates at the length of the characteristic length scale of the velocity field. 
The time it takes for  $|\delta \bm{X}(t)|$ to saturate in backward evolving time
is here referred to as the {\it stir-down time} and is denoted by 
$T_{\delta X}$.
By choosing $|\delta \bm{X}(t)|$ to be sufficiently small,
an approximate expression for $T_{\delta X}$ is given by 
\begin{equation}\label{eqn:stirring}
T_{\delta X}=\frac{1}{h_0}\log(1/|\delta \bm{X}|),\quad \text{for $|\delta \bm{X}| \ll 1$}. 
\end{equation}                                                                     
It follows that qualitatively, 
the evolution of a typical separation between two fluid parcels
can be divided into two parts: 
the first one corresponding to the period that it 
exponentially increases  
and 
the second one to the rest of the time during which its value remains saturated.
Therefore,     
\begin{equation}\label{eqn:distance_asymptotic}
|\delta \bm{X} (t')| 
\sim    
\begin{cases}
\,|\delta \bm{X} (t)| \, e^{h_0 (t-t')}, & \text{ for $0<t-t'\leq T_{\delta X}$},\\ 
\,1, & \text{ for $t-t'>T_{\delta X}$}.
\end{cases} 
\end{equation}

The asymptotic behavior of the chemical difference between any two fluid parcels,
and thus from (\ref{eqn:LagrangianEulerian2}), between any two neighboring points, may be deduced by substituting expression  (\ref{eqn:distance_asymptotic}) into Eq. (\ref{eqn:chemicaldifference_delay}) 
(replacing $\bm{X}$ with $\bm{x}$ and $\delta\bm{X}$ with $\delta\bm{x}$). 
After making a change of variables from $t'$ to $\Delta t=t-t'$ and taking the limit of $t\rightarrow\infty$, the small-scale behavior ($|\delta \bm{x}\ll 1|$) is given by
\begin{equation}\label{eqn:delayconcdiff} 	
\delta c_\infty(\delta\bm{x})\sim 
\int_0^\infty Y(\Delta t)\,\text{min}\{|\delta \bm{x}|e^{h_0 \Delta t},1\}\, d\Delta t
\end{equation}                                                                                         
where a number of space- and time- factors are omitted since they do not affect the scaling laws.   
Note that within the approximation made here, the rate of exponential increase of the separation between fluid parcels, $h_0$,
is taken to be independent of the individual trajectories and
therefore the dependence of $\delta c_\infty$ on $\bm{x}$ is dropped.
In reality, this rate will depend on the trajectory thus modifying the average scaling behavior of the field. (See \cite{Neufeld_etal2000a} for discussion of the implication of this 
for a linearly decaying reactive scalar. The extension of this discussion for the delay case is left for future work.)

\subsection*{Transition length scale}
An expression equivalent to (\ref{eqn:delayconcdiff}) was obtained by \cite{Neufeld_etal1999,Hernandez-Garcia_etal2002} in the context of an ordinary reactive scalar whose reactions involve no delay time. 
In both cases, delay and ordinary, the  asymptotic behavior of the concentration field is governed by 
the convolution in time of the fundamental solution associated with the chemical subsystem 
with the separation between fluid parcels.
However, a fundamental difference between the delay reactive scalar and the ordinary reactive scalar will significantly affect the asymptotic scaling behavior of the delay scalar field and modify it with respect to the scaling behavior of the ordinary reactive scalar.
This difference lies in  the fundamental solution.

As discussed in \S\ref{subsec:KeyProperties}, the behavior of $Y(t)$  associated with a linear DDE is distinctly different depending on whether $t/\tau$ is larger than or less than $1$.
It follows that  
the asymptotic behavior of $\delta c_\infty(\delta\bm{x})$  must differ according to whether   
$T_{\delta x}/\tau$ is larger than or less than $1$.                                    
Since the value of $T_{\delta x}$ depends on  $|\delta \bm{x}|$, 
this transition must  occur
at a certain length scale,  denoted by $\delta x_c$, 
here named the {\it transition  length scale}. 
An approximate expression  for $\delta x_c$ may be obtained by considering
the value of  $|\delta \bm{x}|$ for which
\begin{align} 
T_{\delta x_c}&\sim\tau,\\	
\intertext{from where it can be deduced that  $\delta x_c$  must then approximately be equal to}
\delta x_c&\sim e^{-h_0\tau}.\label{eqn:characteristiclength scale}	
\end{align}    
Thus, the magnitude of the transition length scale is controlled by the product of the delay time with the flow Lyapunov exponent  
while it is independent of the parameter details of the  reactions.    
Expression (\ref{eqn:characteristiclength scale}) represents the first key theoretical result of the paper. 

\subsection*{Scaling regimes}
The scaling behavior of the field is now separately examined for length scales less than and larger than the transition length scale.  
A good way to gain insight into this behavior   
is to consider the absolute value of the integrand of Eq. (\ref{eqn:delayconcdiff}). 
The first function has an exponential decay (perhaps oscillatory) while the second function initially increases exponentially and then saturates. Thus the absolute value of the integrand has a distinct maximum and the dominant contribution to the integral comes from the neighborhood of this maximum. The corresponding dependence of the integral on the value of $\delta x$  implies up to three possible scaling regimes (depending on $\delta x$ and on the other parameters in the problem).

\begin{figure}[!]
\centering 
\begin{minipage}{\linewidth}   
\centerline{$|Y(t)\,\text{min}\{|\delta \bm{x}| e^{h_0 t},1\}|$}    
\centerline{\includegraphics[width=6cm]{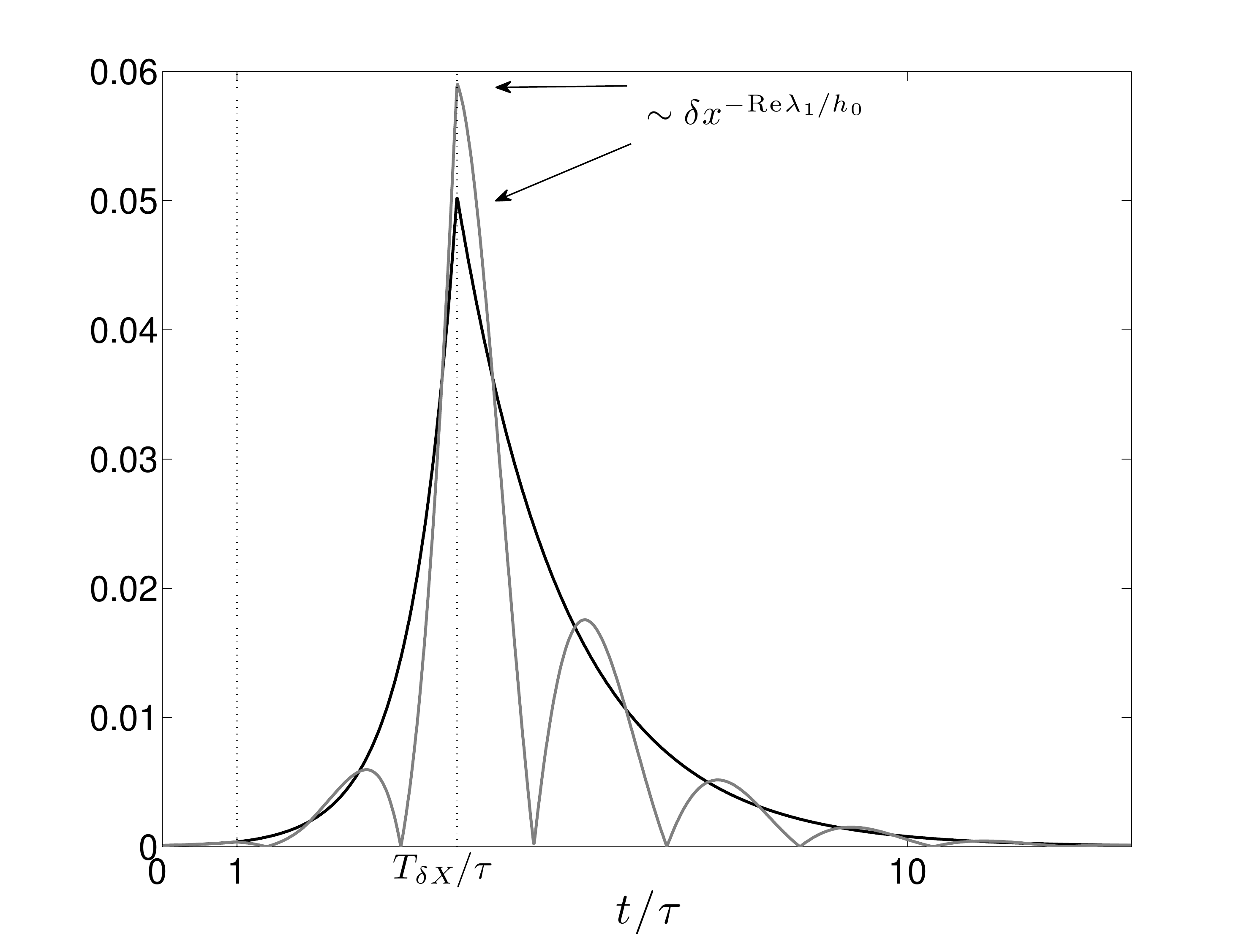}}
\centerline{(a) $T_{\delta X}>\tau$}   
\end{minipage}
\vfill   
\begin{minipage}{\linewidth}
\centerline{\includegraphics[width=6cm]{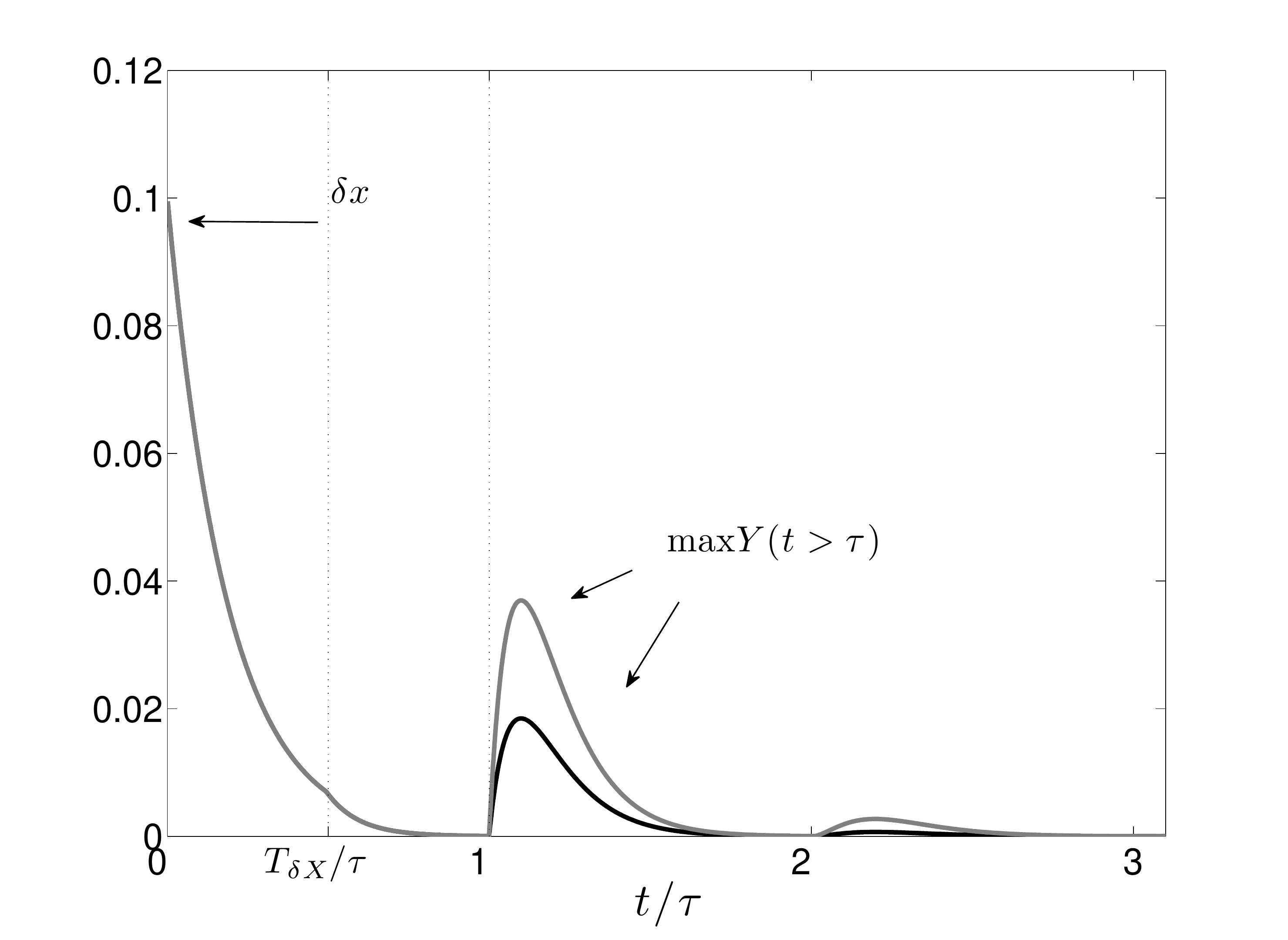}}
\centerline{(b) $T_{\delta X}<\tau, \; |b|/ae \lesssim \delta x_c$}
\end{minipage}
\vfill
\begin{minipage}{\linewidth}
\centerline{\includegraphics[width=6cm]{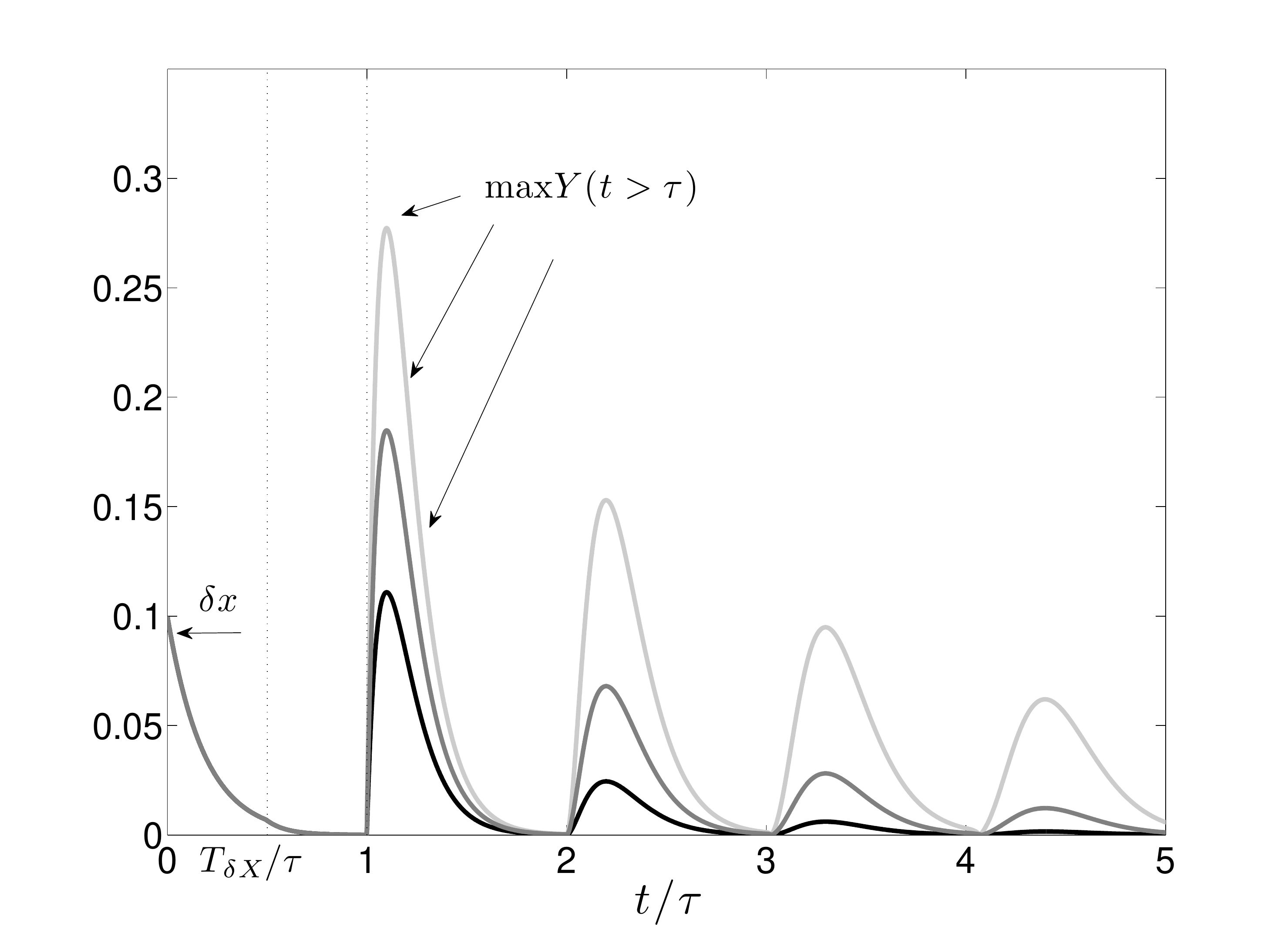}}
\centerline{(c) $T_{\delta X}<\tau, \; |b|/(ae) \gg \delta x_c$}
\end{minipage}
\vfill
\caption{(a) $|Y(t)\,\text{min}\{\delta x e^{h_0 t},1\}|$ 
 plotted for  $\delta X=10^{-4}$ ($T_{\delta X}\approx 4\tau$) 
 for the two sets of parameters $(a,b,\tau)$ previously considered in Fig. \ref{fig:SeriesFundamental1}: 
$(1,-0.16,1)$ (black) and $(1, 0.9,1)$ (gray). 
(b) The same as (a) this time $(1,0.05,10)$ (black) and $(1, 0.1, 10)$ (gray)
with $\delta X=10^{-1}$ so that $T_{\delta X}\approx \tau/2$. 
(c) The same as (b) this time $(1,0.3,10)$ (black), $(1, 0.5, 10)$ (dark gray)
and $(1, 0.75, 10)$ (light gray).  
} 
\label{fig:Product}
\end{figure}

 \subsection*{Regime I \qquad $\bm{|\delta x|<\delta x_c}$}
The first scaling regime, Regime I, concerns length scales that are smaller than $\delta x_c$.
For these length scales, the stir-down times are  larger than the delay time and thus  the chemical dynamics converge at a  rate
which, for the linear case considered here, is exactly given by $-\re\lambda_1$ (see expressions (\ref{eqn:Y_fundamental_alt3}) and (\ref{eqn:YLongShort}) where the `+' sign is omitted since $\re\lambda_1=\re\lambda_1^+$). 
In analogy to the flow Lyapunov exponent that controls the strength of the flow dynamics, 
this rate is called the chemical Lyapunov exponent \cite{Neufeld_etal1999}. 

It therefore follows that within this regime, 
the scaling behavior of the delay reactive scalar field 
is no different to the scaling behavior of an ordinary reactive scalar. 
For both delay and ordinary scalars, the small-scale structure is controlled by 
the relative strength of the chemical to the flow dynamics:   
If $-\re\lambda_1/h_0<1$, the chemical processes 
are too slow to forget the different spatial 
histories experienced by the fluid parcels. 
In this case, the maximum of  $|Y(t)\,\text{min}\{|\delta \bm{x}|e^{h_0 t},1\}|$ occurs at $t=T_{\delta x}$ (see Fig. \ref{fig:Product}(a)) 
and its value depends on
$|\delta \bm{x}|^{-\re\lambda_1/h_0}$.
Thus, in this case, the field's 
spatial structure is 
filamental
i.e. non-differentiable in every direction except the direction along which the filaments grow \cite{Neufeld_etal1999}.
On the other hand, for $-\re\lambda_1/h_0>1$ the chemical processes converge faster
to their equilibrium value than the trajectories diverge from each other.
The maximum of  $|Y(t)\,\text{min}\{|\delta \bm{x}|e^{h_0 t},1\}|$ occurs at $t=0$
from where it can be deduced that the field's structure is everywhere smooth.
Thus, the H\"older exponent within Regime I
is  equal  to $\gamma_1=\text{min}\{-\re\lambda_1/h_0,1\}$.

\subsection*{Regimes II \& III\qquad $\bm{|\delta x|> \delta x_c}$}  
Consider now length scales that are larger than $\delta x_c$.
The corresponding stir down times are smaller than the delay time and thus the chemical dynamics converge at a rate given by $-a$, i.e. the decay rate obtained once the delay term is ignored  
(see expression (\ref{eqn:YLongShort})).

There exist two local maxima for  $|Y(t)\,\text{min}\{|\delta \bm{x}|e^{h_0 t},1\}|$; the first one  is scale-dependent, the second one is a constant (see Figs. \ref{fig:Product}(b-c)). 
The value of the first local maximum is given by
$\underset{t}{\text{max}}\,|e^{-at}\text{min}\,\{|\delta \bm{x}|e^{h_0 t},1\}|=|\delta \bm{x}|^{\text{min}\{a/h_0,1\}}$ 
(where  $Y(t\leq \tau)=e^{-a t}$ was employed (see  expression (\ref{eqn:YLongShort})).
It therefore follows that if this first local maximum is a  global maximum, the field's scaling behavior is  described by a H\"older exponent that satisfies $\gamma_2=\text{min}\{a/h_0,1\}$.
This scaling regime is denoted by Regime II. 
Now focus on the second local maximum which is  given by $\text{max}\,|Y(t\geq\tau)|$. 
Since this is a constant, a flat scaling regime will ensue if the second local maximum is larger than the first local maximum. This scaling regime is denoted by Regime III. However, if the second local maximum is smaller than the first local maximum, Regime III does not appear.
 
To investigate the range of length scales for which Regime III appears, consider in more detail $\text{max}\,|Y(t\geq\tau)|$. Now $|Y(t\geq\tau)|$ has a maximum at either $t=\tau$ 
or at some $t=t^\ast$, where $t^\ast$ is defined as 
the value of $t$ for which $dY(t)/dt$ is first equal to $0$ and thus 
$aY(t^\ast)=-bY(t-t\ast)$.
First consider the local maximum of $|Y(t\geq\tau)|$ to occur for $\tau<t^\star\leq 2\tau$.
Within this time period and using 
the method of steps (see \S \ref{subsec:KeyProperties}), 
$Y(t)$ can be exactly expressed as $Y(t)=e^{-a t}-b(t-\tau)e^{-a(t-\tau)}$.  
Combining this expression with  expression (\ref{eqn:relationa_eigenval}) 
for $Y(t)$ for $0<t\leq \tau$, $t^\star$ must satisfy
$e^{-a t^\star}-b(t^\star-\tau)e^{-a(t^\star-\tau)}=-b/ae^{-a(t^\star-\tau)}$  from where 
we can deduce that a local maximum of $|Y(t\geq\tau)|$ occurs if
$0<1/a+1/be^{-a\tau}<\tau$. 
In this case, 
$\text{max}\,|Y(t>\tau)|=|b|/a\,e^{-1-a/be^{-a\tau}}$. 
Now consider $t^\star>2\tau$. In this case $Y(t\leq 2\tau)$ is monotonically decreasing and thus 
$\text{max}\,|Y(\tau\leq t\leq 2\tau)|=e^{-a\tau}$. 
 Since 
$|Y(t^\ast)|=|b|/a|Y(t-t\ast)|$,  
 $\text{max}\,|Y(t\geq 2\tau)|\leq |b|/a\,e^{-a\tau}$ and therefore $\text{max}\,|Y(t\geq \tau)|=e^{-a\tau}$. Finally, if no $t^\star$ exists, then again $\text{max}|Y(t\geq \tau)|=e^{-a\tau}$. 
All three cases can be summarized by  
\begin{equation}
\label{secondmax}
\text{max}|Y(t\geq\tau)|=\text{max}\{e^{-a\tau},\frac{|b|}{a}e^{-1-a/be^{-a\tau}}\}. 
\end{equation}

Comparing the value of the first local maximum, $|\delta \bm{x}|^{\text{min}\{a/h_0,1\}}$, with the value of the second local maximum, given by Eq. (\ref{secondmax}), we are able to obtain the following  estimate for $\delta x_2$, the length scale that separates  Regime II and III:
\begin{equation}\label{eqn:estimatedx2}
\delta x_2\sim\text{max}\,|Y(t\geq\tau)|^{\text{max}\{h_0/a,1\}}.
\end{equation}
The following points can be noted: 
\begin{enumerate}
\item If $|b|/(ae)\ll \delta x_c$, $\delta x_2\ll\delta x_c$ and thus there appears no Regime III. 
\item If $|b|/(a e) \sim 1$,  $\delta x_2\gg\delta x_c$ and thus 
Regime III will ensue at all length scales larger than $\delta x_c$. 
\end{enumerate}

\subsection*{H\"older Exponents} 
To summarize, the following set of scaling laws describe the 
spatial structure of the stationary-state delay reactive scalar field 
as  $|\delta \bm{x}|$ varies:
\begin{subequations}\label{eqn:delay_holders}     
\begin{equation}\label{eqn:corrected_char_length} 
	|\delta c_\infty(\delta\bm{x})|\sim
	\begin{cases} 
	|\delta \bm{x}|^{\gamma_1}, &\;\text{for $|\delta \bm{x}|<\delta x_c$}\\ 
		\text{flat},  &\;\text{for $\delta x_c<|\delta \bm{x}|<\text{max}\{\delta x_2,\delta x_c\}$}\\
  	|\delta \bm{x}|^{\gamma_2}, &\;\text{for $|\delta \bm{x}|>\text{max}\{\delta x_2,\delta x_c\}$}
		\end{cases}
\end{equation}
where the H\"older exponents $\gamma_1$ and $\gamma_2$ are given by
\begin{align}
\gamma_1&=\text{min}\{1,-\re\lambda_1/h_0\}, \label{eqn:Holder1}\\ 
\gamma_2&=\text{min}\{1,a/h_0\} \label{eqn:Holder2}.    
\end{align}	
\end{subequations}
Therefore, Regime II occurs for $|\delta \bm{x}|>\delta x_2$ and Regime III for 
 $\delta x_2>|\delta \bm{x}|>\delta x_c$. It happens that Regime III will not be present if $\delta x_2$ and $\delta x_c$ are not well separated. Similarly, Regime II will not be present if $\delta x_2$ is not sufficiently small.

Expression (\ref{eqn:delay_holders}) represents the second key theoretical result of this paper. 
The more general case for which several interacting chemical species are present is shown in the Appendix to be a slight variant of this expression. 
Special cases for which the species are not symmetrically coupled with each other may give rise to structures that are characterized by different H\"older exponents for different species.  Such a case is the delay plankton model whose behavior is examined in \S\ref{subsec:numerics_delayplanktonmodel}.

\section{Numerical Results: Two Examples}\label{sec:DelayNumericalResults}
To complement the theoretical results obtained in the previous section, a set of numerical simulations are here performed, 
firstly for the single linear  delay reactive scalar whose evolution within a fluid parcel 
was introduced in Eq.  (\ref{eqn:1Ddelay}),
and secondly for the delay plankton model that \cite{Abraham1998} first used for his numerical investigations.  
This model, shortly to be described, serves not only as a test-bench of the theory presented in Sec. \ref{sec:DelayTheory}
but also as an interesting application of it.

In both examples, the fluid parcels are 
advected by a model strain flow  
whose velocity field is
given by
\begin{equation}\label{eqn:v}
\bm{v}(\bm{x},t)= \left[ \begin{array}{rl}
 -\displaystyle{\frac{2 }{T}}\Theta(T/2-t\mod T)\cos(2\pi y+\phi) \\[0.2 cm]
 -\displaystyle{\frac{2 }{T}}\Theta(t\mod T-T/2)\cos(2\pi x+\theta) 
 \end{array} \right],
\end{equation}  
where $\Theta(t)$ is the Heaviside step function defined to be equal to unity for $t\geq0$ and zero otherwise 
and  $x$ and $y$ are the domain's horizontal and vertical axis respectively.     
The phase angles $\theta$ and $\phi$ change randomly at each period $T$, 
varying the directions of expansion and contraction and hence ensuring that all parts of the flow are equally mixed \cite{Bohr_etal1998,Ott1993}.     
Variation of $T$ has an effect on the magnitude of the flow Lyapunov exponent, $h_0$, 
without changing the shape of the trajectories and the spatial structure of the flow.
It may be shown that $h_0$ is inversely proportional to $T$ with 
\begin{equation}\label{eqn:Lyapunov_T}
h_0\approx 2.33/T, 
\end{equation}    
where the constant is numerically determined.

A large-scale inhomogeneity is injected into the system by introducing a spatially smooth forcing
\begin{equation}\label{eqn:force}
C_0(\bm{x})=1-1/2\cos[2\pi(x+y)]),
\end{equation}
oriented along the diagonal of the domain to avoid having the same preferred alignment to the flow. 
The space-dependence of the force couples the reaction dynamics with  the flow  dynamics  
and results in the formation of complex spatial patterns. 

A statistical steady state is reached after approximately $20T$. 
To reconstruct the stationary distributions of the corresponding  reactive scalars,
an ensemble of fluid parcels whose  final positions are fixed onto a grid 
are followed.
Using Eq. (\ref{eqn:v}), the parcels are tracked backwards in time up to a point when their initial concentrations are known. 
Thereafter, knowing their trajectory, their final concentration is determined by integrating the reaction equations forward in time using a second order Runge-Kutta method.
This way, 
to obtain the concentration fields along a one-dimensional transect, it is not necessary to determine the whole two-dimensional field.
The absence of interpolation permits greater accuracy at smaller length scales.
The initial concentrations are chosen to be equal to their mean equilibrium values, though as long as the reaction dynamics are stable, 
the final result should be independent of this choice.

\begin{figure}[!]     
\centering    
\begin{minipage}{\linewidth}       
\centerline{\includegraphics[width=6cm]{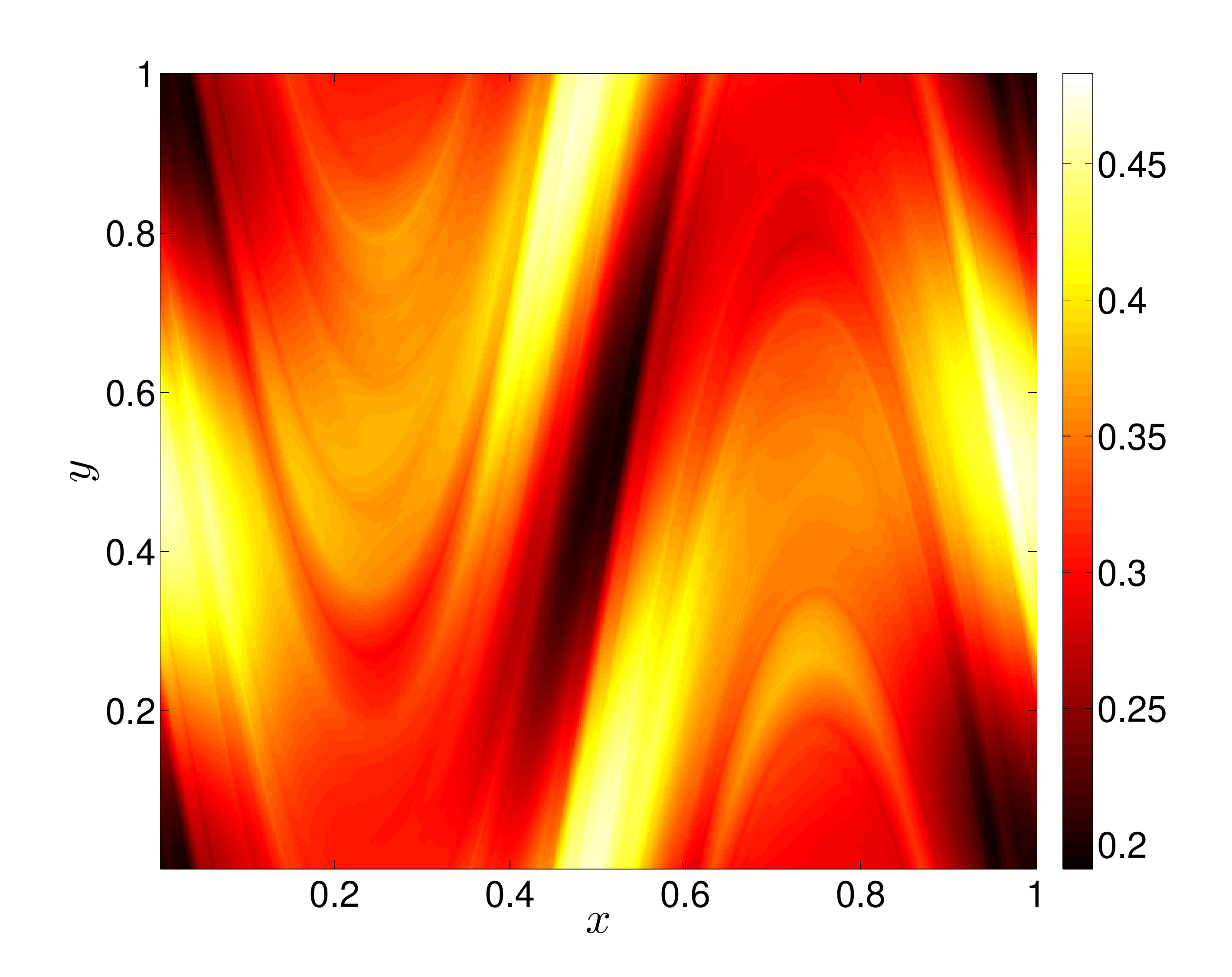}}      
\centerline{(a) $\tau=0$}    
\end{minipage}
\vfill    
\begin{minipage}{\linewidth} 
\centerline{\includegraphics[width=6cm]{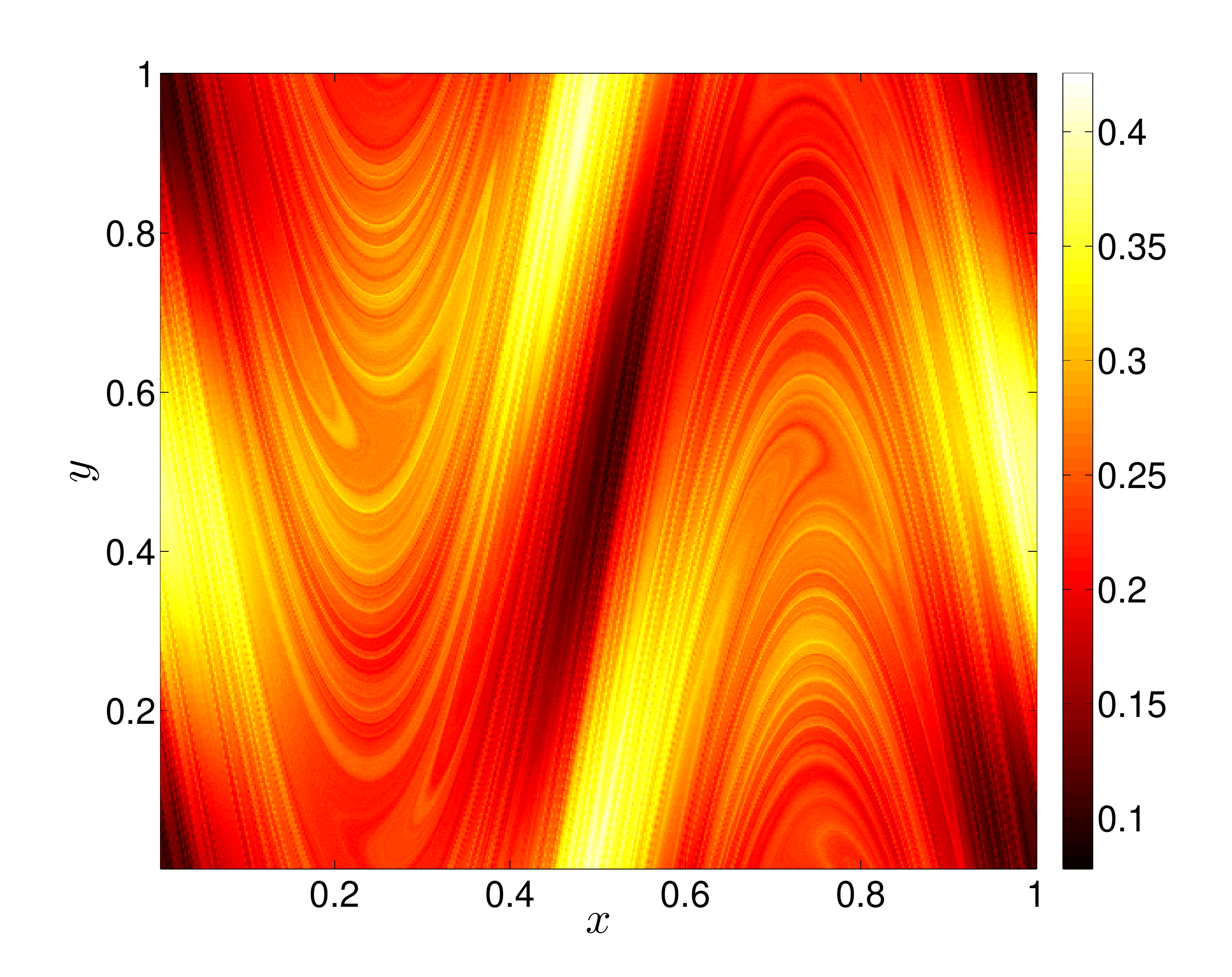}}       
\centerline{(b) $\tau=1$}   
\end{minipage} 
\vfill
\begin{minipage}{\linewidth}
\centerline{\includegraphics[width=6cm]{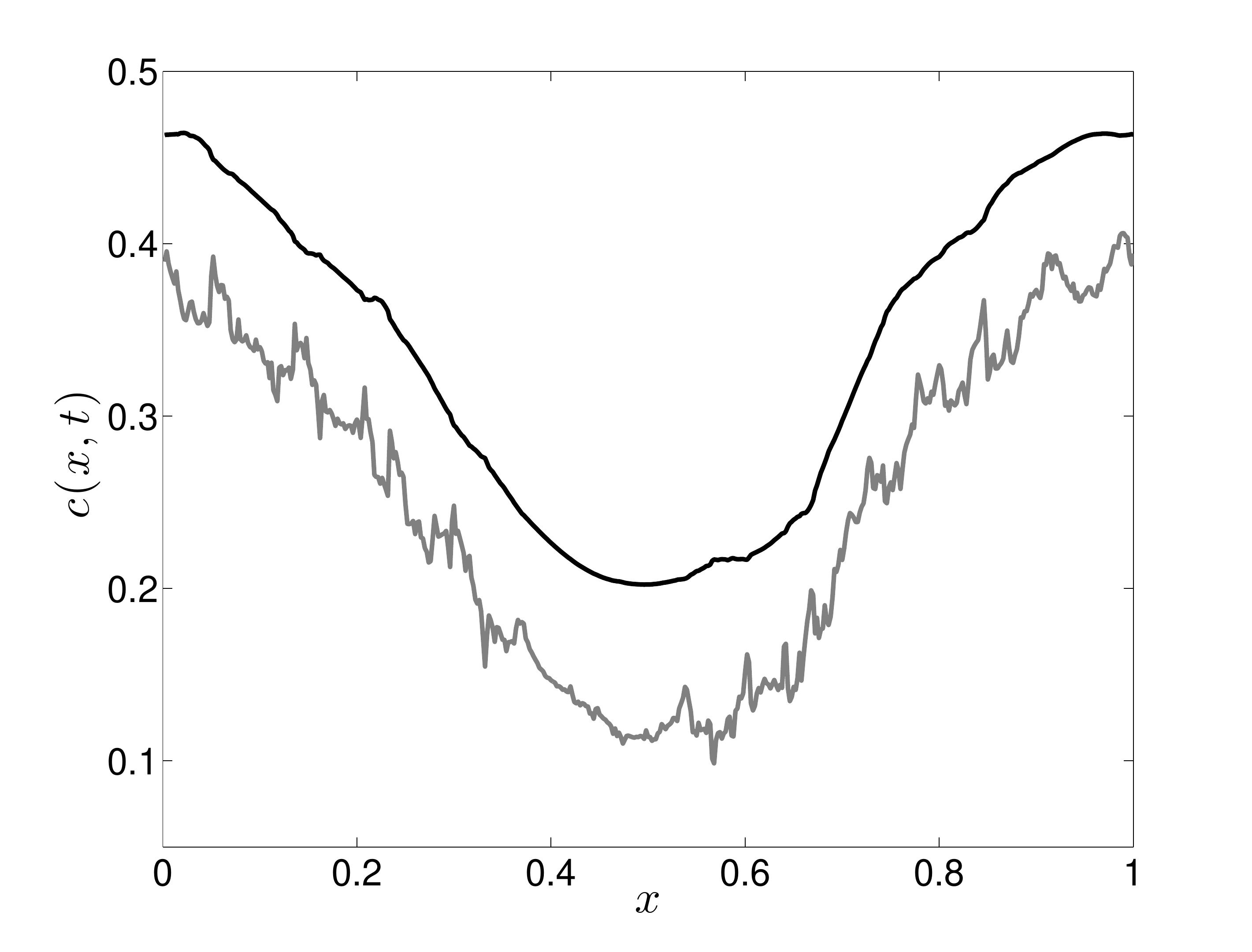}}  	 
\centerline{(c) Intersection}     
     \caption{(Color online) 
	Snapshots of reactive scalar distributions whose reactions evolve according to Eq. (\ref{eqn:1Ddelay}) at statistical equilibrium ($t=20T$). 
    The two cases depict  
	(a) a linearly decaying reactive scalar ($a=3$, $b=0$) for which no delay time is present  
	and   
	(b) a linear delay reactive scalar ($a=3$, $b=1$, $\tau=1$). 
    The period $T=1$ such that $h_0\approx 2.33$ with  
	$a>h_0$.
	The smoothly varying force  is diagonally oriented given by Eq. (\ref{eqn:force}). 
	The bars on the right give the concentration values. 
        (c) One-dimensional transects ($y=0.5$) for the linearly decaying reactive scalar (black line) and 
	the delay reactive scalar (gray line) 
    }
	\label{fig:1DSnapshots} 
 \end{minipage} 
\end{figure}

The stationary distributions of a linearly decaying reactive scalar 
and a linear delay reactive scalar, 
with reactions evolving according to Eq. (\ref{eqn:1Ddelay}),  
are  depicted respectively in Figs. \ref{fig:1DSnapshots}(a) and (b). 
Notice the distinct difference between the two distributions:  
Fig. \ref{fig:1DSnapshots}(a) contains no delay term whereas Fig. \ref{fig:1DSnapshots}(b) contains a delay term 
and it is this delay term that is responsible for the filamental behavior of the concentration field. 
This difference is more easily observed in the corresponding one-dimensional transects shown in Fig. \ref{fig:1DSnapshots}(c).

The most common method to characterize the scaling behavior of the distributions
 is to consider their Fourier power spectra. 
An alternative method is to consider the concentration difference between points separated by a fixed distance. 
The latter is called the structure function \cite{MoninYaglom1975} and 
it is the method we employ here since it allows an easy comparison  
between the theoretical results of the previous section with the numerical results of this section. 
The first-order structure function associated with the field $c(\bm{x},t)$ is defined as
\begin{equation}\label{eqn:def_structurefunction}
S(\delta x)\equiv\langle |\delta c(\delta \bm{x};\bm{x},t) |\rangle\sim \delta x^\gamma, 
\end{equation}
where 
$\langle\ldots\rangle$ denotes averaging over different values of  $\bm{x}$ and $\delta x\equiv|\delta \bm{x}|$. 
Recall that $\delta c(\delta \bm{x};\bm{x},t)\equiv \delta c(\bm{x}+\delta \bm{x},t)-\delta c(\bm{x},t)$.    
For the time being we assume that the $\gamma$ appearing in (\ref{eqn:def_structurefunction}) is precisely the H\"older exponent as predicted by previous theoretical arguments.

For both the delay reactive scalar and the delay plankton model, the parameters are chosen in such a way that all three scaling regimes, described by expression (\ref{eqn:delay_holders}),  emerge within the range of length scales considered. 
To control this range, the magnitude of the characteristic length scale 
that separates the Regime I from Regimes II and III, denoted by $\delta x_c$,   needs to be considered.
Substituting expression (\ref{eqn:Lyapunov_T}) into (\ref{eqn:characteristiclength scale}),     
the expression for $\delta x_c$ for  the  model strain flow (\ref{eqn:v}) becomes 
\begin{equation}\label{eqn:estimate_char}
\delta x_c\approx  \exp(-2.33 \, \tau/T).  
\end{equation}	                      
Thus, the value of $\delta x_c$ is modified by 
varying the value of  $\tau/T$ (see Table \ref{table:characteristic_length scale} where the value of $\delta x_c$ is calculated for some key values of  $\tau/T$).

\begin{table}[!]
\topcaption{ An estimate for the characteristic length scale, calculated for the model strain flow (\ref{eqn:v}) for $L=1$ using expression (\ref{eqn:estimate_char}).}
\label{table:characteristic_length scale}
\begin{tabular*}{\textwidth}{@{\extracolsep{\fill}}  c  c  c c  c }
  \hline \hline
  $\tau/T$ & $1$ &  $ 2$ &  $ 3$ &   $4$  \\ 
  $\delta x_c$ & $\approx 10^{-1}$  & $\approx 10^{-2}$  & $\approx 10^{-3}$  & $\approx 10^{-4}$ \\
  \hline \hline
\end{tabular*}
\end{table}

\subsection{The Linear Delay Reactive Scalar}\label{subsec:numericsscalarfield}
We now  examine the scaling behavior of the delay reactive scalar distribution  as the value of
$\tau/T$ varies.
In each case, the first-order structure function  is calculated  over $10$ evenly spaced intersections. The scaling exponent is obtained from the 
slope of the first-order structure function and it is then compared to the set of scaling laws (\ref{eqn:delay_holders}).

\begin{figure}[!]    
\centering
\begin{minipage}{\linewidth}
\centerline{Regime I}      
\centerline{\includegraphics[width=6cm]{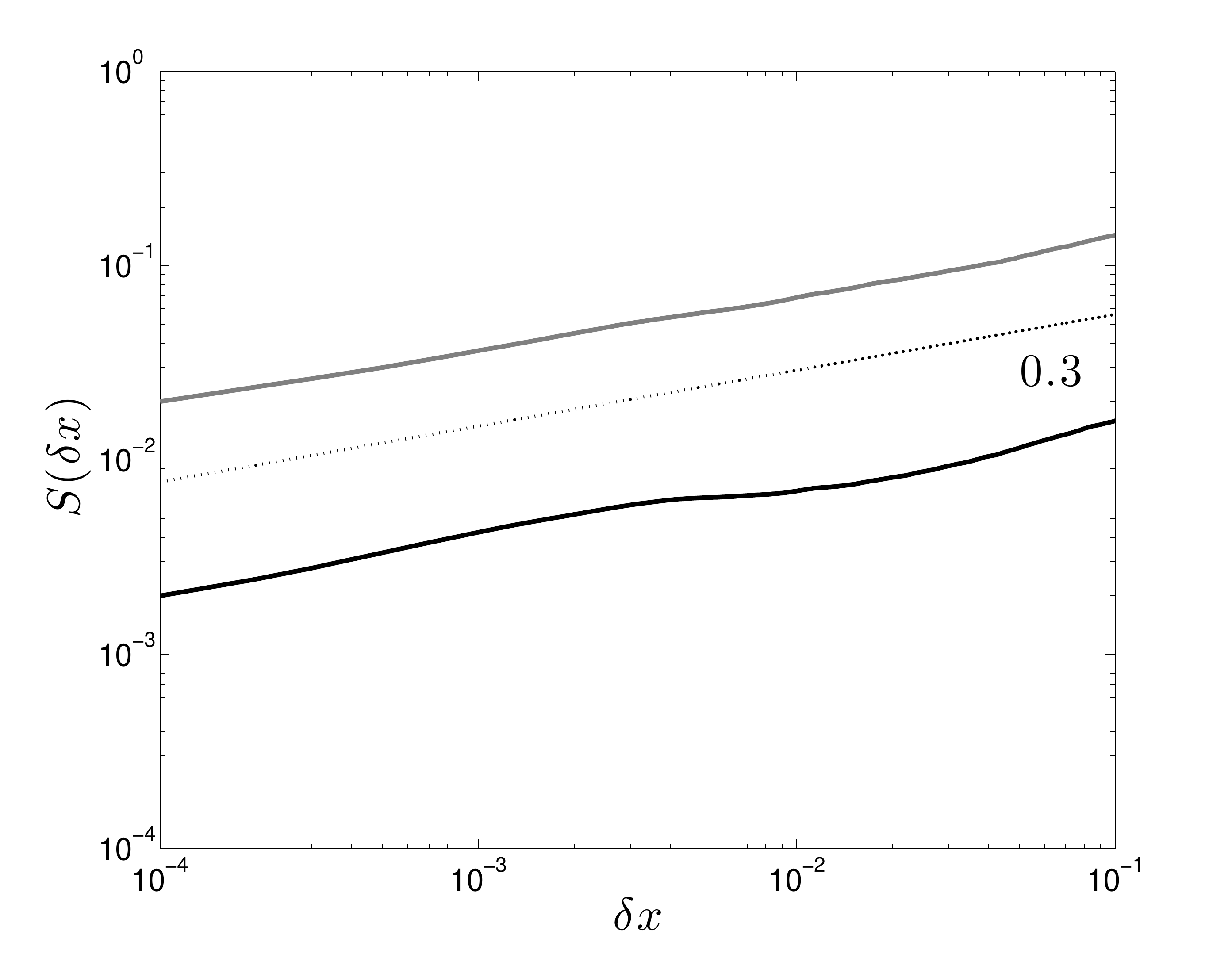}}  
\centerline{(a) $-\re\lambda_1/h_0=0.3$}  
\vfill  
\centerline{\includegraphics[width=6cm]{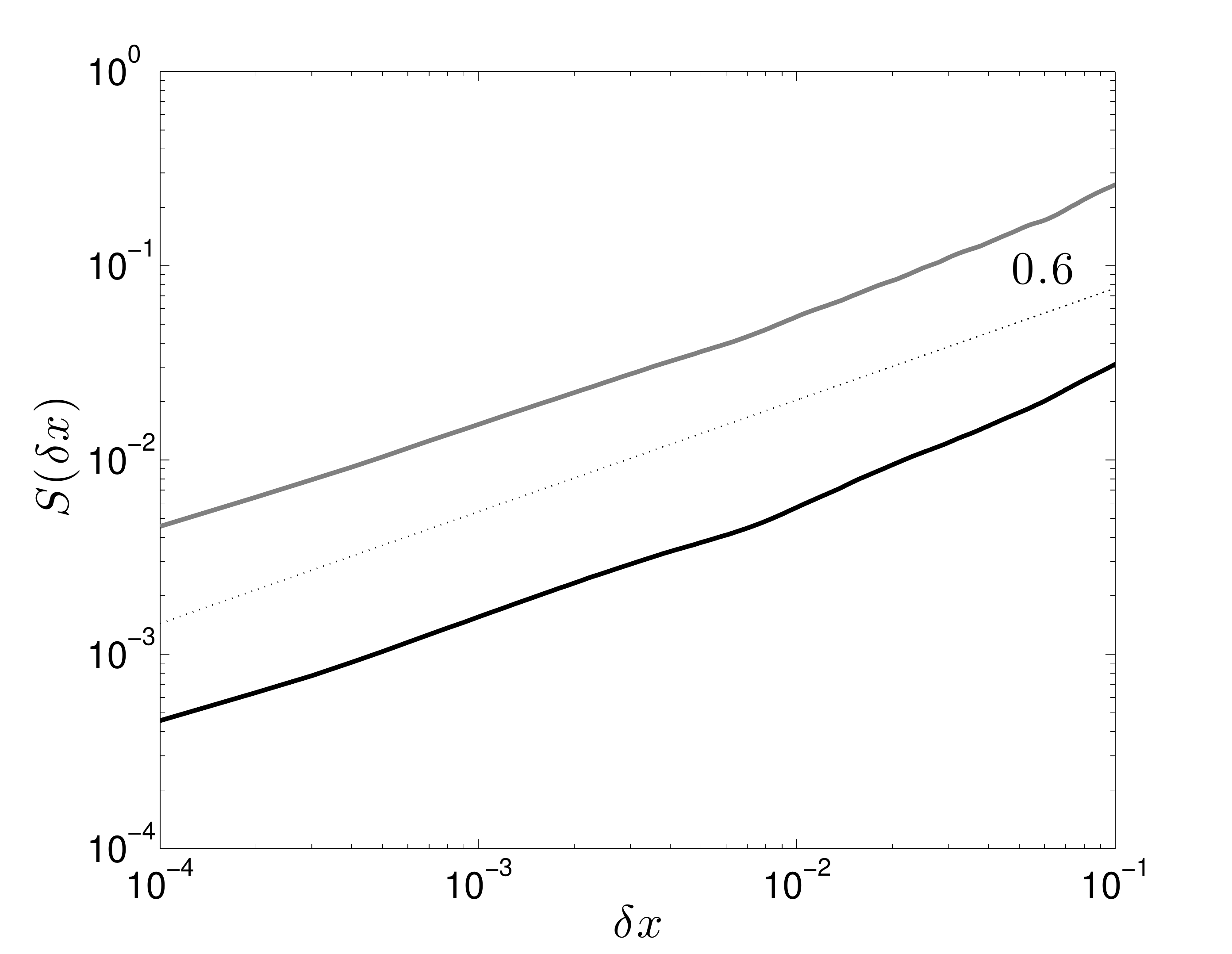}}
\centerline{(b) $-\re\lambda_1/h_0=0.6$}        
\vfill           
\centerline{\includegraphics[width=6cm]{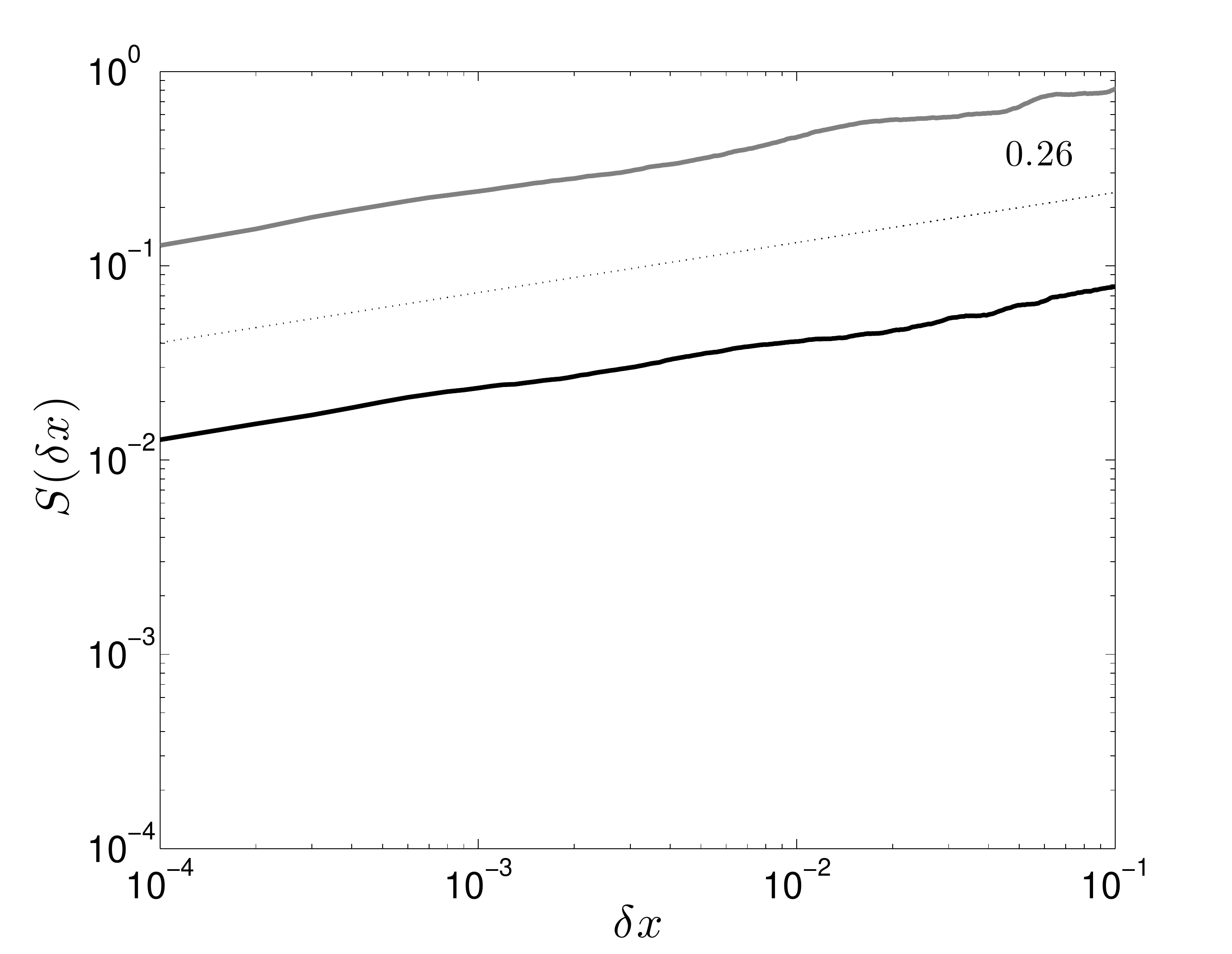}} 	 
\centerline{(c) $-\re\lambda_1/h_0=0.26$}     
     \caption{
	(a) First-order structure functions for the linear delay reactive scalar (\ref{eqn:1Ddelay}) averaged over $10$ evenly spaced transects (parallel to the $x$-axis). 
	These are  
    calculated at statistical equilibrium ($t=20T$) for the two sets of parameters $(a,b,\tau)$ that were considered in Fig. \ref{fig:SeriesFundamental1},  both with $\re\lambda_1=-0.68$ but different $\im\lambda_1$: for
	$(1,-0.16,1)$ (gray solid line) $\lambda_1$ is real while for $(1, 0.9,1)$ (black solid line) $\lambda_1$ is complex. 
    Flow constant is $T=1$.  The theoretical prediction is depicted by the dotted line.      
      	 (b) Same  as (a) but $T=2$. 
	(c) Same as (a) but this time $\re\lambda_1=-0.03$ with   
	different $b$  and $\tau$ and $T$:
	$(1,0.92,5)$ (gray solid line), $(1, 0.7,10)$ (black solid line) and $T=20$.  
	In all cases, $\tau/T\leq 1$. 
    Black dotted lines  correspond to theoretical prediction (\ref{eqn:Holder1}).}
	\label{fig:RegimeIrealComplexRegimeI5tau10tau20T} 
 \end{minipage} 
\end{figure}

\subsection*{Regime I}  
Initially, $\tau/T\leq 1$ so that $\delta x_c\gtrsim 0.1$ (see Table \ref{table:characteristic_length scale}).
This way only Regime I will appear within the range of length scales considered (recall that finite-size effects become important for $\delta x>0.1$). The validity of $-\re\lambda_1/h_0$, the ratio associated with the
H\"older exponent within Regime I (see (\ref{eqn:Holder1})), is tested.
Three different aspects are examined:
the first aspect investigates the impact that the imaginary part of $\lambda_1$
may have on the  scalar field. Recall that $\lambda_1$ denotes the root of the characteristic equation (\ref{eqn:1Dchar}) that has the least negative real part.   
According to expression (\ref{eqn:Holder1}), $\im\lambda_1$ does not contribute to the field's scaling behavior.
This is confirmed in the numerical results that are shown in 
Fig.  \ref{fig:RegimeIrealComplexRegimeI5tau10tau20T}(a). 
There, the first-order structure functions obtained from two  parameter sets, chosen so that both share the same $\re\lambda_1$  but different $\im\lambda_1$, are found to 
share the same scaling exponent (their slopes are equal).
In particular, for the first set of parameters, $\lambda_1$ is real while for the second, $\lambda_1$ is complex.

The second aspect investigates how the scaling behavior varies as  the value of $T$  (and therefore $h_0$) varies.         
We consider the same set of parameters as the ones in Fig.  \ref{fig:RegimeIrealComplexRegimeI5tau10tau20T}(a), where this time $T=2$ thus 
leading to a larger value for the H\"older exponent (double than before). 
The corresponding scaling exponents 
are in good agreement with the theoretical prediction (\ref{eqn:Holder1}) 
(see Fig. \ref{fig:RegimeIrealComplexRegimeI5tau10tau20T}(b)).
Finally, the third aspect explores larger values for both $\tau$ and $T$.   
For two sets of parameters, both of which share the same $\re\lambda_1$, 
the scaling exponents are in good agreement with the theoretical prediction (\ref{eqn:Holder1}).

\subsection*{Regimes I \& II } 
The coexistence of the  Regimes I and II is now investigated by setting 
$\tau/T=2$ so that $\delta x_c\approx 10^{-2}$.   
At the same time, $\delta x_2\sim |b|/(ae)$ is chosen to be of the same order of magnitude as $\delta x_c$.
This way, Regime III, whose appearance depends on the value of $\delta x_2$ relative to $\delta x_c$ (see Eq. \ref{eqn:delay_holders}) is limited. 
Note that because $-\re\lambda_1$ increases as $|b|/a$ increases (to verify consider Eq. (\ref{eqn:1Dchar})), a smaller value of $|b|/(ae)$ results in a larger value for the H\"older exponent within Regime I. Therefore, to obtain an interesting change of behavior from Regime I to Regime II, we are limited on how small we can choose  $|b|/(ae)$ to be.

To test  the validity of the set of scaling laws (\ref{eqn:delay_holders}),  we examine
the structure functions obtained from two sets of parameters, with different value for 
$|b|$, shown in Fig. \ref{fig:RegimeI_II}(a) 
(see also Fig. \ref{fig:Product}(b) for comparison with theory). 
For the first parameter set  
the value of $|b|$ is smaller than for the second parameter set which implies that 
 the first parameter set has  a larger $-\re\lambda_1$ than the 
second parameter set.
Thus  within Regime I, the first parameter set has a larger H\"older exponent.
At the same time, $a/h_0>1$ for both parameter sets   
and thus the H\"older exponent within Regime II is equal to $1$. 

Comparing the theory to the numerics, we can deduce that there is 
good agreement.  $\delta x_c$ captures sufficiently well the transition
between Regimes I and  II. This transition occurs for slightly larger length scales for the second parameter set since it possesses a larger value of $\delta x_2$. 
Within Regime I, the field's 
scaling exponent   is close enough to its theoretical value, though this agreement is expected to become better for smaller length scales (see e.g. Figs. (\ref{fig:RegimeIrealComplexRegimeI5tau10tau20T})).
Within Regime II,  the scaling exponent is, as expected, equal to $1$.

A flatter structure than predicted by theory appears for the intermediate length scales ($10^{-3}<\delta x<10^{-2}$) for which Regime I should continue to hold.            
It appears that this intermediate structure can be explained
by noticing that
the rate of exponential increase of the separation between neighboring fluid parcels 
is distributed. 
A complete development of this argument is left for future work.

\subsection*{Regimes I \& II \& III } 
\begin{figure}[!]                   
\begin{minipage}{\linewidth}                            
\begin{minipage}{0.48\linewidth}
\centerline{Regimes I \& II}     
\centerline{\includegraphics[width=7cm]{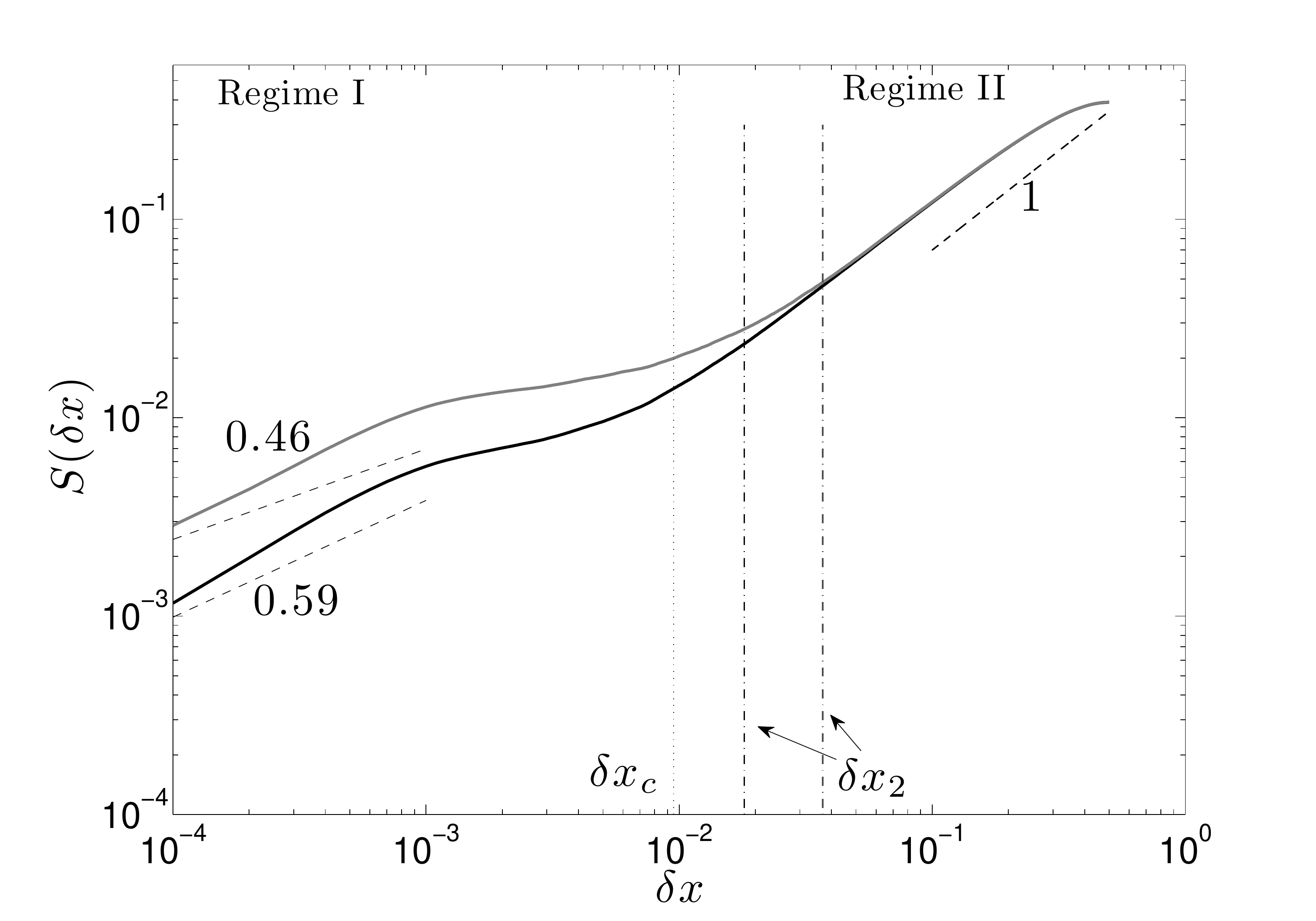}} 
\centerline{(a) $|b|/(ae)\sim \delta x_c$} 
\end{minipage}
        \hfill
\begin{minipage}{0.48\linewidth}
\centerline{Regimes I \& II \& III}     
  \centerline{\includegraphics[width=7cm]{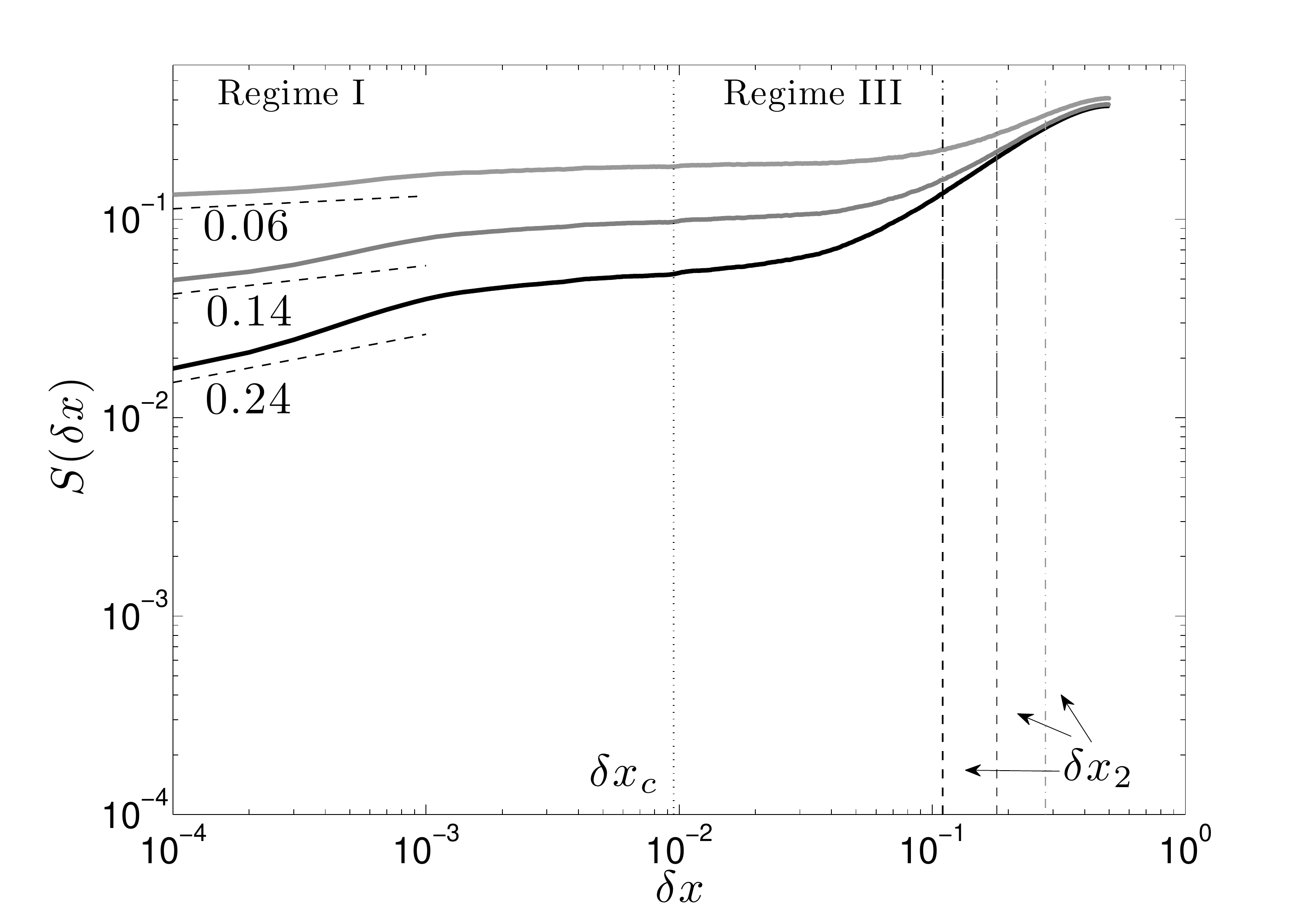}}
  \centerline{(b) $|b|/(ae) \gg \delta x_c$}
\end{minipage}   
\end{minipage}       
\caption{Same as Fig. \ref{fig:RegimeIrealComplexRegimeI5tau10tau20T} but this time 
 $\delta x_c\approx 0.01$ ($\tau/T=2$).
The set of parameters $(a, b, \tau)$ are:
(a) $(1, 0.05, 10)$ (black) with  $\delta x_2\approx 0.02$ and 
$(1, 0.1, 10)$ (black) with  $\delta x_2\approx 0.04$. 
(b) $(1, 0.3, 10)$ (black) with $\delta x_2\approx 0.11$ , $(1, 0.5, 10)$ (dark gray) 
with $\delta x_2\approx 0.18$  and $(1,0.75,10)$ (light gray) $\delta x_2\approx 0.27$.
In all cases $T=5$ which leads to $a>h_0$. 
Also shown are the  predictions for the 
 H\"older exponents (black dashed lines), $\delta x_c$ (black dotted line) and $\delta x_2$, based on the estimate given by (\ref{eqn:estimatedx2}) (dashed-dotted line with different shades for each parameter set). 
}
\label{fig:RegimeI_II}
\end{figure}

The coexistence of the Regimes I, II and III is now investigated
by keeping $\tau/T=2$ while increasing the value of   
$\delta x_2\sim |b|/(ae)$ by an order of magnitude larger than $\delta x_c$.
This is achieved by considering the same set of parameters as in Fig. \ref{fig:RegimeI_II}(a) but 
increasing the value of $|b|$. This increase results in an increase in
 the value  of $\delta x_2$ and a decrease in the value of $-\re\lambda_1$ 
 (the value for 
 $\delta x_c$ remains the same).

The structure functions corresponding to three sets of parameters, 
shown in Fig. \ref{fig:RegimeI_II}(b) (see also Fig. \ref{fig:Product}(c) for comparison with theory),
are now examined. 
As expected,  Regime III appears within a wide range of length scales, whose range increases as the value of $|b|$ increases. 
The value of $\delta x_2$ provides a good estimate for the length scale separating Regime II from Regime III. 
When $|b|\sim0.75$, 
$\delta x_2\approx 0.27$ in which case the Regime III
appears for all length scales larger than 
$\delta x_c$, thus displacing Regime II (see Fig. \ref{fig:RegimeI_II}(b)). 
Similarly to the numerical results shown in Fig. \ref{fig:RegimeI_II}(a), a good agreement 
between theory and numerics 
is obtained within Regime I, the agreement being better for smaller length scales 
(the flat regime also appearing here). 
As before, the field's scaling behavior within Regime II is smooth.

\subsection{The Delay Plankton Model}\label{subsec:numerics_delayplanktonmodel}   
Having  investigated the scaling behavior of the linear delay reactive scalar field, 
the focus now turns to the delay plankton model.
This is a typical nutrient-predator-prey system \cite{Murray1993} 
where the effect of the former is parameterised by the prey carrying capacity, denoted by $C$.
The interactions among the biological species are given by the following set of nonlinear delay-differential equations
\begin{subequations}\label{eqn:biology_convenience}
\begin{align}
\frac{dC}{dt}&=\alpha(C_0(\bm{x})-C),\label{eqn:biology_conveniencea}\\
\frac{dP}{dt}&=P(1-P/C)-PZ,\label{eqn:biology_convenienceb}\\
\frac{dZ}{dt}&=P(t-\tau)Z(t-\tau)-\delta Z^2,\label{eqn:biology_conveniencec}
\end{align}
\end{subequations} 
where  $P$ stands for phytoplankton and $Z$ for zooplankton, 
$t$ is a dimensionless time scaled to the phytoplankton production rate $r$ ($t/r$ is the real time) and 
$\alpha$ denotes the rate at which the carrying capacity relaxes to the background  source $C_0(\bm{x})$.             
The phytoplankton growth is logistic and grazing takes place according to a simple $PZ$ term.
Zooplankton death occurs at a  rate $\delta$ and is 
described by a quadratic in $Z$ term, representing grazing due to higher trophic levels. 
The key feature of this model is the introduction of the time $\tau$ that 
represents the time it takes for the zooplankton to mature ($\tau/r$ in real time).   
For although it is reasonable to assume  an instantaneous change in the prey population 
once prey and predator are encountered, it is not reasonable to assume an instantaneous change in the predator population.
                                                                           
\begin{figure}[!]     
\centering
\begin{minipage}{\linewidth}  
\centerline{\includegraphics[width=6cm]{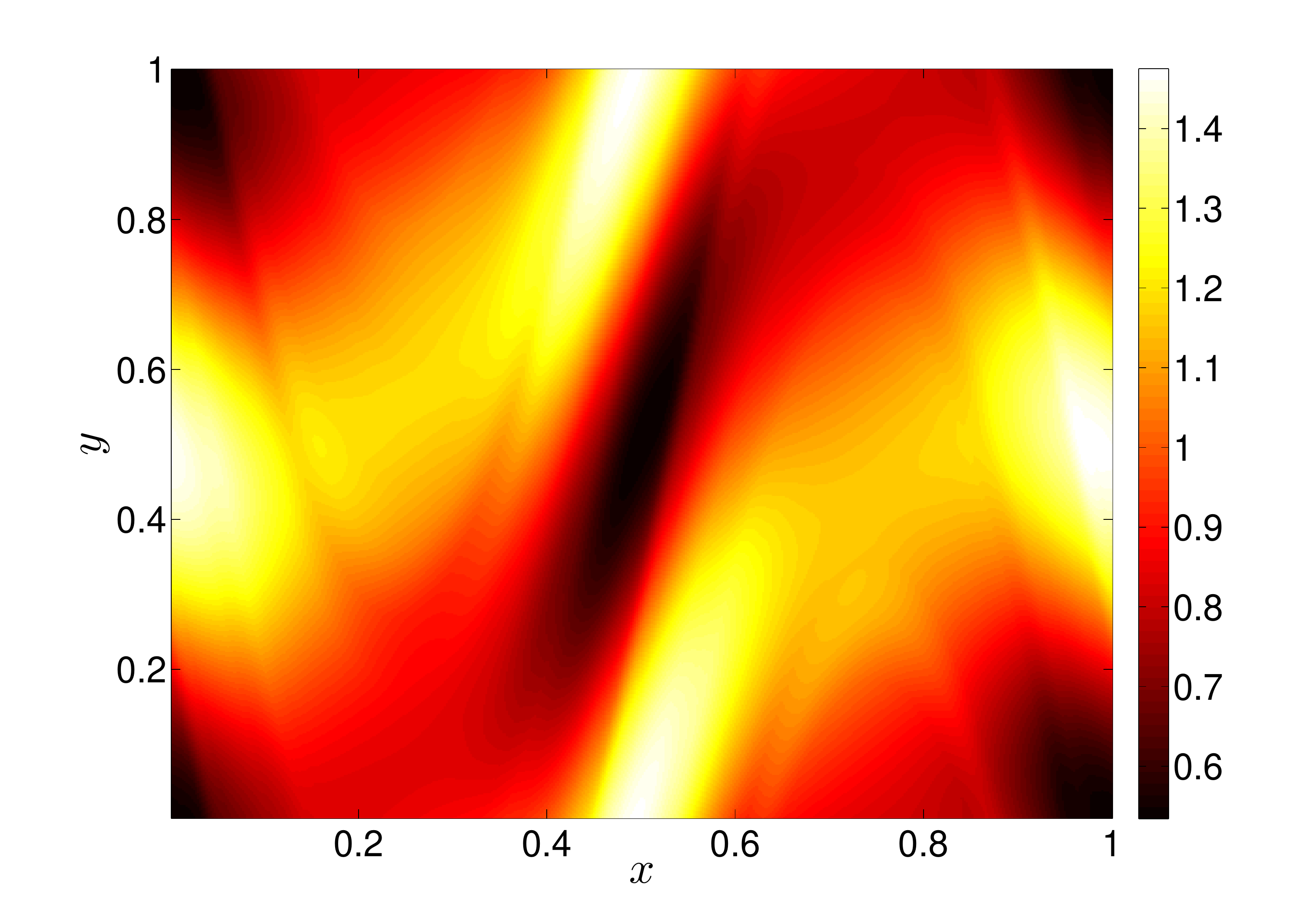}}  
\centerline{(a) Carrying Capacity} 
\vfill  
\centerline{\includegraphics[width=6cm]{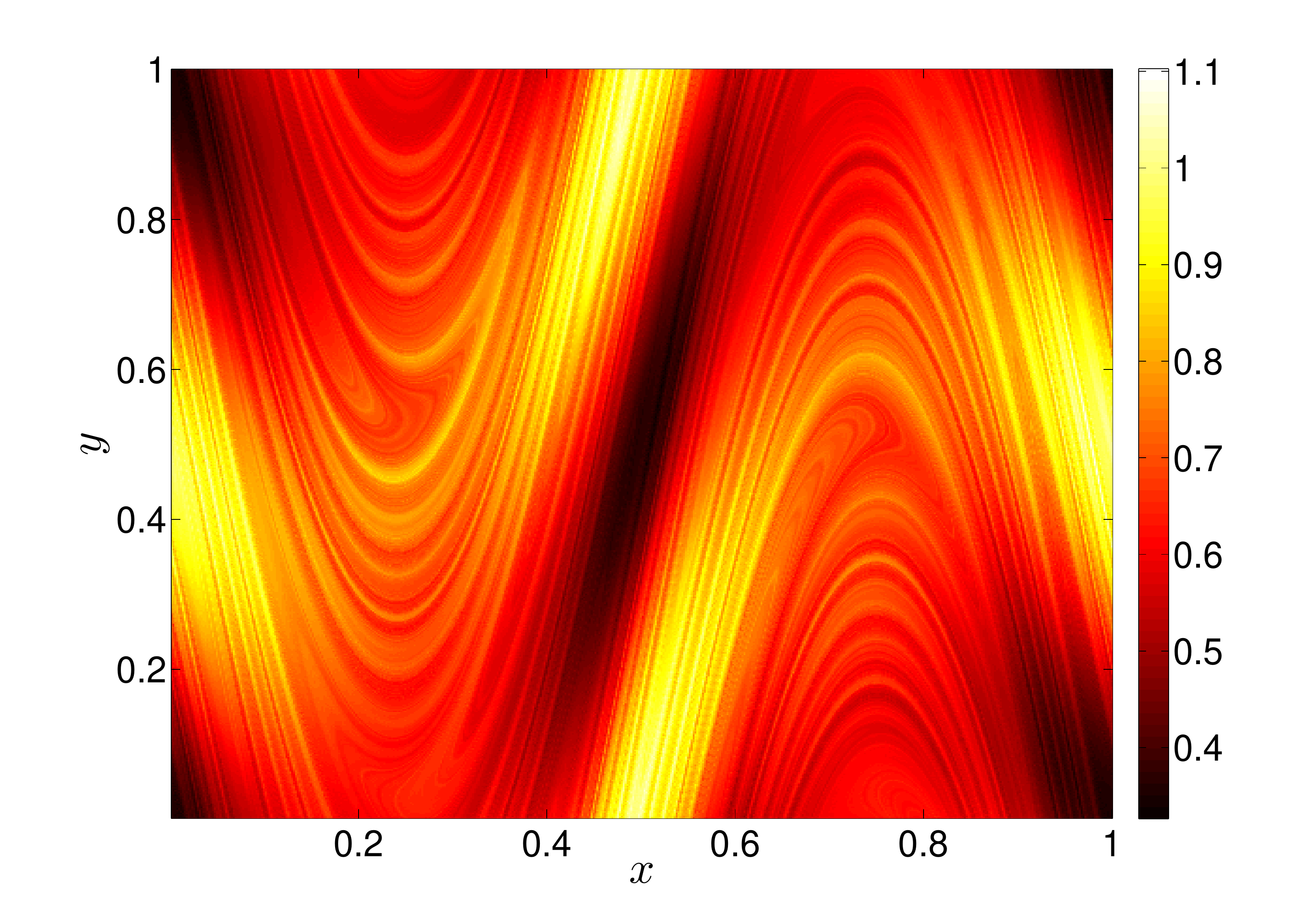}}
\centerline{(b) Phytoplankton}        
\vfill           
\centerline{\includegraphics[width=6cm]{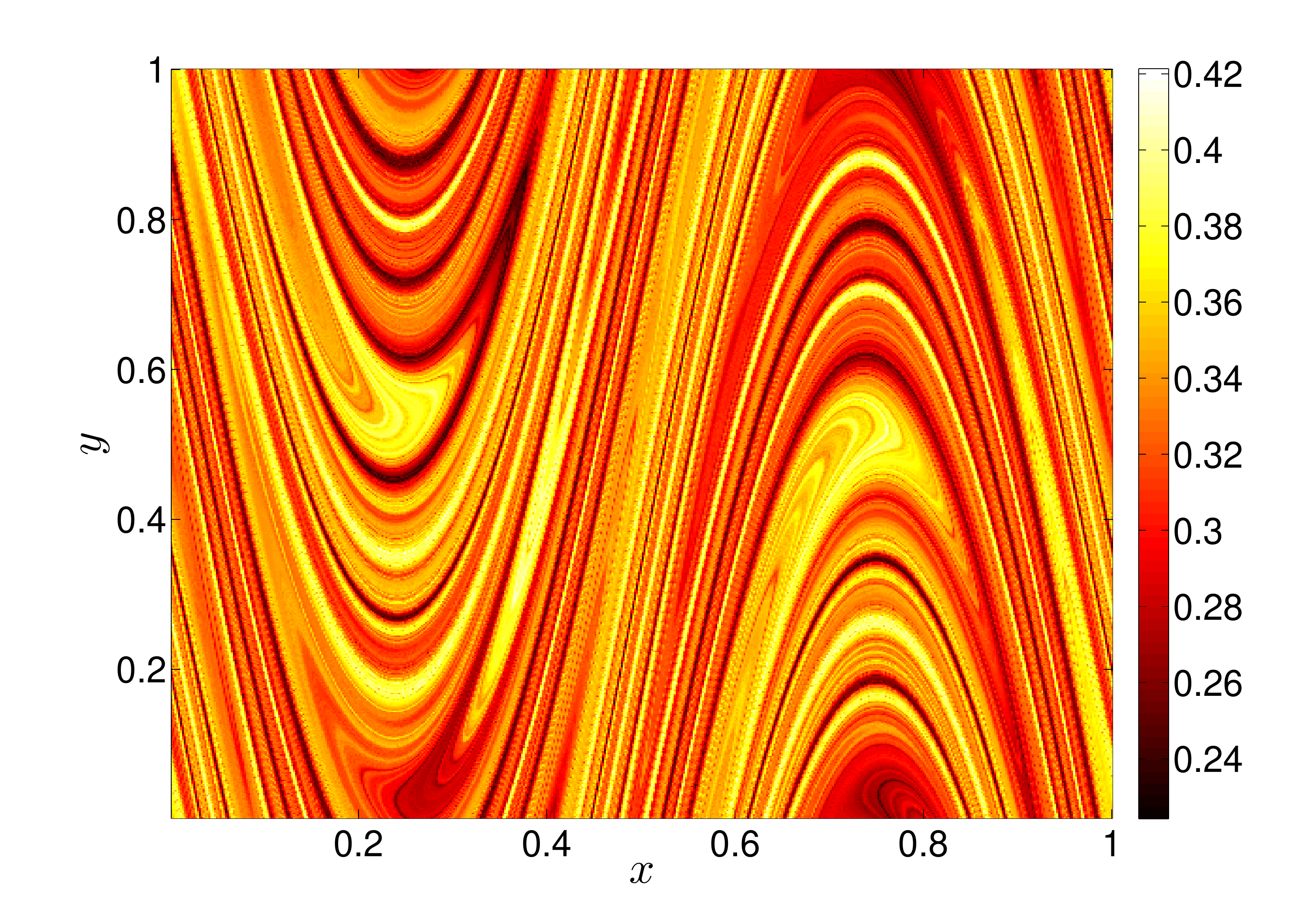}} 	 
\centerline{(c) Zooplankton}     
     \caption{(Color online) 
        Snapshots of the biological distributions at statistical equilibrium ($t=20T$) for the delay plankton model
         (\ref{eqn:biology_convenience})  stirred by the model strain flow (\ref{eqn:v}) with $\tau=20$, $\alpha=0.25$, $\delta=2$ and $T=20$.
        As before the force is diagonally oriented, described by (\ref{eqn:force}).       
	 }
	\label{fig:SnapshotsDelayPlanktonModel} 
 \end{minipage} 
\end{figure}

The stationary distributions for $C$, $P$ and $Z$,  attained when coupled to 
the strain model flow (\ref{eqn:v}),  
are depicted for a particular set of parameters in Fig. \ref{fig:SnapshotsDelayPlanktonModel}. 
Before analyzing any numerical simulations, the particular plankton dynamics need first to be examined. 
While the scaling behavior  of a  general system of delay reactive scalar fields 
has been set out in the Appendix, certain non-generic features are easier to address for each model in question.
 
For the delay plankton model,  the non-generic feature is the existence of asymmetrical couplings between 
the phytoplankton's carrying capacity and the subsystem comprising of the phytoplankton and the zooplankton.  \cite{Hernandez-Garcia_etal2002} considered the case of a zero delay time and deduced that the phytoplankton and zooplankton  should always share the same small-scale structure. 
The numerical results that \cite{TzellaHaynes2007} obtained show that the same holds for a non-zero delay time,   
provided the length scales remain sufficiently small. 
However, on larger scales, a second scaling regime appears in which  
the zooplankton structure is flat while the phytoplankton has a structure  similar to  its carrying capacity.
Although the appearance of a second scaling regime 
is inherent to any system of delay reactive scalar fields, 
the decoupling among the species is particular to the delay plankton model.

\begin{figure}[!]
\begin{minipage}{\linewidth}
\begin{minipage}{0.45\linewidth}
\centerline{\includegraphics[width=7cm]{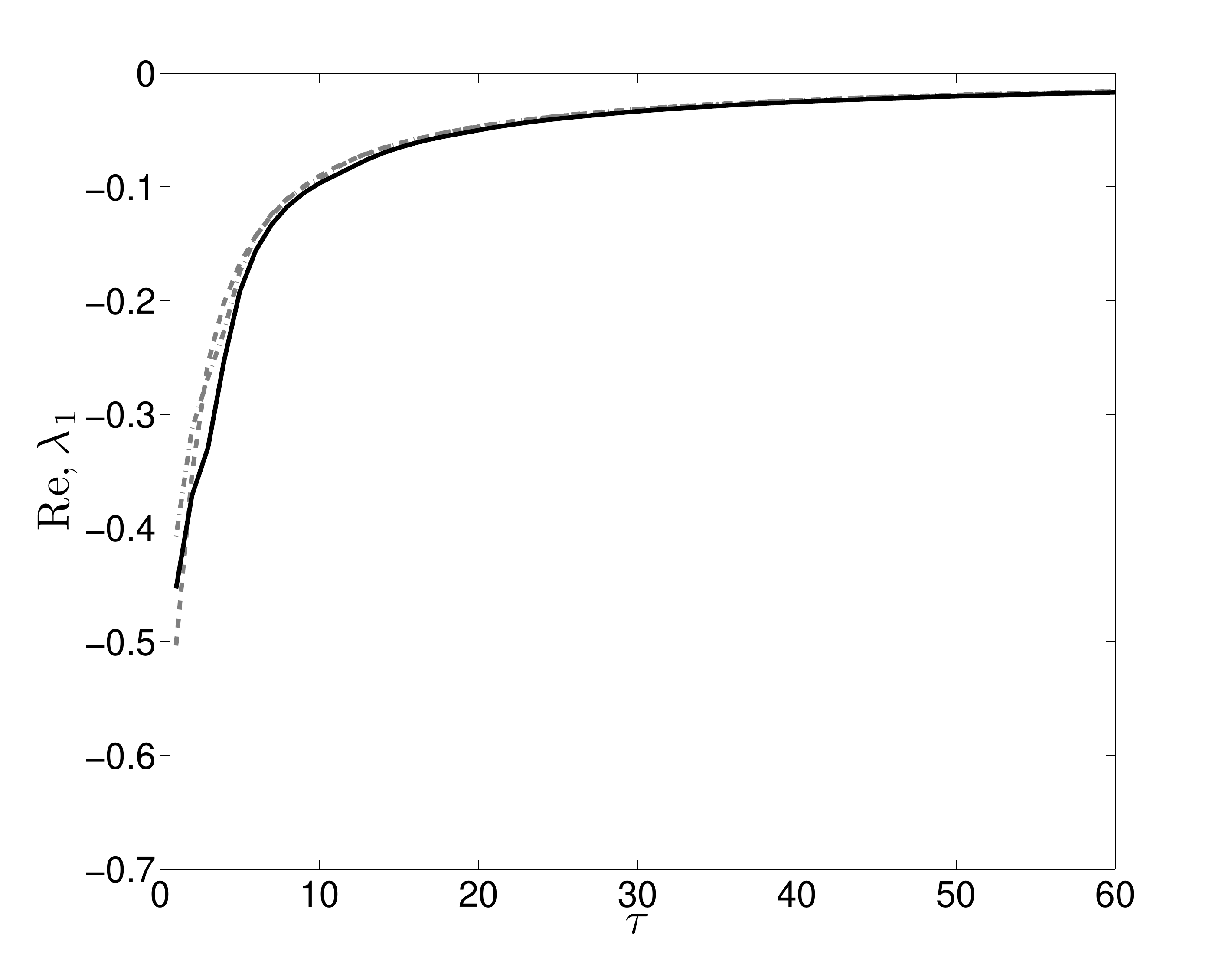}}  
\end{minipage}
\hfill
\begin{minipage}{0.45\linewidth}
\centerline{\includegraphics[width=7cm]{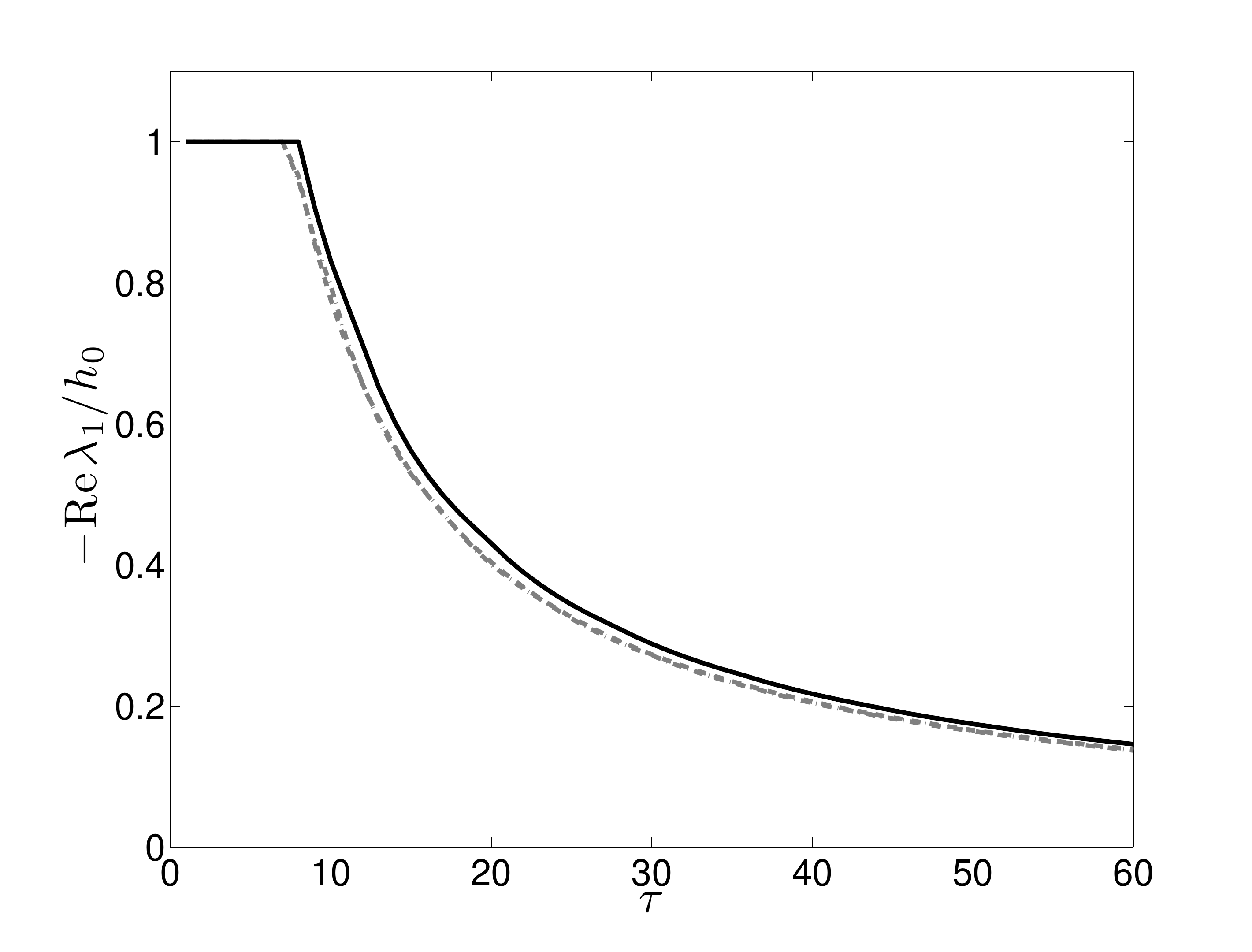}}
\end{minipage}
      \caption{(a) The value of $\re\lambda_1$,  
       associated with the rate of the slowest decaying eigenfunction of the linearized phytoplankton-zooplankton subsystem, 
       calculated and plotted as a function of $\tau$ for  $C_0=1$ (black solid line), 
       $C_0=0.5$ (gray dashed line) and $C_0=1.5$ (gray dashed-dotted line) 
      ($\delta =2$).  
       Its value is determined by considering the roots of the characteristic Eq. (\ref{eqn:delayplankton_char}).
       (b) The value of $\text{min}\{-\re\lambda_1/h_0,1\}$, the theoretical value for the H\"older exponent  shared between the phytoplankton and the zooplankton, plotted as a function of $\tau$   
       for $h_0\approx 0.117$ ($T=20$) and  $C_0=1$ (black solid line), 
       $C_0=0.5$ (gray dashed line) and $C_0=1.5$ (gray dashed-dotted line).  }
      \label{fig:RealLambda1}
      \end{minipage}
\end{figure}

To fully explain the scaling behavior of the delay plankton model, the theory of Sec. \ref{sec:DelayTheory} 
 must be extended in order to accommodate the particularities of this model.
In the absence of advection and  within a certain range of parameters, 
the delay plankton model has a single fixed point of equilibrium, given by 
\begin{equation}\label{eqn:fixedpoint}
C^{\star}=C_0(\bm{x}), \; P^{\star}=\delta C^{\star}/(\delta+C^{\star}) \; \text{and} \; Z^{\star}=P^{\star}/\delta.     
\end{equation}
This point is stable for $\tau=0$. 
For $0.5\leq C_0(\bm{x})\leq 1.5$ and $\delta=2$, as in the simulations performed here, this point remains stable for any $\tau> 0$. 
Linearizing the delay plankton model around this point of equilibrium 
results in the following expressions for the matrices $\bm{A}$ and $\bm{B}$:   
\begin{subequations}\label{eqn:matricesAB}
\begin{align}
\bm{A}=
\left(\begin{matrix} 
	\alpha & 0 & 0 \\
	-(P^\ast/C^\ast)^2 & P^\ast/C^\ast & P^\ast\\
	0 & 0 & 2P^\ast
\end{matrix}
\right)&\\ 
\intertext{and}                    		  
\bm{B}=-P^\ast\left(\begin{matrix} 
	0 & 0 & 0 \\
	0 & 0 & 0\\
	0 & 1/\delta & 1
\end{matrix}
\right)&, 
\end{align} 
\end{subequations} 
where $\bm{A}$ and $\bm{B}$ are the matricial equivalents of $a$ and $b$ for the one-dimensional linear delay reactive scalar (\ref{eqn:1Ddelay}) (for further details see App. (\ref{app:key})). 
Certain matrix coefficients (i.e. $-(P^\ast/C^\ast)^2$, $P^\ast/C^\ast$, $2P^\ast$) were simplified using (\ref{eqn:fixedpoint}).

From Eq. (\ref{eqn:charsystem}), the characteristic matrix is given by $\bm{H}(\lambda)=\lambda\bm{I}+\bm{A}+\bm{B}e^{-\lambda \tau}$. 
Thus, using Eq. (\ref{eqn:matricesAB}), 
\begin{equation}\label{eqn:matrices_char_matNPZ}     
\begin{split}
&\bm{H}(\lambda)=\\
&\left(\begin{matrix} 
	\lambda+\alpha & 0 & 0 \\
	-\left(P^\ast/C^\ast \right)^2 & \lambda+P^\ast/C^\ast & P^\ast\\
  	0 & -e^{-\lambda\tau}P^{\ast}/\delta & \lambda-P^{\ast} e^{-\lambda\tau}+2P^\ast
\end{matrix}
\right).   
\end{split}
\end{equation}
It follows that the characteristic equation  corresponding to the linearized delay plankton model satisfies (see Eq. (\ref{eqn:charsystem}))
\begin{subequations}\label{eqn:delayplankton_char}
\begin{equation}
h(\lambda)\equiv\text{det}\bm{H}(\lambda)=
(\lambda+\alpha)\,g(\lambda)=0	,
\end{equation}	
where $g(\lambda)=0$ is the characteristic equation associated with the phytoplankton-zooplankton subsystem. 
\begin{equation}
g(\lambda)=	
\left|\begin{matrix} 
	\lambda +P^{\ast}/C^{\ast} & P^{\ast}\\
	-e^{-\lambda\tau}P^{\ast}/\delta & \lambda-P^{\ast} e^{-\lambda\tau}+2P^{\ast}
\end{matrix}
\right|.
\end{equation} 
\end{subequations}     
As in the one-dimensional case, the number of roots are infinite for $g(\lambda)=0$ (and therefore for $h(\lambda)=0$).
At the same time, 
the magnitude of $\re\lambda_1$, the root with the least negative real part, 
decreases as $\tau$ increases
with $\re\lambda_1\rightarrow 0$ as $\tau\rightarrow\infty$. 
Its value is determined for fixed $C_0$ and $\delta$ and plotted 
in Fig. \ref{fig:RealLambda1}(a) as a function of $\tau$ for
$\delta=2$ and three key values of $C_0(\bm{x})$: $1$, $1.5$ and $0.5$ i.e.   
its average, maximum and minimum (see Eq. (\ref{eqn:force})).
Notice that the difference between the $\re\lambda_1$ calculated for these
three values of $C_0(\bm{x})$ is minor. 
It is therefore expected that the value for the least negative chemical Lyapunov exponent
associated with the nonlinear dynamics of the delay plankton model is close to $-\re\lambda_1$.

In the theoretical considerations made in Sec. \ref{sec:DelayTheory},  
the scaling behavior of a linear delay reactive scalar 
was described by
the set of scaling laws (\ref{eqn:delay_holders}).
A similar set of scaling laws holds for a system of nonlinearly interacting scalars 
(see App. \ref{app:scaling}): 
the H\"older exponent within Regime  I 
is governed by the ratio of the least negative chemical Lyapunov exponent, -$\re\lambda_1$, to the flow Lyapunov exponent, $h_0$; 
within Regime II,  the  H\"older exponent  is governed by $-a_1/h_0$, where
$a_1$ is the slowest decay rate associated with the reduced system that is obtained once all delay terms 
are ignored.
As for the single delay reactive scalar, the appearance of a flat scaling regime, Regime III,  depends on whether $\delta x_2$, the length scale associated with this regime, is larger than $\delta x_c$, the transition lengthscale. 
Note that the value of $\delta x_2$ is not necessarily the same for each species 
(see \S\ref{app:scaling}).

This set of scaling laws was deduced for the general case  
in which the  product    of     
the {\it fundamental matrix}, the matricial equivalent of the fundamental solution 
(see \S\ref{app:key}),
with the  direction of the forcing in the chemical space  has non-zero entries 
(see Eq. (\ref{eqn:chemicaldifference_delay})).      
If that is not the case, the set  of scaling laws (\ref{eqn:delay_holders}) may need to be modified and
different regimes for different species are expected. 
Note however that in all cases the value of $\delta x_c$ is not affected 
as its value only depends on $\tau$ and not on the particular chemical dynamics.
                
To examine the existence of zero entries for the linearized delay plankton model, consider first the form of the eigenfunctions  that comprise its fundamental matrix. 
This matrix,  denoted by $\bm{M}_{\bm{Y}}(t)$,
can be written as an infinite sum of eigenfunctions, each proportional to 
$e^{\lambda_i t}\text{adj}\bm{H}(\lambda_i)$, where $\text{adj}$ is short for adjoint 
(see Eq. (\ref{eqn:fundamental_matrix2})). 
In the delay plankton model, the forcing is given by the source $C_0(\bm{x})$. 
Since this is applied only to the carrying capacity,  
the product of $\text{adj}\bm{H}(\lambda_i)$ with the forcing direction is given by
\begin{equation}\label{eqn:adj_char_matNPZ}
\text{adj}\bm{H}(\lambda_i)\cdot 
\left(
\begin{matrix} 
1\\
0\\
0	 
\end{matrix}
\right)	
=  
\left(  
\begin{matrix} 
\phantom{m}g(\lambda_i)\\
m_1(\lambda_i)\\
m_2(\lambda_i)	 
\end{matrix} 
\right)   
\end{equation}
with  
\begin{subequations}
\begin{align}
m_1(\lambda_i)&=(\lambda_i-P^{\ast}[e^{-\lambda_i\tau}-2])(P^{\ast}/C^{\ast})^2,\\	
m_2(\lambda_i)&=e^{-\lambda_i\tau}{P^\ast}^3(\delta C^\ast)^{-1},     
\end{align}	   
\end{subequations}     
where to deduce the above, Eqs. (\ref{eqn:matrices_char_matNPZ})  and (\ref{eqn:delayplankton_char}) were employed. 
   
Examining the behavior of Eq. (\ref{eqn:adj_char_matNPZ})  as a function of  $\lambda_i$ where $h(\lambda_i)=0$ and $i=1\ldots\infty$, it can be deduced that as long as these are distinct (achieved by appropriately choosing the parameter range), 
the only $\lambda_i$ for which $g(\lambda_i)\neq 0$ is  $\lambda_i=-\alpha$.
Therefore,  a single eigenfunction governs the scaling behavior of $C$
from where it can be inferred that   
a single H\"older exponent characterizes its spatial structure. 
Its 
value is given by
\begin{equation}\label{eqn:HolderNdelay}
\gamma_C=\text{min}\{1,\alpha/h_0\}.
\end{equation}     
This result is hardly surprising as it is easy to observe that the carrying capacity evolves independent of the rest of the species and as the much studied 
linearly decaying scalar with chemical Lyapunov exponent equal to $-\alpha$.

On the other hand  $m_1(\lambda_i)$, $m_2(\lambda_i)\neq 0$ for all $\lambda_i$, where $i=1\ldots\infty$ and thus 
no special considerations are necessary  for the phytoplankton and zooplankton;
their scaling behavior within 
Regime I is shared and governed by the least negative chemical Lyapunov exponent, $\re\lambda_1$.

However, within  Regime II a different scenario takes place.
The fundamental matrix corresponding to this regime may be exactly written as 
$\bm{M}_{\bm{Y}}(t)=\exp[-\bm{A}t]$ (see App. \ref{app:key}).  
Using a singular value decomposition, $\bm{M}_{\bm{Y}}(t)$ can be re-written in terms of 
$\bm{\hat{a}}_i e^{a_it}\bm{\hat{a}}_i^{\dag}$                             
where $\bm{\hat{a}}_i$ and $\bm{\hat{a}}_i^{\dag}$ 
correspond to the  normalized right and left eigenvectors of $\bm{A}$ with eigenvalue $a_i$ 
where $i=1\ldots 3$. 
To understand the scaling behavior of the phytoplankton and zooplankton, it is necessary to consider for each $i$, the product of $\bm{\hat{a}}_i e^{a_it}\bm{\hat{a}}_i^{\dag}$ with the forcing direction.  
The eigenvalues of $\bm{A}$ are given by   
\begin{subequations}
\begin{align} 
\{a_1, a_2, a_3\}&=\{-\alpha, -P^{\ast}/C^{\ast}, -2P^{\ast}\},  \\
\intertext{while the product of $\bm{\hat{a}}_i e^{a_it}\bm{\hat{a}}_i^{\dag}$ with the forcing direction is given by}  
\left\{\bm{\hat{a}}_i e^{-a_i t} \bm{\hat{a}}_i^{\dag}\cdot
\left(
\begin{matrix} 
   	 1 \\ 
     0  \\
	 0 
\end{matrix} 
\right)   
\right\}
&=
\left\{
e^{-\alpha t} 
\left(
\begin{matrix} 
   	 \cdot \\ 
	 \cdot \\
	 0   
\end{matrix} 
\right)
, 
e^{- P^\ast/C^\ast t}
\left(
\begin{matrix} 
   	 0 \\ 
	 \cdot\\
	 \cdot  
\end{matrix} 
\right)
, 
e^{- 2 P^\ast t}
\left(
\begin{matrix} 
   	 0 \\ 
     0 \\
	 0   
\end{matrix} 
\right)
\right\}.
\end{align}	
\end{subequations}	
where $(\cdot)$ indicates that the dependence on (non-zero) constants has been suppressed 
for brevity as their magnitude does not increase or decrease in a systematic way and therefore they do not contribute to the fields' scaling laws 
(see \S\ref{app:scaling}).  
       
It follows that, within Regime II, two terms that are decaying exponentially with rates   
$-\alpha$ and $-\lambda_{P}=-P^{\ast}/C^{\ast}$
contribute to the scaling behavior of the phytoplankton.
The term that corresponds to the smallest decay rate   
will dominate the scaling behavior of the phytoplankton (see also App. \ref{app:scaling}).  
Note that in all the simulations performed here, $\alpha<P^\ast/C^\ast$ where
the value of $P^\ast/C^\ast$ is calculated for $0.5\leq C_0(\bm{x})\leq 1.5$, the range of values of $C_0(\bm{x})$ (see Eqs. (\ref{eqn:force}) and (\ref{eqn:fixedpoint})).
It is therefore expected that the scaling behavior of the phytoplankton is dominated by $-\alpha$,
the same rate that determines its carrying capacity.  
Conversely, none of these terms contribute to the scaling behavior of the zooplankton,
implying that within Regime II, the zooplankton is decoupled from the biological forcing and 
thus evolves like a passive tracer. 
Therefore, within this regime, its spatial structure is flat. 

As a consequence  
Regime III  appears in the scaling behavior of the phytoplankton only. 
The value for $\delta x_2$ separating Regimes II and III maybe estimated 
using Eq. (\ref{eqn:estimatedx2system}).
In all numerical simulations performed here, $\delta x_2\lesssim \delta x_c$ and therefore 
this flat regime is not prevalent in the scaling behavior of  the phytoplankton.

The following set of expressions for the  H\"older exponents associated with the phytoplankton, $\gamma_P$, and 
the zooplankton, $\gamma_Z$, describe the distributions scaling behavior within the two regimes:
\begin{list}{}{\topsep 10pt}
   {\usecounter{Lcount}
    \setlength{\rightmargin}{\leftmargin}}
\item
For Regime I,     
\begin{subequations}\label{eqn:HolderPZRegimeI_II}
\begin{equation}\label{eqn:HolderPZRegimeI}  
\phantom{X}
\gamma_{PZ}=\gamma_{P}=\gamma_{Z}=\text{min}\{\gamma_C,-\re\lambda_1/h_0\}.              
\end{equation}                                                                                     
\item For Regime II,  
\begin{equation}\label{eqn:HolderPZRegimeII} 
\begin{split}	
\quad
\gamma_{P}\neq\gamma_{Z}\;\text{with}\;
&\gamma_{P}=\text{min}\{\gamma_{C},-\lambda_{P}/h_0\},\\
&\gamma_{Z}=0.
\end{split}               
\end{equation}      
\end{subequations} 
\end{list}

To summarize, within Regime I, $P$ and $Z$ share the same small-scale structure 
characterized by  the H\"older exponent $\gamma_{PZ}$. 
This structure is either shared by $C$ i.e. $\gamma_{PZ}=\gamma_{C}$ (see Eq. (\ref{eqn:HolderNdelay}) for $\gamma_C$), 
or is more filamental than $C$ i.e. $\gamma_{PZ}<\gamma_{C}$. 
Within Regime II, the small-scale structure of $Z$ is flat (zero H\"older exponent)
while that of $P$ is either shared with  $C$ or is more filamental than $C$.

The numerical results obtained from a set of simulations  
performed 
firstly for a varying value of $\tau$, 
and secondly for a varying value of $T$ are now analyzed.   
    
\begin{figure}[!]    
\begin{minipage}{\linewidth} 
	 \centerline{Varying $\tau$}    
  \begin{minipage}{0.48\linewidth}
  \centerline{\includegraphics[width=5.5cm]{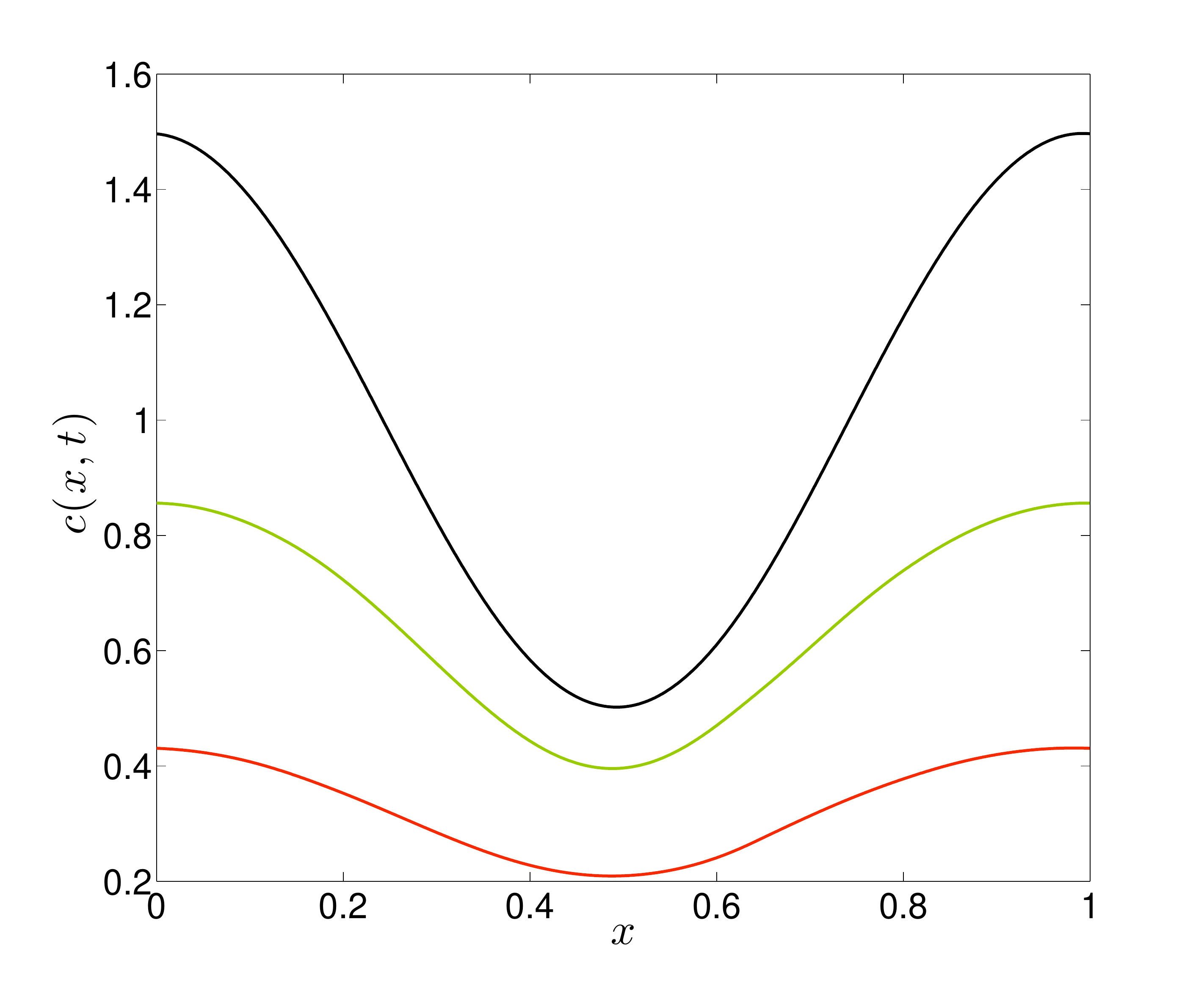}}  
  \end{minipage}
   \hfill
  \begin{minipage}{0.47\linewidth}
  \centerline{\includegraphics[width=6cm]{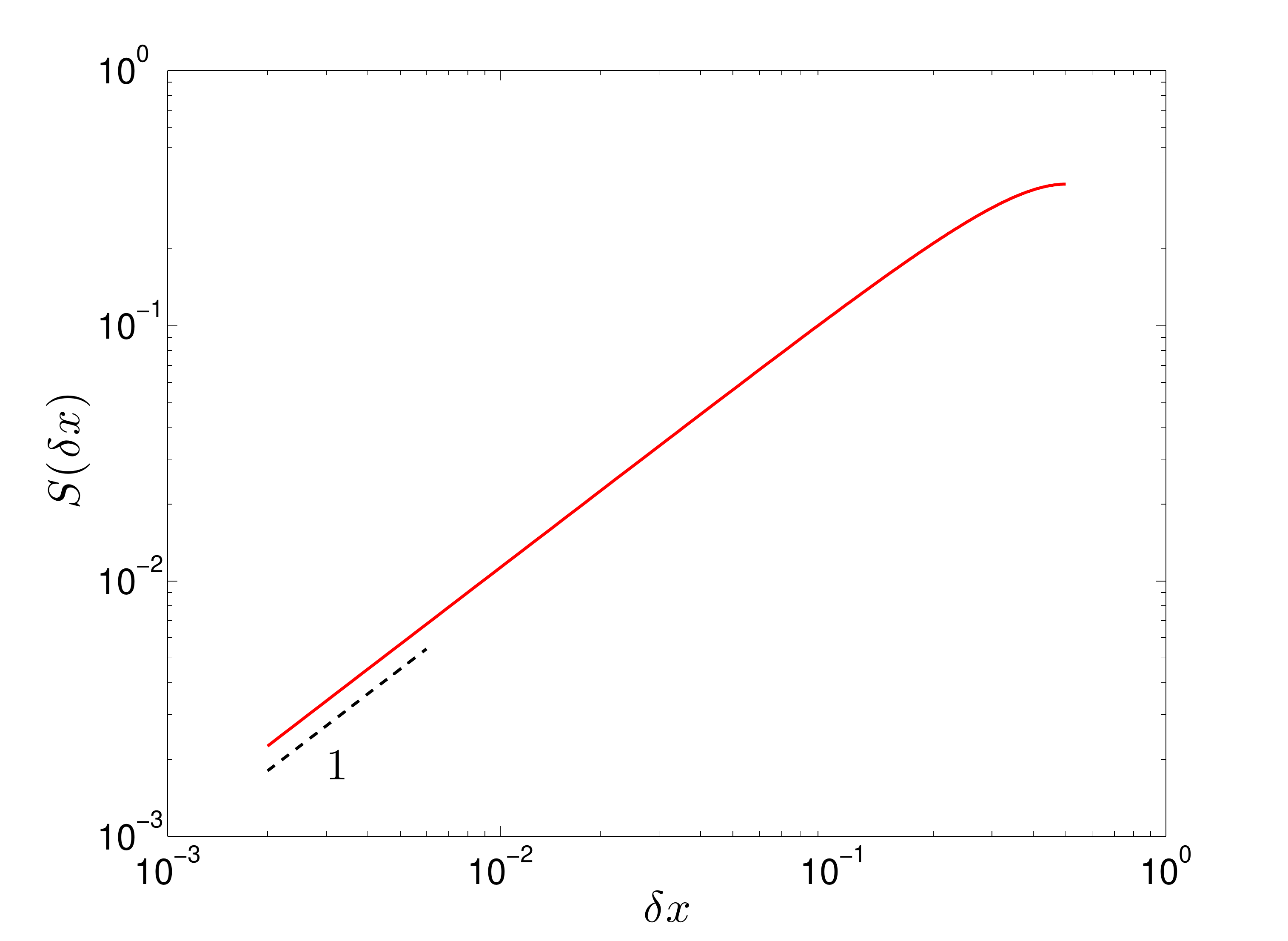}}
  \end{minipage}
  \vfill
  \begin{minipage}{\linewidth}
  \centerline{(a) $\tau=1$}  
  \end{minipage}      
   \begin{minipage}{0.47\linewidth}
  \centerline{\includegraphics[width=5.3cm]{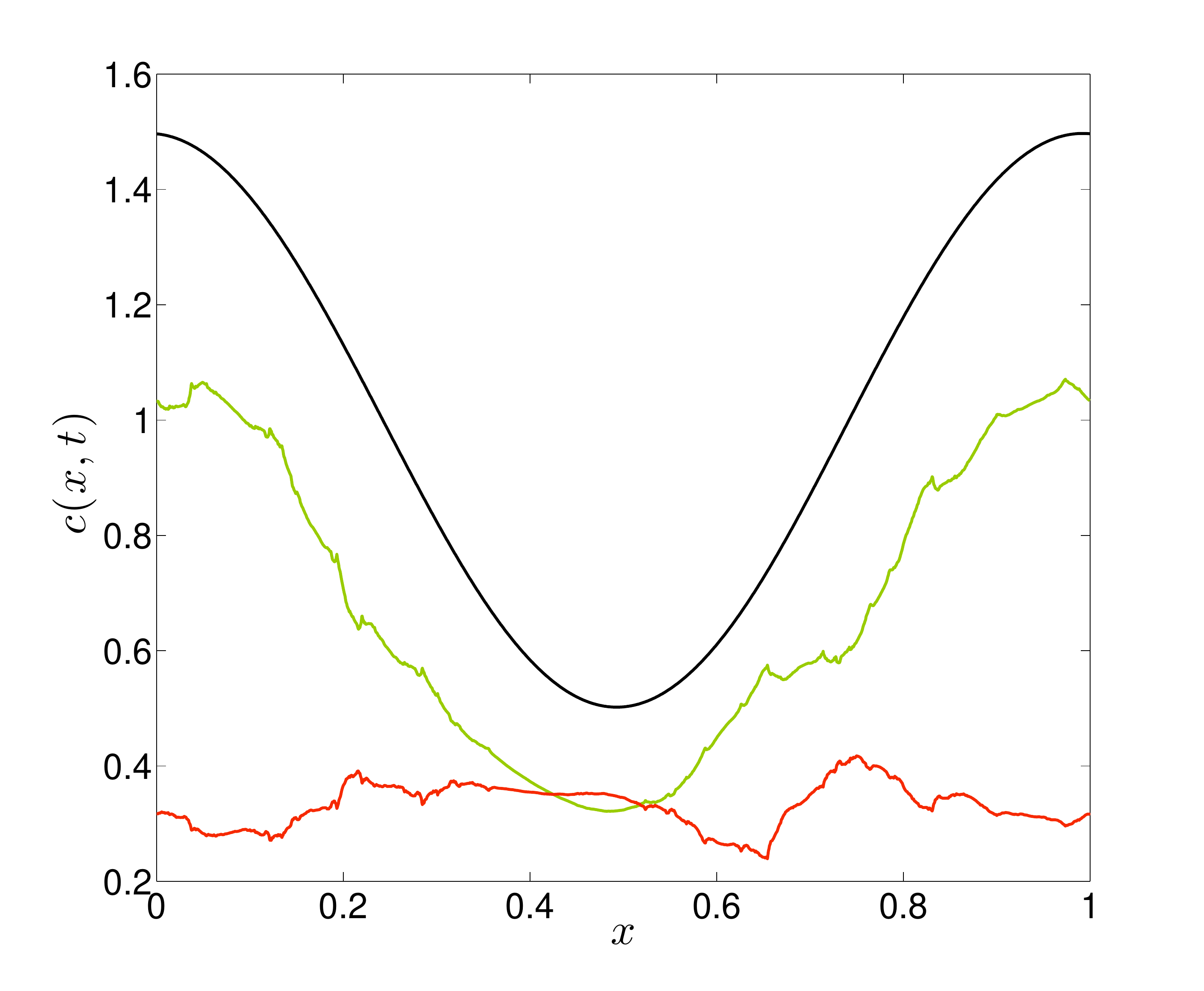}}  
  \end{minipage}
        \hfill
  \begin{minipage}{0.47\linewidth}
  \centerline{\includegraphics[width=5.8cm]{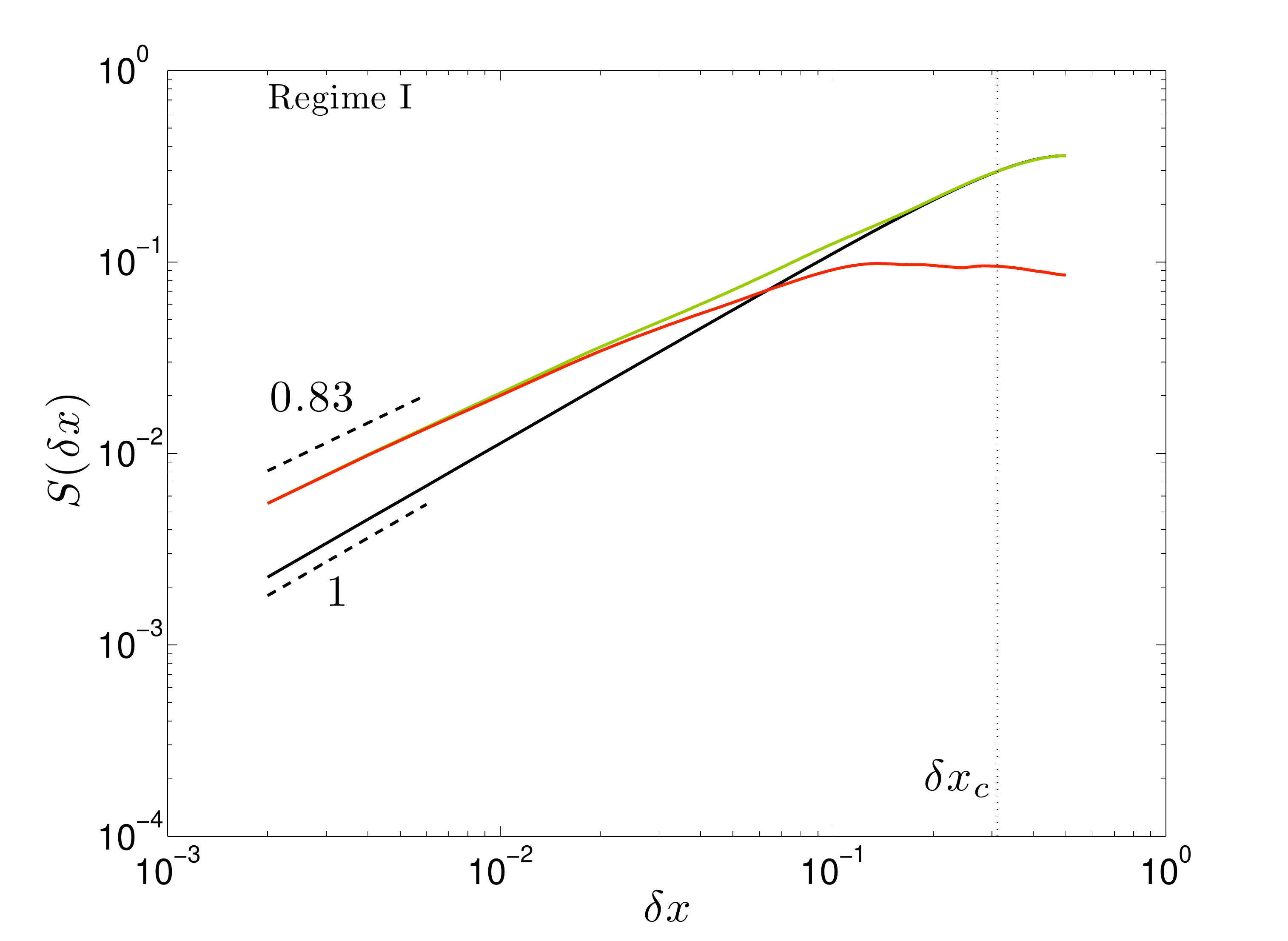}}
  \end{minipage}
  \vfill
  \begin{minipage}{\linewidth}
  \centerline{(b) $\tau=10$}  
  \end{minipage}              
  \begin{minipage}{0.48\columnwidth}
  \centerline{\includegraphics[width=5.3cm]{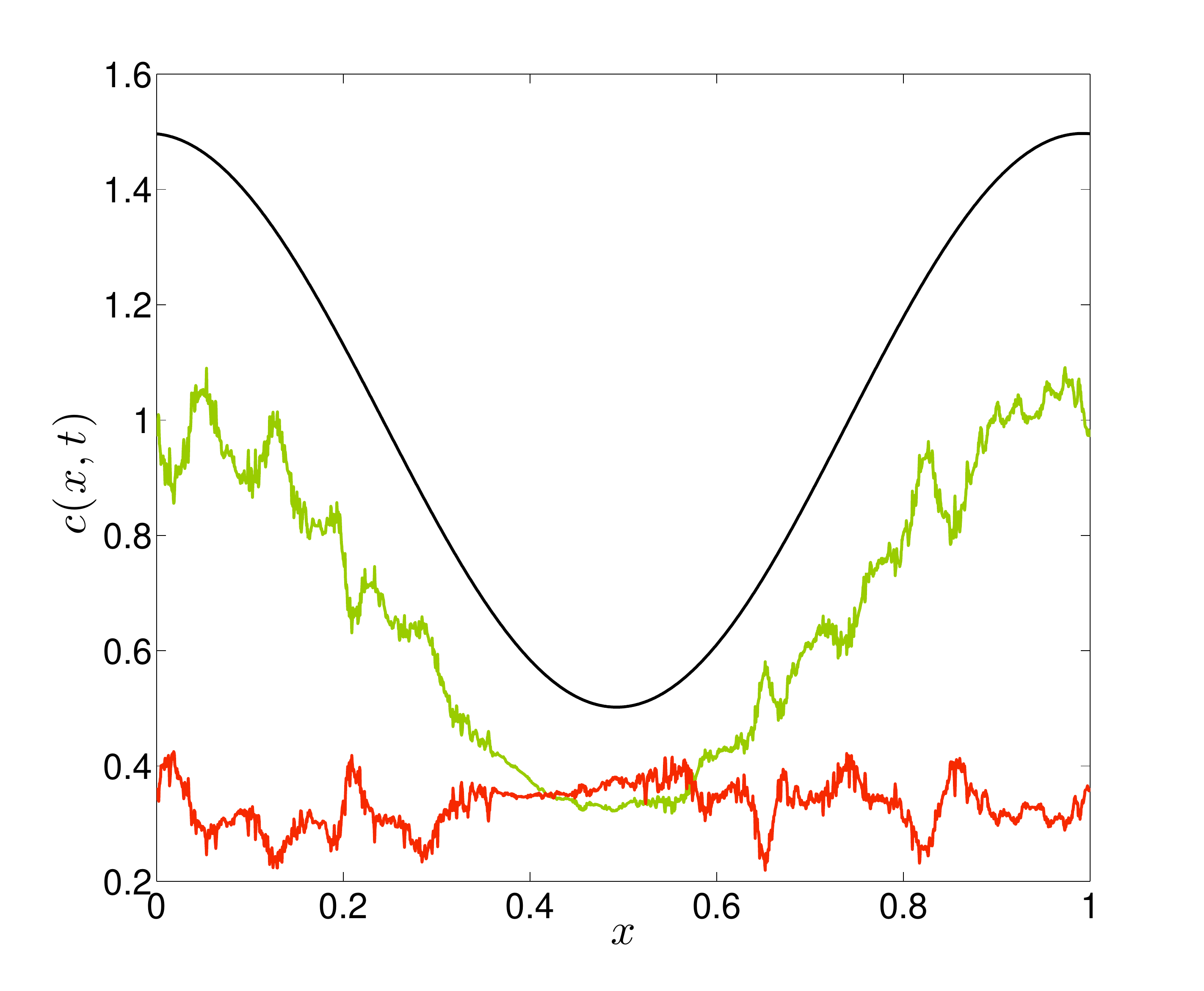}}  
  \end{minipage}
        \hfill
  \begin{minipage}{0.48\columnwidth}
  \centerline{\includegraphics[width=5.8cm]{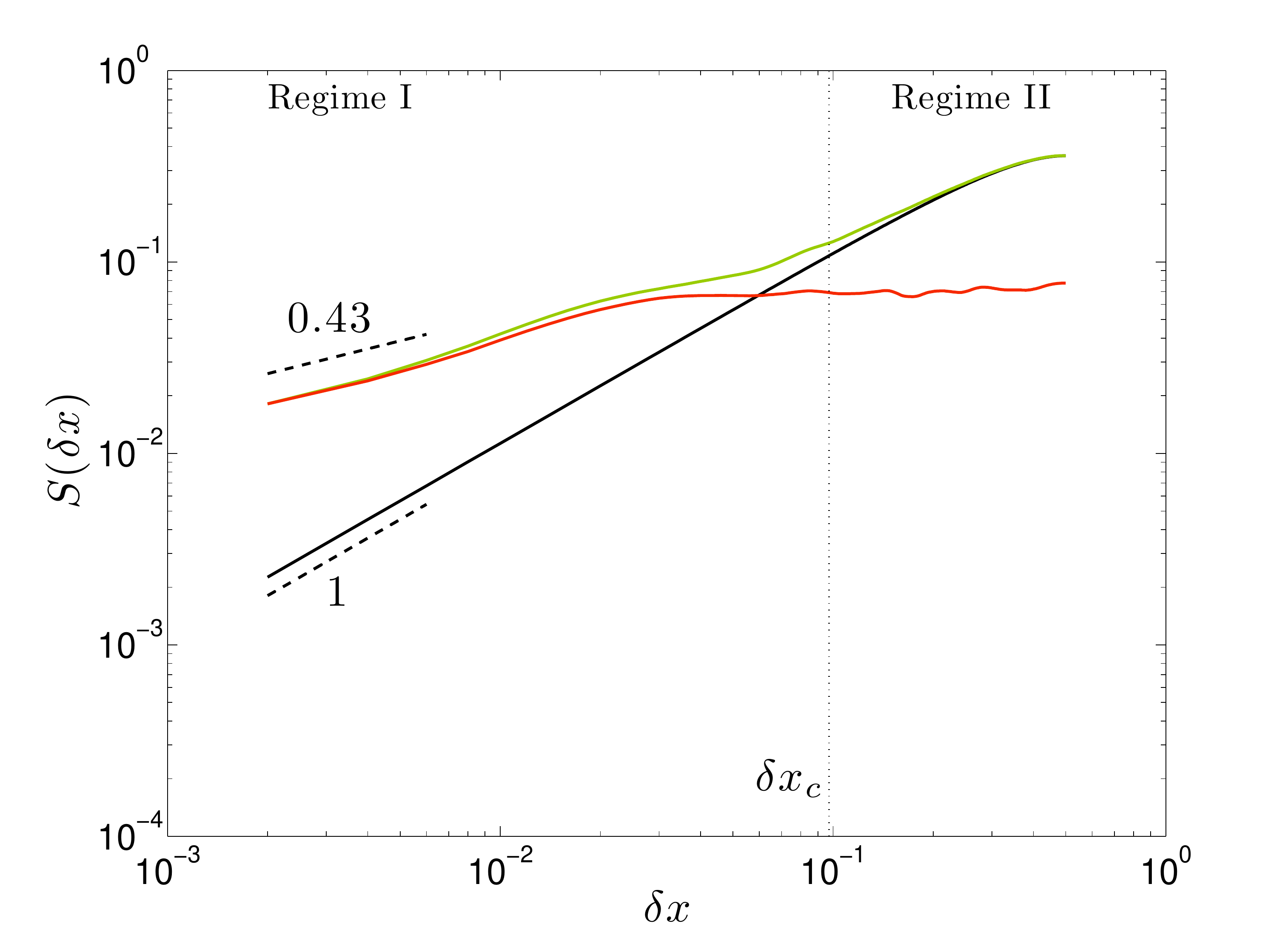}}
  \end{minipage}
  \vfill
  \begin{minipage}{\linewidth}
  \centerline{(c) $\tau=20$}  
  \end{minipage}               
  \begin{minipage}{0.48\linewidth}
 \centerline{\includegraphics[width=5.3cm]{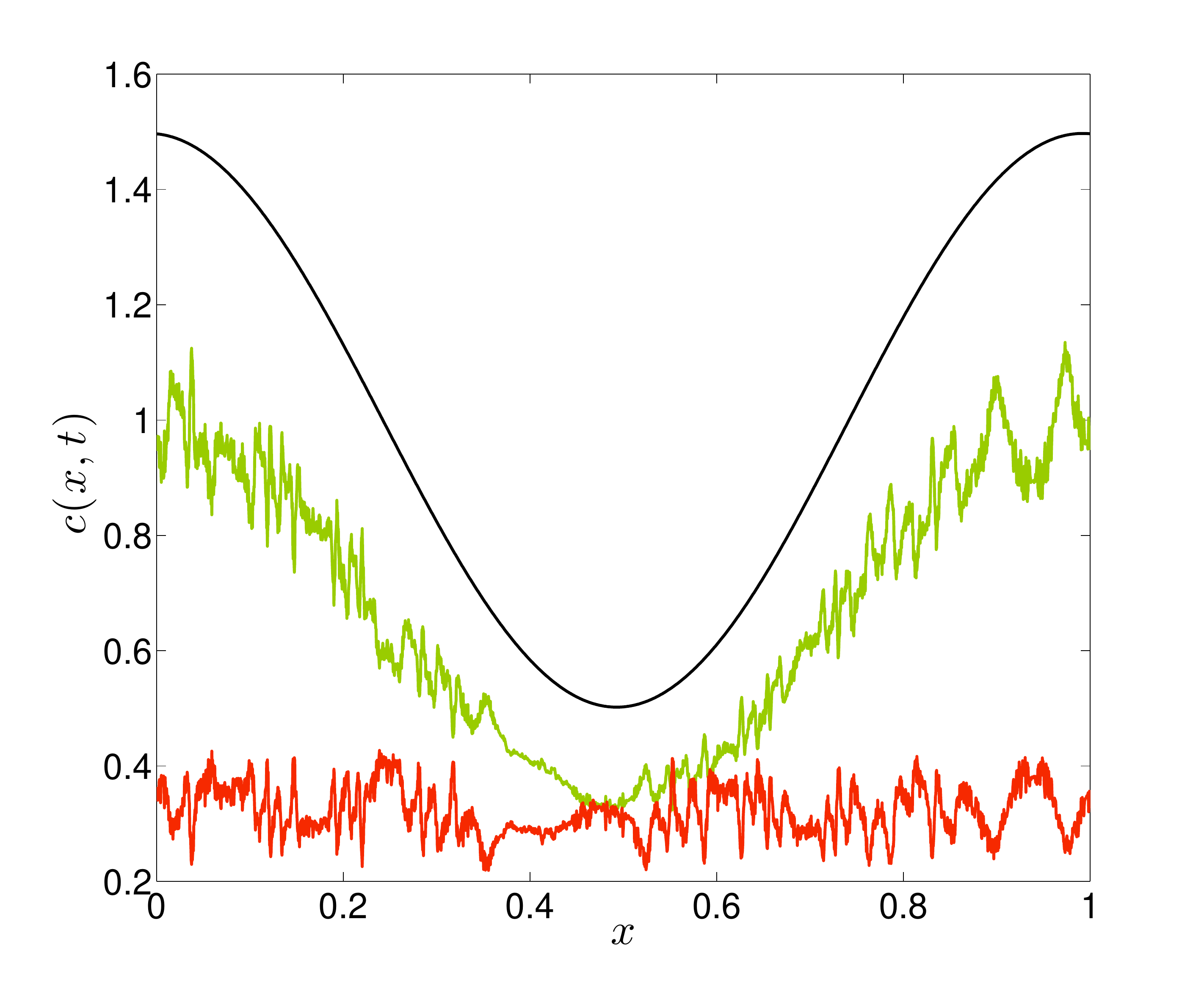}}  
  \end{minipage}
        \hfill
  \begin{minipage}{0.48\linewidth}
  \centerline{\includegraphics[width=5.8cm]{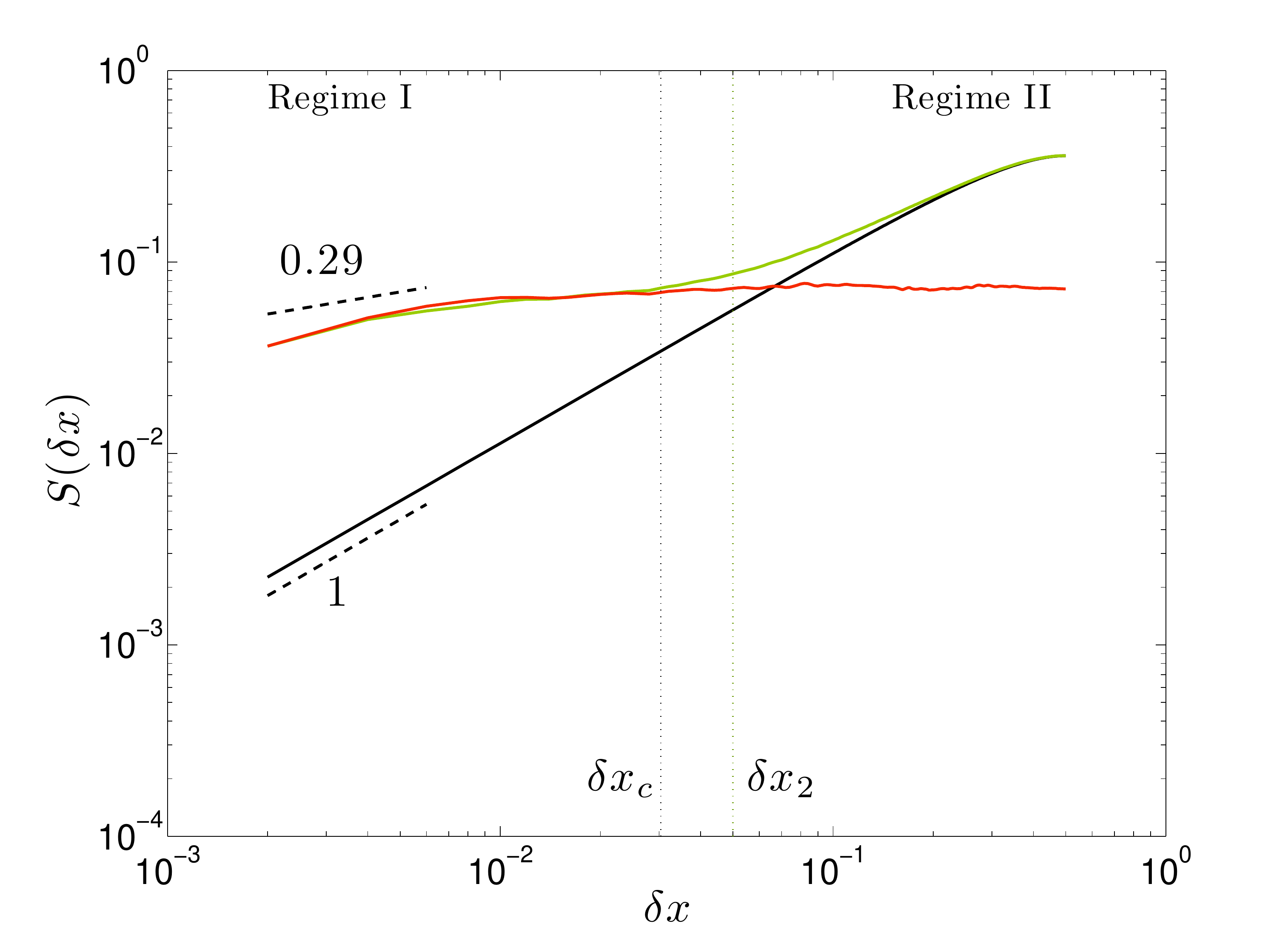}}
  \end{minipage}
  \vfill
  \begin{minipage}{\linewidth}
  \centerline{(d) $\tau=30$}  
  \end{minipage}
     \caption{(Color online) Representative intersections ($y=0.5$) (left) and their corresponding first-order structure functions averaged over 500 evenly spaced intersections (right) at statistical equilibrium ($t=20T$) for the delay plankton model (\ref{eqn:biology_convenience}) advected by (\ref{eqn:v}) for $\delta=2$, $T=20$ and  $\alpha=0.25>h_0\approx 0.117$. Graphs show carrying capacity in black, phytoplankton in light gray (green) and zooplankton in dark gray (red). The value of $\delta x_c$ is marked by a vertical dotted black line (if $\delta x_c<0.5$) and the value of $\delta x_2$ is marked if it is larger than $\delta x_c$. A dotted line for each slope of gradient equal to the theoretical value of the H\"older exponent is drawn for reference.  
	}\label{fig:VariationTau} 
\end{minipage}
\end{figure}

\subsection*{Variation of $\bm{\tau}$}
In the set of numerical results shown in Fig. \ref{fig:VariationTau}, 
the evolution of the concentration fields (calculated over an intersection) and their first-order structure functions
(calculated over $500$ evenly spaced horizontal intersections)
corresponding to  
the zooplankton, phytoplankton and its carrying capacity,
are examined as a function of $\tau$. 
Note that  the structure functions have been offset  to emphasise that for small $\tau$ all species share the same behavior at all length scales. For larger $\tau$, 
the phytoplankton and zooplankton share the same structure at small length scales 
while at larger length scales the phytoplankton shares the same structure as its carrying capacity.

Starting from a small value for $\tau$ for which only Regime I appears 
 and in which all the planktonic distributions are smooth (Fig. \ref{fig:VariationTau}(a)),    
the behavior of both the phytoplankton and the zooplankton becomes increasingly filamental as the value of $\tau$ increases (Fig. \ref{fig:VariationTau}(b-d)).
This behavior is in agreement with the prediction that 
the magnitude of their shared chemical Lyapunov exponent 
decreases as $\tau$ increases, approaching zero for large values of $\tau$ (see Fig.\ref{fig:RealLambda1}(a)). 
A comparison between theory and numerics within Regime I may be made by consulting Fig. \ref{fig:RealLambda1}(b) where 
the H\"older exponent, given by 
$\text{min}\{-\re\lambda_1/h_0,1\}$,  is calculated 
and plotted as a function of $\tau$ for the same values of $C_0(\bm{x})$
as in Fig. \ref{fig:RealLambda1}(a).  
As a reference, a line of the same slope as the theoretical value for the H\"older exponent  is drawn for each case depicted in Fig. \ref{fig:VariationTau}(a-d). 
The agreement between theory and numerics is very close. 
  
At the same time as the value of $\tau$ increases,
the value of the transition length scale 
decreases  according to the theoretical expression (\ref{eqn:estimate_char}).
This leads                                                                                                                   
 to the appearance of Regime II.
Within the latter regime,  the theoretical prediction is confirmed: 
the distribution of the phytoplankton is smooth 
and similar to the distribution of its carrying capacity
while that of
the zooplankton is flat, equivalent to the distribution of a passive (non-reactive) tracer. 
The theoretical value for $\delta x_c$ predicts sufficiently well the transition between the first and second scaling regime.

For Figs.  \ref{fig:VariationTau}(a-c)), 
$\delta x_2<\delta x_c$, thus explaining why no Regime III is observed for the phytoplankton
(where to estimate $\delta x_2$, Eq. (\ref{eqn:estimatedx2system}) was used). 
The only exception is shown in Fig. \ref{fig:VariationTau}(d) for which $\tau=30$. 
For this case $\delta x_2\sim\delta x_c$ and thus 
within a  short region of length scales, 
a flat regime is predicted to appear for the phytoplankton. 
Indeed a flat regime is observed but
as in the case of the single delay scalar 
(see \S\ref{subsec:numericsscalarfield}), 
this flat regime is extended to length scales that lie within Regime I (though 
still close to $\delta x_c$).  
A larger value of $\delta x_2$ is obtained by further increasing the value of $\tau$. 
This is clearly depicted  in  Fig. \ref{fig:VariationT}(a) where Regime III appears for a substantial range of length scales.
For scales larger than $\delta x_2$, Regime II appears.

\subsection*{Variation of $\bm{T}$}
\begin{figure}[!]    
\begin{minipage}{\linewidth} 
	 \centerline{Varying $T$}    
  \begin{minipage}{0.48\linewidth}
  \centerline{\includegraphics[width=5.3cm]{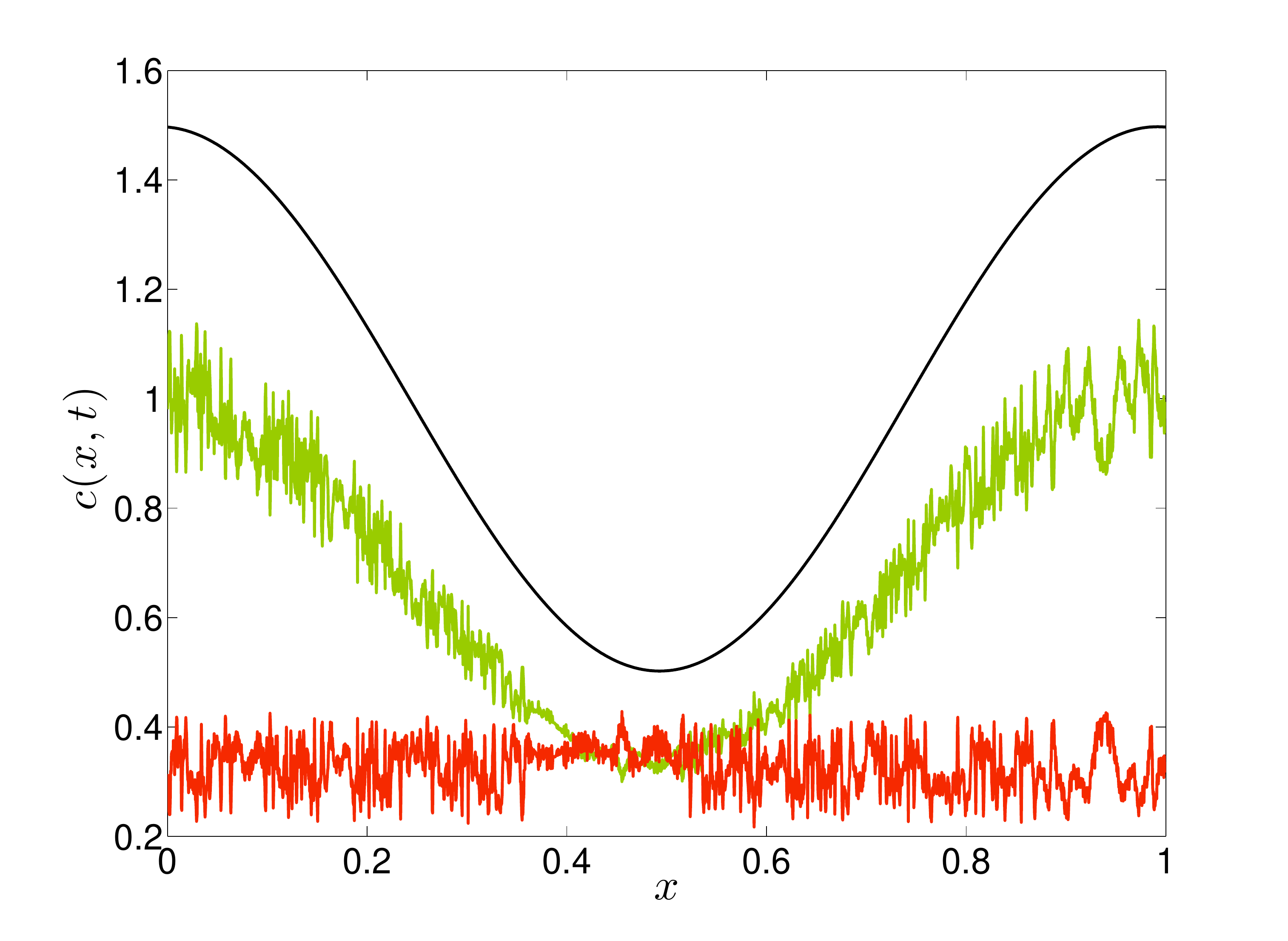}}  
  \end{minipage}
   \hfill
  \begin{minipage}{0.48\linewidth}
  \centerline{\includegraphics[width=5.8cm]{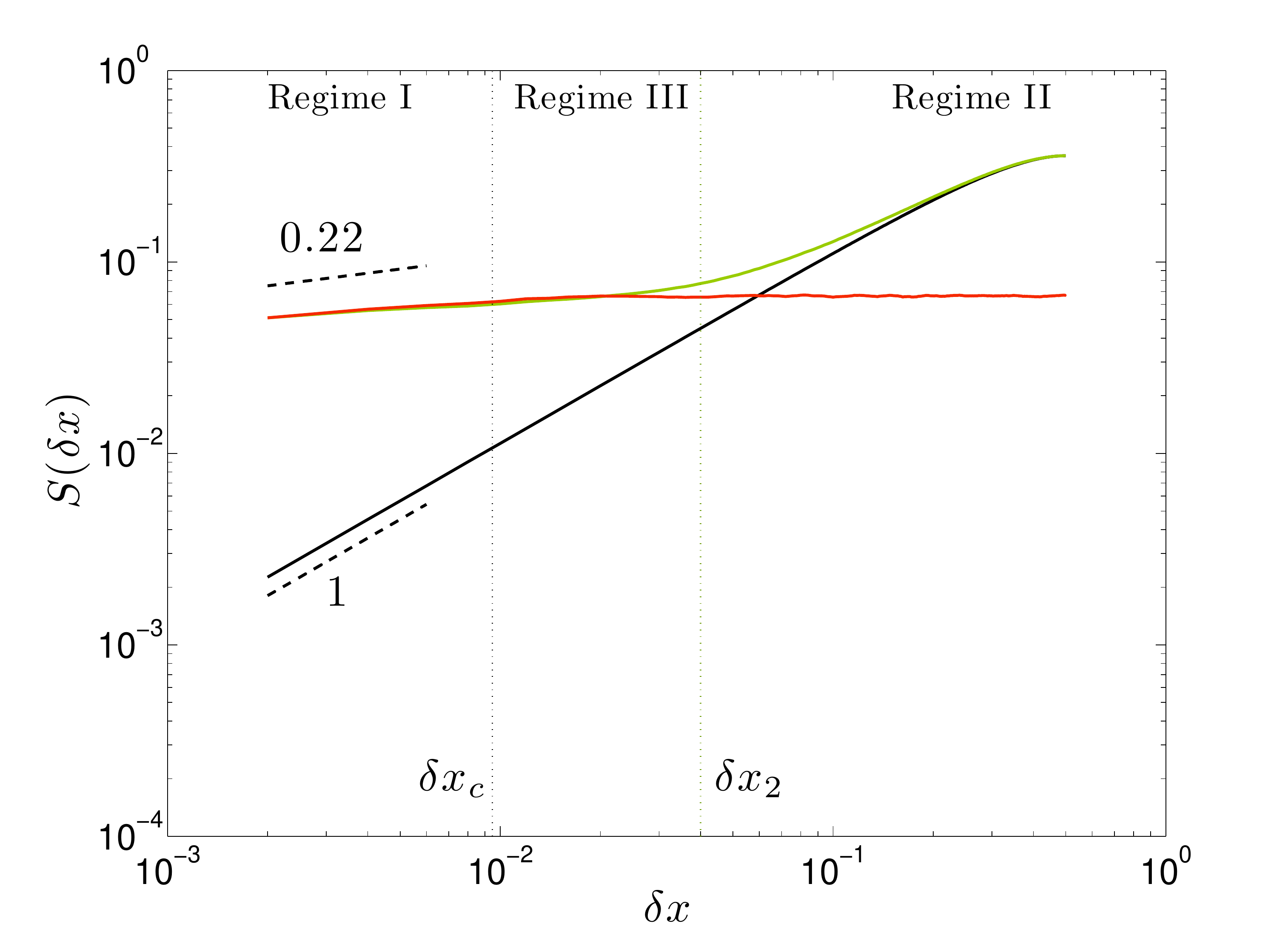}}
  \end{minipage}
  \vfill
  \begin{minipage}{\linewidth}
  \centerline{(a) $T=20$}  
  \end{minipage}  
   \begin{minipage}{0.47\linewidth}
  \centerline{\includegraphics[width=5.3cm]{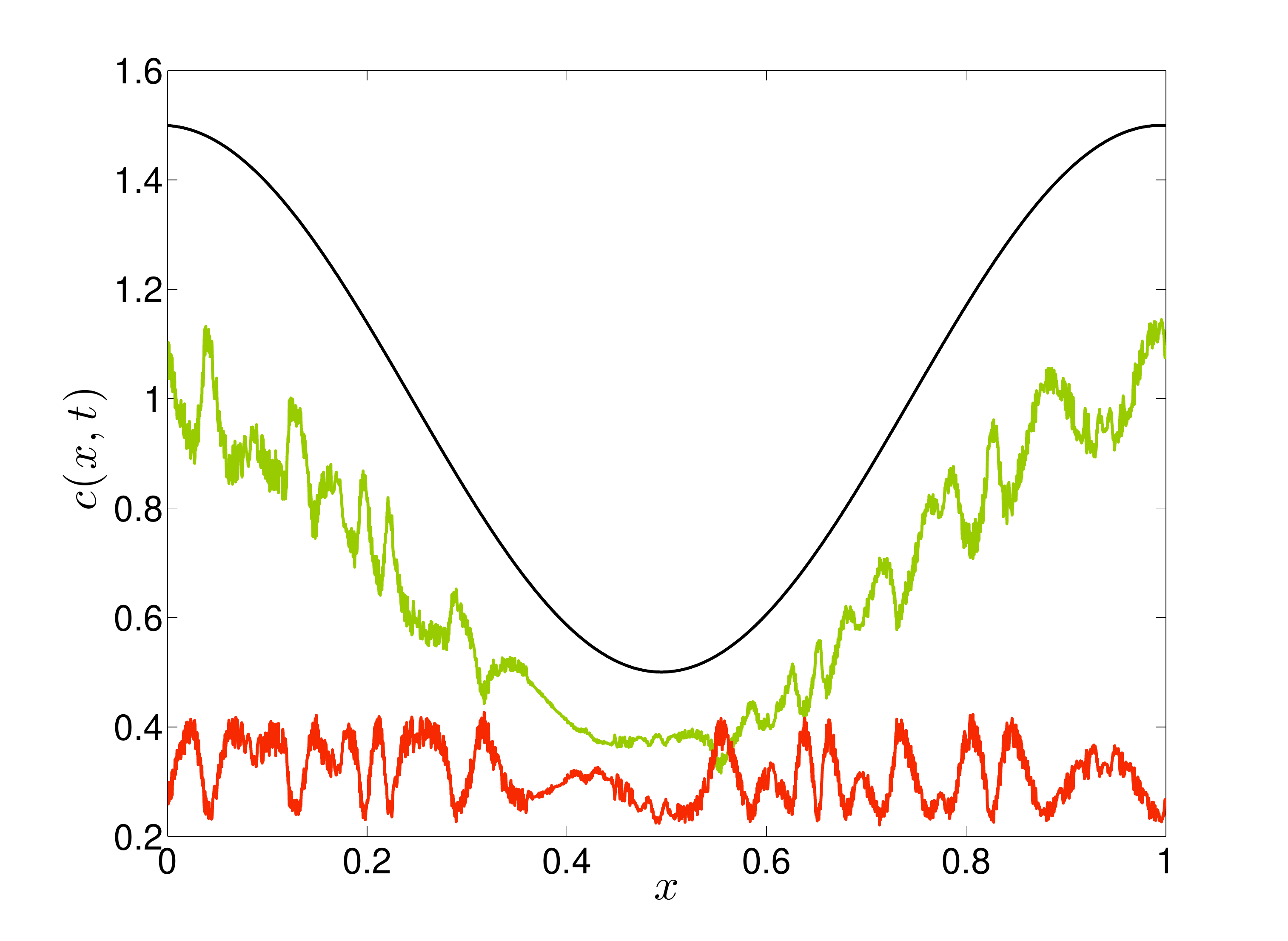}}  
  \end{minipage}
        \hfill
  \begin{minipage}{0.47\linewidth}
  \centerline{\includegraphics[width=5.8cm]{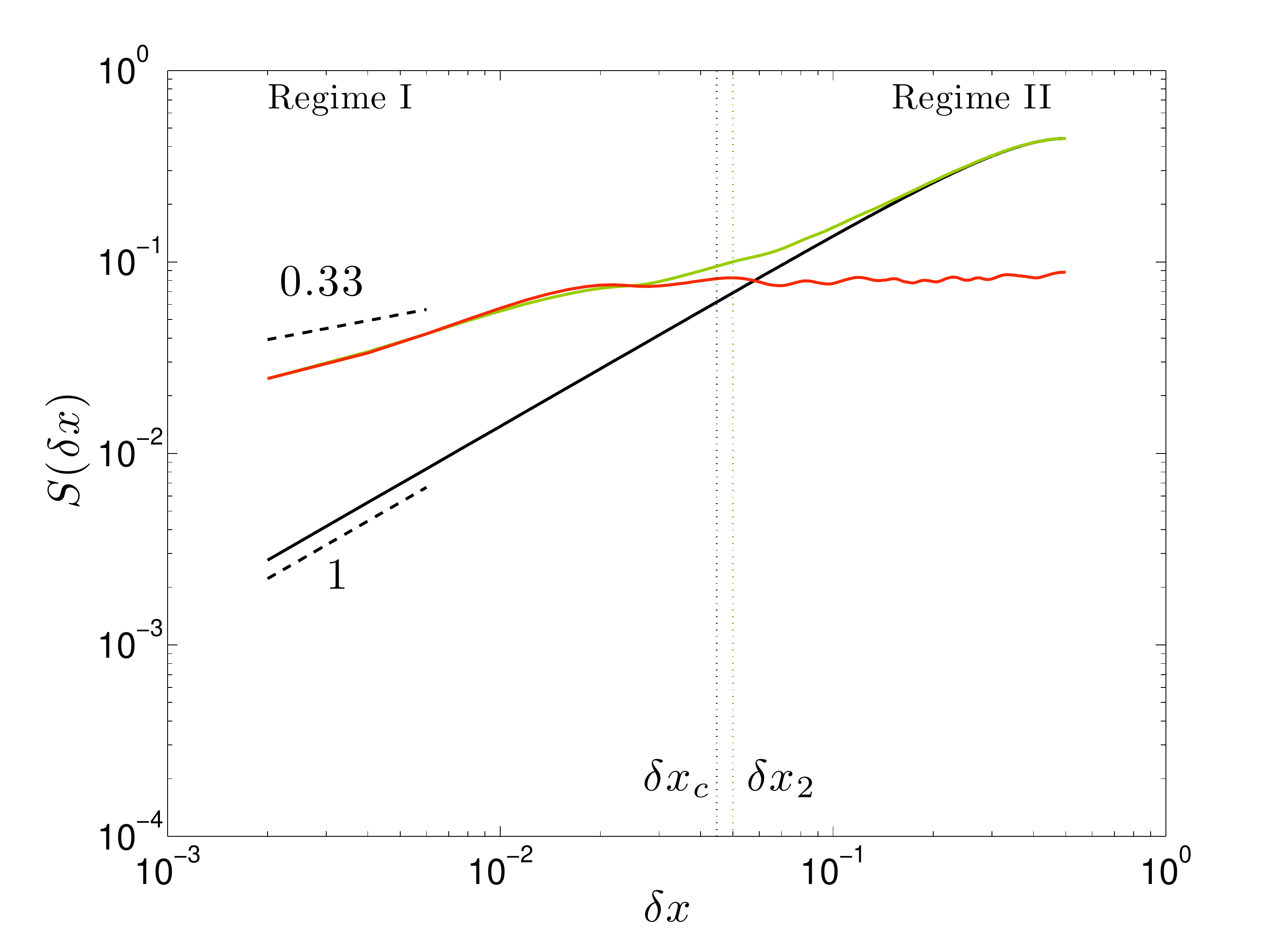}}
  \end{minipage}
  \vfill
  \begin{minipage}{\linewidth}
  \centerline{(b) $T=30$}  
  \end{minipage}              
  \begin{minipage}{0.48\linewidth}
  \centerline{\includegraphics[width=5.3cm]{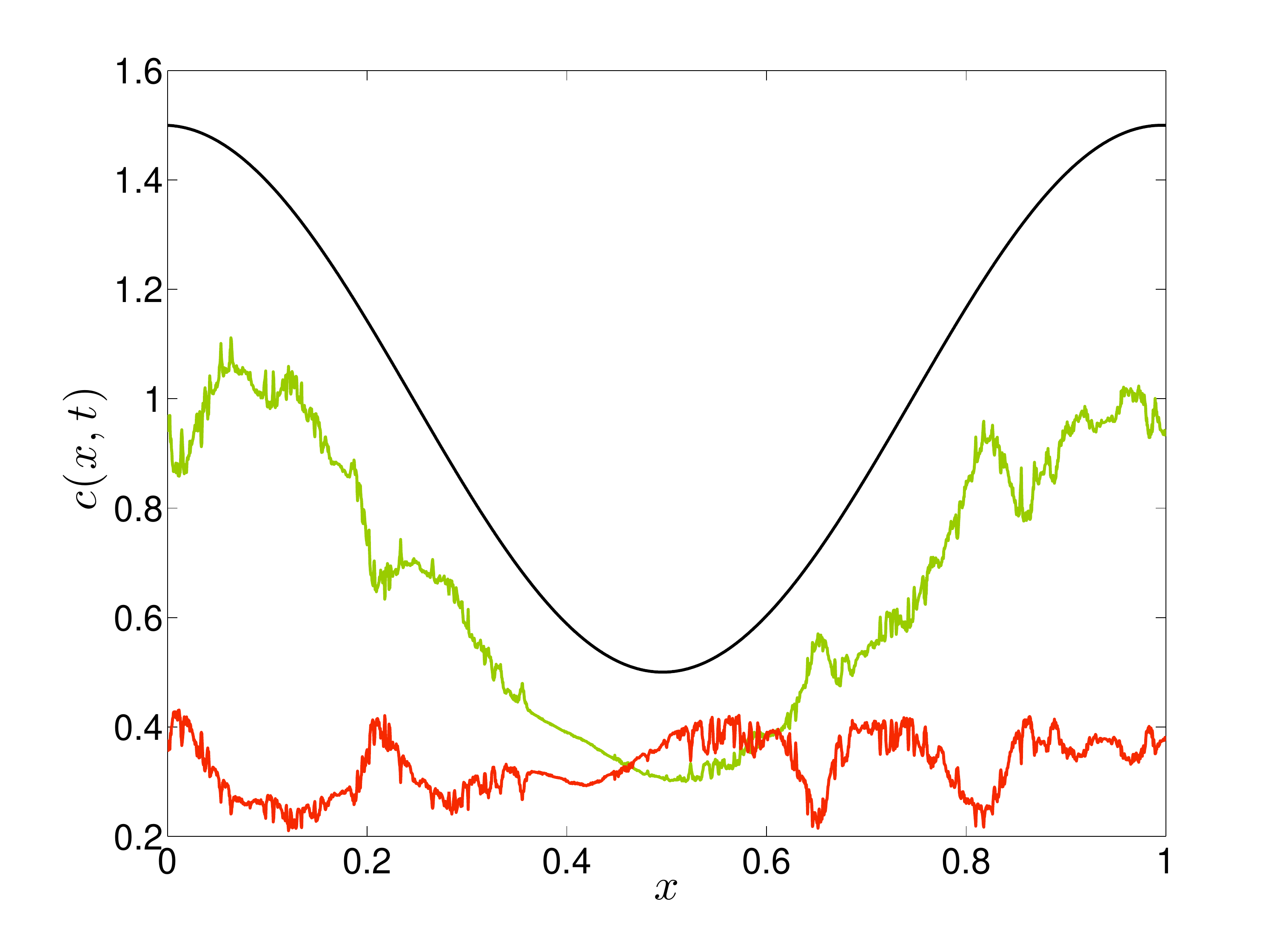}}  
  \end{minipage}
        \hfill
  \begin{minipage}{0.48\linewidth}
  \centerline{\includegraphics[width=5.8cm]{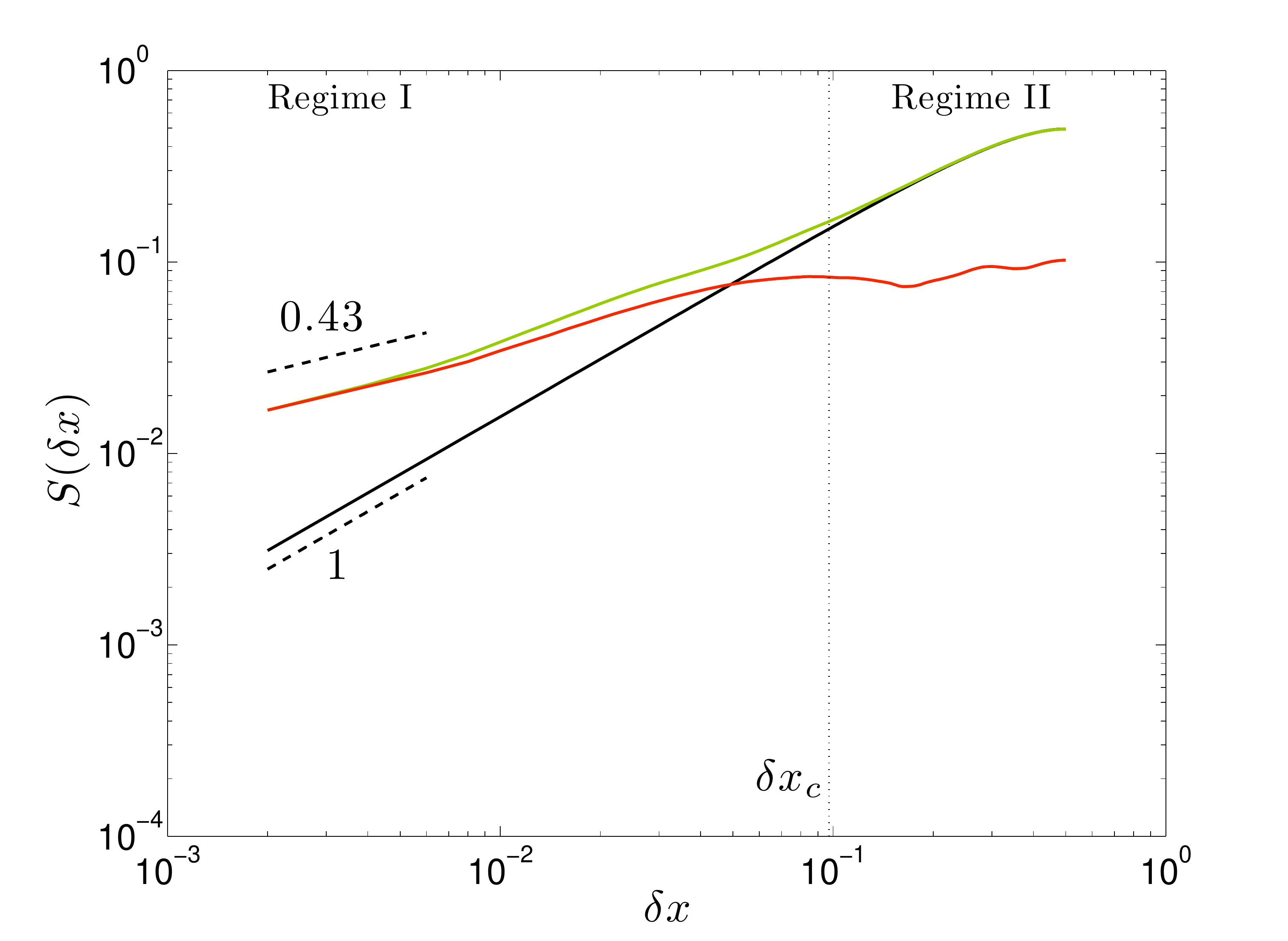}}
  \end{minipage}
  \vfill
  \begin{minipage}{\linewidth}
  \centerline{(c) $T=40$}  
  \end{minipage}               
     \caption{(Color online) Same as Fig. \ref{fig:VariationTau} but this time the flow parameter $T$ varies ($\tau=40$, $\delta=2$).         
     }      
	\label{fig:VariationT} 
\end{minipage}
\end{figure}

The evolution  of the concentration fields and their first-order structure functions,
are now examined as a function of the stirring strength of the flow, the latter parameterised by the value of $T$, and shown in Fig. (\ref{fig:VariationT}).
Starting from Fig. (\ref{fig:VariationT}(a)), as the value of $T$ increases,  so does  the value for $\delta x_c$  (see Eq. (\ref{eqn:estimate_char})) along with the range of length scales for which Regime I appears.
Again, the agreement between theory and numerics is close   
with  $\delta x_c$ providing a good prediction  for when the transition from Regime I to either Regime III (see Fig.  \ref{fig:VariationT}(a))  or Regime II (see Figs.  \ref{fig:VariationT}(b-c))  occurs.

\section{Summary and Conclusions}\label{sec:DelayConclusions}  
This paper has considered the spatial properties of chaotically
advected delay reactive scalar fields, i.e. scalar fields whose
reactions explicitly contain a delay time.  The investigation was
motivated by the need for a theoretical explanation for previous
numerical results obtained for a delay plankton model \cite{Abraham1998,TzellaHaynes2007} but the results are relevant to other 
chemical and biological systems \cite{Roussel1996,Murray1993}.

The system considered had stable reaction dynamics in which
spatial inhomogeneity is forced by a spatially smooth source and in
which the reacting species are advected by a two-dimensional, unsteady
and incompressible flow. The case of reactions described by a single linear delay equation was considered in detail as a simple prototype and the results were then extended to a reaction described by a system of nonlinearly interacting delay equations. 
Two main conclusions were drawn concerning the scaling behavior of
the delay reactive scalar fields. The first was that, no matter how
large the value of the delay time, at sufficiently small length scales
the scaling behavior is characterized by a H\"older exponent whose
value depends on the ratio of the slowest decay rate associated with
the reaction dynamics, i.e. the least negative chemical
Lyapunov exponent, to the  flow Lyapunov exponent.  Thus,
within this scaling regime, denoted as Regime I, the
introduction of a delay time into the reactions results in a scaling
behavior that is a straightforward generalization of that for which 
there is no delay time. For the particular case of the
delay plankton model, this implies that the phytoplankton and
zooplankton share the same scaling behavior at small scales. 

On the other hand, when the stirring of the flow is sufficiently strong or the delay time is sufficiently large, the scaling behavior undergoes a change beyond a transition lengthscale. 
The expression for the transition length scale was deduced to depend on both the stirring strength and the delay time, exponentially decreasing as a function of their product. 
This change of behavior is inherent to the delay system and may be described by 
three different scenarios: 
The first scenario occurs when a second scaling regime, denoted as Regime II,  
is created to accompany the first scaling regime. This new scaling regime appears at all small-scales that are larger than the transition length scale. 
The scaling behavior within this second regime is essentially captured by a reduced reaction system in which all reaction terms that contain a delay time are ignored. 
The value of the corresponding H\"older exponent depends on the ratio of the slowest decay rate associated with the reduced reactive processes to the  flow Lyapunov exponent. For the particular case of the delay plankton model, this result explains why the zooplankton assumes a similar distribution to a passive (non-reactive) scalar while the phytoplankton assumes a different less-filamental distribution. 
A second scenario occurs when the second scaling regime is preceded by a flat scaling regime, denoted as Regime III.  In this case there are three scaling regimes present: Regimes I, II and III. 
For this to happen, the transition length scale needs to be small compared to 
the ratio of the  
reaction terms that contain a delay time
to those terms that do not. 
As this ratio increases, so does  the range of length scales for which Regime III appears. 
When this ratio reaches the order of unity,
a last scenario occurs in which 
the Regime III appears at   
all small-scales that are larger than the transition length scale. In this case Regime II does not appear.

We believe that the investigation presented here resolves the main
issues concerning the small-scale spatial structure of chaotically
advected delay reactive scalar fields. Although the models under
consideration are highly simplified, they can be readily extended to
include any number of interacting species or space-dependent
productivity and death rates. As long as the reactions are stable, the
above conclusions remain unchanged.

There are, however,  details that need further examination. 
This paper has avoided the implications of a 
distribution of finite-time flow and chemical Lyapunov exponents. 
Some of the implications of a
distribution of finite-time flow Lyapunov exponents have been
addressed by \cite{Neufeld_etal2000a}. The implications of a distribution of
finite-time chemical Lyapunov exponents, avoided in this paper by
basing discussion on solutions of model chemical systems with constant
coefficients, could be incorporated in a similar way. 
It is believed that including these effects may give a better description of the 
fields' scaling behavior  
within Regime I for length scales close  to the transition length scale.

The primary theoretical predictions of this paper are the parameter dependence of the scaling behavior in three different regimes
and the transition length scales between those regimes. 
This makes it possible to develop a quantitative evaluation of the theory, for example, as applied to observations of ocean plankton distributions at the mesoscale;
one of
the principal motivations for the line of investigation in this
paper. Depending on the time it takes for the zooplankton to mature
and the stirring induced by the straining activity of the mesoscale
eddies, three, instead of one, scaling regimes may characterize the
plankton distributions. Given the differing spatial distributions
exhibited by the plankton at the open mesoscale ocean \cite{MackasBoyd1979,Tsuda1995,MartinSrokosz2002}, it is
worth taking into account the existence of these two new scaling regimes
when trying to interpret oceanic measurements at a large range of
length scales. A degree of care should be taken however as the ocean is
highly complex and the presence of small-scale forcing is ubiquitous
in the ocean, reflecting not only the individual zooplankton behavior
but also the presence of strong localized upwelling. Because the
impact that these processes have on larger scales may be significant \cite{MahadevanArcher2000,Martin_etal2002},
it is important to build the complexity of the idealised models
considered here by including both more realistic dynamics, in which
vertical effects and frontal circulation are taken into account, as
well as some of the characteristics of the individual zooplankton
behavior such as diurnal vertical migration. Finally, the distinct
role that a delay time plays on the formation of structures in
reactive scalar distributions is expected to prompt further research
on the subject. But it should be emphasized the results presented in
this paper have potential application beyond the field of ocean
sciences, to any system involving fluid flow and chemical or biological interactions.\\

\noindent
\textbf{Acknowledgments.} The authors are grateful to B. Legras, J. H. P. Dawes and A. P. Martin for their useful comments as well as A. Iserles for his insight. 
AT is currently supported by the  Marie Curie Individual fellowship HydraMitra No. 221827.

\appendix

 \setcounter{equation}{0}
 \def\theequation{A\arabic{equation}}
  \def\thesubsection{\Alph{subsection}}

\section*{Appendix: System of Delay Reactive Scalars}\label{app:appendix}
In this appendix 
we extend the theoretical results obtained in Sec. \ref{sec:DelayTheory}  
for a single delay reactive scalar 
to a system of such fields. 

\subsection{Key Properties of a System of Forced Linear Delay Equations}\label{app:key}    
Consider a system of forced, linear DDEs,
\begin{equation}\label{eqn:forcedsystem}
\dot{\bm{y}}=-\bm{A}\,\bm{y}(t)-\bm{B}\,\bm{y}(t-\tau)+\bm{f}(t),
\end{equation}  
where $\bm{A}$, $\bm{B}$ $\in \mathbb{R}^{n\times n} $ and $\bm{y}$, $\bm{f}$ $\in \mathbb{R}^{n}$.  
Retracing the same steps as for the single case (\ref{eqn:forced1D}), 
the characteristic equation that corresponds to the homogeneous part of Eq. (\ref{eqn:forcedsystem}), 
\begin{equation}\label{eqn:systemhom}
\dot{\bm{y}}=-\bm{A}\,\bm{y}(t)-\bm{B}\,\bm{y}(t-\tau),
\end{equation}
is obtained by looking for solutions of the form $\bm{c}e^{\lambda t}$, where  $\bm{c}\in \mathbb{R}^{n}$, 
and $\lambda\in\mathbb{C}$. 
The form of the characteristic equation is given by
\begin{equation}\label{eqn:charsystem}
h(\lambda)\equiv\text{det}\bm{H}(\lambda)\equiv|\lambda\bm{I}+\bm{A}+\bm{B}e^{-\lambda \tau}|=0,
\end{equation}                                                                             
where $\bm{H}(\lambda)$ is defined as the {\it characteristic matrix}.
 
The  {\it fundamental matrix}, is defined as the matricial solution to (\ref{eqn:systemhom}) with  initial conditions
\begin{equation}\label{eqn:initialcds_system}
\bm{M}_{\bm{Y}}(t) = \left\{ \begin{array}{rl}
 \bm{0}, &\mbox{ $t<0$,} \\
 \bm{I}, &\mbox{ $t=0$.}
 \end{array} \right. 
\end{equation}       

For $0\leq t \leq \tau$,
an exact expression for $\bm{M}_{\bm{Y}}(t)$ can be obtained using the method of steps (see \S\ref{subsec:KeyProperties}). 
For $t>\tau$ it is more useful to take the Laplace transform of $\bm{M}_{\bm{Y}}(t)$.
Using 
Eq. (\ref{eqn:systemhom}),
\begin{equation}\label{eqn:fundamental_aux_system}
\mathcal{L}(\bm{Y})(\lambda)=\bm{H}^{-1}(\lambda)\cdot\bm{1}. 
\end{equation}                                    
from where
\begin{equation}\label{eqn:fundamental_formal_system1b}  	
\bm{M}_{\bm{Y}}(t)=\int_{(\gamma)}e^{\lambda t}\bm{H}^{-1}(\lambda)\, d\lambda, 
\quad t>0,
\end{equation}
where $\gamma> \text{max}\{\re\lambda:h(\lambda)=0\}$.
The inverse of $\bm{H}(\lambda)$ can be written in terms of its matrix of cofactors, 
$\text{adj}\bm{H}(\lambda)$, and its determinant $h(\lambda)$.    
Integrating  $e^{\lambda t}\bm{H}^{-1}(\lambda)$  along a suitably chosen contour results in  
\begin{equation}\label{eqn:fundamental_matrix2}
\bm{M}_{\bm{Y}}(t)=
\sum_{j=0}^\infty\underset{\lambda=\lambda_j}{\text{Res}}
e^{\lambda t}\frac{\text{adj}\bm{H}(\lambda)}{h(\lambda)},
\quad t>0,
\end{equation}
where, similarly to the single case, 
the infinite series (\ref{eqn:fundamental_matrix2}) is proved \cite{Lunel1995} to be uniformly convergent in $t$. 
Because $\bm{A},\bm{B}$ are real, the roots of (\ref{eqn:charsystem}) are either real or come in complex conjugate pairs.  
For parameters chosen so that all roots  are distinct,      
$e^{\lambda t}\bm{H}^{-1}(\lambda)$ only has simple poles. 
By combining the contributions from each complex conjugate pair, (\ref{eqn:fundamental_matrix2})  becomes 
\begin{subequations}\label{eqn:fundamental_formal_system_TOTAL}
\begin{equation}
\bm{M}_{\bm{Y}}(t)=\lim\limits_{N\rightarrow\infty}\bm{M}_{\bm{Y}_N}(t), \quad t>0,    
\end{equation}       
where  $\bm{M}_{\bm{Y}_N}(t)$ is equal to
\begin{equation}\label{eqn:fundamental_formal_system3_bis}  
\bm{M}_{\bm{Y}_N}(t)=  \sum_{\substack{j=1\\ \{\lambda_j^+:\,\im\lambda_{j}\geq 0\}}}^N  
e^{\re \lambda_j^+ t}\bm{\hat{H}}(\lambda_j^+,t) 
\end{equation}      
with $\re \lambda_j^+>\re \lambda_{j+1}^+$. 
$\bm{\hat{H}}(\lambda_j^+,t)$  is a real matrix equal to
\begin{equation}\label{eqn:Hhatlambda}
\bm{\hat{H}}(\lambda_j^+,t)=2^{\mathcal{H}(\im\lambda_j^+)}\,
\re \left(e^{i\im\lambda_j^+ t}\frac{\text{adj}\bm{H}(\lambda_j^+)}{h'(\lambda_j^+)}\right),	                 
\end{equation}                                                                                                                            
\end{subequations}
with  $\mathcal{H}(x)$ previously defined in Eq. (\ref{eqn:H(x)_back2}).   
Hence, for sufficiently large $t$,  the behavior of $\bm{M}_{\bm{Y}_N}(t)$ is dominated by 
$\bm{M}_{\bm{Y}_1}(t)$.
Therefore, $\bm{M}_{\bm{Y}_N}(t)$  satisfies the following approximate expression 
\begin{equation}\label{eqn:YLongShort_system}
\bm{M}_{\bm{Y}}(t) = \left\{ \begin{array}{rll}
&e^{-\bm{A}t},  &\mbox{ $0\leq t\leq\tau$,} \\
\sim& \bm{M}_{\bm{Y}_1}(t)  , &\mbox{ $t>\tau$.}
 \end{array} \right.
\end{equation} 
Thus, similarly to the fundamental solution corresponding to a single  DDE (see Eq. (\ref{eqn:YLongShort})), 
the  behavior of the fundamental matrix of a system of linear DDEs is distinctly different to the  
behavior of the fundamental matrix of a system of linear ODEs.  
At the same time,  for $t\leq\tau$, the fundamental matrix 
obtained by setting $\bm{B}=\bm{0}$ in  Eq. (\ref{eqn:forced1D})  is identical to 
$\bm{M}_{\bm{Y}}(t)$.

The general solution to  Eq. (\ref{eqn:forcedsystem})  depends on $\bm{M}_{\bm{Y}}(t)$.   
Let this solution  be denoted by $\bm{y}(\bm{\phi},\bm{f})(t)$, 
where  $\bm{\phi}(t)$ represents the initial conditions given by
\begin{equation}
\bm{y}(t)=\bm{\phi}(t),\quad t\in[-\tau,0].	
\end{equation}	
Provided that the forcing, $\bm{f}(t)$, is exponentially bounded, 
$\bm{y}(\bm{\phi},\bm{f})(t)$, 
 is obtained by considering the Laplace transform of Eq. (\ref{eqn:forcedsystem}).
This leads to 
\begin{equation} 
\begin{split}	
\bm{H}(\lambda)\mathcal{L}(\bm{y})(\lambda)=
\bm{\phi}(0)-e^{-\lambda\tau}\bm{B}\cdot&\int_{-\tau}^0e^{-\lambda\theta}\bm{\phi}(\theta) d\theta \\
+&\int_0^\infty e^{-\lambda t} \bm{f}(t) 
dt,
\end{split} 	
\end{equation} 
from where it can be deduced that 
\begin{subequations}\label{eqn:variation_system_formula}
\begin{equation}\label{eqn:variation_system}
\bm{y}(\bm{\phi},\bm{f})(t)=\bm{y}(\bm{\phi},\bm{0})(t)+\int_{0}^{t}\bm{M}_{\bm{Y}}(t-t')\cdot\bm{f}(t')\,dt',
\end{equation}                                       
with $\bm{y}(\bm{\phi},\bm{0})(t)$ the solution to Eq. (\ref{eqn:systemhom}), 
\begin{equation}\label{eqn:variation_homogeneous_system}   
\bm{y}(\bm{\phi},\bm{0})(t)=\bm{M}_{\bm{Y}}(t)\cdot\bm{\phi} (0)-
\int_{-\tau}^{0}\bm{M}_{\bm{Y}}(t-\theta-\tau)\cdot\bm{B}
\cdot\bm{\phi}(\theta)\,d\theta.	
\end{equation}	  
\end{subequations}
Representation (\ref{eqn:variation_system_formula}) corresponds to the variation of constants formula for a system of forced linear DDEs.

\subsection{Scaling Behavior}\label{app:scaling}  
Consider the chemical activity within a fluid parcel to be given by Eq. (\ref{eqn:chem})
repeated now,
\begin{equation}\label{eqn:parceldelay1system}
\frac{d}{dt}\bm{C}_{\bm{X}(t)}=
\bm{\mathcal F}_{-\tau}(\bm{C}_{\bm{X}(t)},\bm{X}(t)),    
\end{equation}	                                                                                          
where once again, $\bm{C}_{\bm{X}(t)}$ 
represents the fluid parcel's chemical concentration at a time $t$, with the fluid parcel
trajectory evolving according to Eq. (\ref{eqn:traj}). 

The evolution of the chemical difference between a pair of fluid parcels may be obtained by linearizing 
(\ref{eqn:parceldelay1system}) around a fluid parcel.   
This gives
\begin{equation}\label{eqn:evolution_chemical_difference_delaysystem}
\frac{d}{dt}\,\delta \bm{C}(t)=
 \frac{\partial\bm{\mathcal F}_{-\tau}}{\partial\bm{C}}\cdot{\delta \bm{C}(t)}
+\frac{\partial\bm{\mathcal F}_{-\tau}}{\partial\bm{C}_{-\tau}}\cdot\delta \bm{C}(t-\tau)
+\frac{\partial\bm{\mathcal F}_{-\tau}}{\partial\bm{xX}}\cdot{\delta \bm{X}(t)},	
\end{equation}	                                                                                                                    
where again $\{\delta\bm{X}(t)\}$, the label on the fluid parcel difference is suppressed for brevity of notation.
The gradient matrices 
$\partial\bm{\mathcal F}/\partial\bm{C},\partial\bm{\mathcal F}/\partial\bm{C}_{-\tau}\in \mathbb{R}^{n\times n}$ 
while $\partial\bm{\mathcal F}/\partial\bm{X}\in\mathbb{R}^{n\times d}$, where $d$ is the system's spatial dimension.

To analyze the scaling behavior of the fields, we first consider that both matrices $\partial\bm{\mathcal F}/\partial\bm{C}$ and $\partial\bm{\mathcal F}/\partial\bm{C}_{-\tau}$ are  constant 
such that expression (\ref{eqn:evolution_chemical_difference_delaysystem}) assumes a similar form to Eq. (\ref{eqn:forcedsystem}), with 
\begin{equation}\label{eqn:evolution_chemical_difference_delaysystem2}
\frac{d}{dt}\delta \bm{C}(t)=-\bm{A}\cdot\delta \bm{C}(t)
-\bm{B}\cdot\delta \bm{C}(t-\tau)
+\frac{\partial\bm{\mathcal F}}{\partial\bm{C}}\cdot{\delta \bm{C}(t)},	
\end{equation}        
where 
$\bm{A}=-\partial\bm{\mathcal F}/\partial\bm{C}$ and $\bm{B}=-\partial\bm{\mathcal F}/\partial\bm{C}_{-\tau}$. (The non-constant case is discussed later.) 

Using the  variation of constants formula  (\ref{eqn:variation_system_formula}), 
the chemical difference  may be expressed in terms of the fundamental matrix, $\bm{M}_{\bm{Y}}(t-t')$, as
\begin{equation}\label{eqn:dcsystem_1}
\begin{split}	
\delta \bm{C}(t)= 
\bm{M}_{\bm{Y}}(t)\cdot&\delta \bm{C}(0)
- \int_{-\tau}^0 \bm{M}_{\bm{Y}}(t-\theta-\tau) \cdot \bm{B}\cdot \bm{\phi}(\theta)d\theta  
\\
&+\int_0^t \bm{M}_{\bm{Y}}(t-t') \cdot \left(\frac{\partial \bm{\mathcal F}}{\partial \bm{X}}\cdot \delta\bm{X}(t')\right) dt',
 \end{split}
\end{equation}    
where for $t\in[-\tau,0]$,
$\delta \bm{C}(t)=\bm{\phi}(t)$.  
In the long-time limit and for $\re\lambda_1<0$,
where $\re\lambda_1=\text{max}\{\re\lambda:h(\lambda)=0\}$,
the first two terms in Eq. (\ref{eqn:dcsystem_1}) 
describing the evolution of the initial conditions vanish. 
Substituting the exact expression  (\ref{eqn:fundamental_formal_system_TOTAL}) for $\bm{M}_{\bm{Y}}(t)$
into (\ref{eqn:dcsystem_1}), the  long-time chemical difference of the $i$th chemical species
is given by 
\begin{equation}\label{eqn:evolution_chemical_difference_delaysystem3}   
\delta C_i(t)\approx
\sum_{j=1}^\infty
 \left(\int_0^{t}     
e^{\re\lambda_j^+ (t-t')}  
\bm{\hat{H}}(\lambda_j^+,t) 
\cdot \delta_{\bm{X}}\bm{\mathcal F}(t') 
dt'\right)_i
\quad t\gg t'.
\end{equation}
Since for $0\leq t\leq \tau$, 
$\bm{M}_{\bm{Y}}(t)=\exp[-\bm{A}t]$, (\ref{eqn:evolution_chemical_difference_delaysystem3}) becomes
\begin{equation}\label{eqn:evolution_chemical_difference_delaysystem4} 
\begin{split}   
\delta C_i(t)\approx 
&\sum_{j=1}^\infty  
\left(
\int_0^{t-\tau}  
e^{\re\lambda_j^+ (t-t')}  
\bm{\hat{H}}(\lambda_j^+,t)
\cdot\delta_{\bm{X}}\bm{\mathcal F}(t') 
dt'\right)_i  \\
&\sum_{k=1}^{n}     
\int_{t-\tau}^t (\bm{\hat a}_k)_i\, e^{-a_k (t-t')} \bm{\hat a}_k^\dag\,\cdot\, \delta_{\bm{x}}\bm{\mathcal F}(t') 
 dt',
 \quad t\gg t',
\end{split}  
\end{equation}  
where $\bm{\hat a}_k$  and $\bm{\hat a}_k^\dag$ are respectively the right and left eigenvectors of $\bm{A}$
that correspond to the eigenvalue $a_k$,  
normalized so that $\bm{\hat a}_k^\dag\bm{\hat a}=1$. 
Because $\bm{\mathcal F}_{-\tau}$ depends smoothly on space, its spatial derivatives do not increase or decrease in a systematic way 
and therefore  
\begin{subequations}
\begin{align}
\bm{\hat{H}}(\lambda_j^+,t) \cdot\delta_{\bm{X}}\bm{\mathcal F}(t')&\sim \bm{c}_j |\delta \bm{X}(t')| \\
\intertext{and}
\bm{\hat a}_k\bm{\hat a}_k^\dag\,\cdot\, \delta_{\bm{X}}\bm{\mathcal F}(t')&\sim \bm{c}_k'|\delta \bm{X}(t')| ,  
\end{align}
\end{subequations}                                                              
where $\bm{c}_j,\bm{c}_k'\in\mathbb{C}^n$ are constant vectors.                 
             
The dominant behavior of  $\delta C_i(t)$ 
 is determined by the slowest decaying 
eigenfunction within each integral.
Thus, Eq. (\ref{eqn:evolution_chemical_difference_delaysystem4}) approximately becomes
\begin{subequations}
\begin{equation}\label{eqn:evolution_chemical_difference_delaysystem5}
	\begin{split} 
	\delta C_i(t)\sim
	&\int_0^{t-\tau} (\bm{\tilde{c}}_1)_i  e^{\re\lambda_1 (t-t')} |\delta \bm{X}(t')| dt'\\
	+& \int_{t-\tau}^t (\bm{\tilde{c}}_1')_i e^{-\re a_1 (t-t')} |\delta \bm{X}(t')| dt, 
	  \end{split}
\end{equation}
where
\begin{align}
\re a_1&=\text{max}\{\re a:\text{det}(\bm{A}-a\bm{I})=0\}\\  
\intertext{and}
\re\lambda_1&=\text{max}\{\re\lambda:\text{det}\bm{H}(\lambda)=0\},
\end{align}                                                        
with $\bm{\tilde{c}}_1,\bm{\tilde{c}}_1'\in\mathbb{R}^n$ constant vectors related to $\bm{c_1}$ and $\bm{c_2}$.
\end{subequations}

In the limit of $t\rightarrow\infty$ and for $\Delta t=t-t'$, 
the chemical difference between a pair of fluid parcels and thus from (\ref{eqn:LagrangianEulerian2}), 
the fields' small-scale behavior 
may  be captured by 
\begin{subequations}\label{eqn:delayconcdiff_system} 
\begin{equation}\label{eqn:delayconcdiff_systema}
(\delta c_i)_\infty(\delta\bm{x})\sim 
\int_0^\infty Y_{\bm{M}}(\Delta t)\text{min}\{\delta \bm{x}e^{h_0\Delta t},1\}d\Delta t
\quad\text{for $|\delta \bm{x}|\ll 1$}, 
\end{equation}
where the evolution of a typical line element 
stirred by chaotic advection flow (see Eq. (\ref{eqn:stirring})) was taken into account.   
The term $Y_{\bm{M}}(\Delta t)$ represents the exponential part of the slowest decaying eigenfunction and is defined as
\begin{equation}\label{eqn:generalsystemY_M} 
Y_{\bm{M}i}(t)=\,
\begin{cases} 
\bm{\tilde{c}}_1\, e^{-\re a_1 t},                  & \text{for $0\leq t\leq\tau$},\\ 
\bm{\tilde{c}}_1'\, e^{\re\lambda_1 t},              &  \text{for $t>\tau$}.
\end{cases}      
\end{equation}	
\end{subequations}  
Expression (\ref{eqn:delayconcdiff_systema}) is essentially equivalent to expression (\ref{eqn:delayconcdiff}) (with $a=\re a_1$).
Thus, the same conclusions obtained for a single delay reactive scalar also  apply for a system of such scalars and thus the same set of scaling laws as  (\ref{eqn:delay_holders})
 characterise the small-scale structure of the fields. 
Namely, for the general case considered here, the fields' stationary state spatial structure is a priori shared and  
can be classified into two scaling regimes: the first regime, Regime I, is governed by 
the least negative chemical Lyapunov exponent that corresponds to the full delay system.
Regimes II and III appear at length scales larger than the transition length scale. The expression for the latter, denoted by $\delta x_c$,
remains unchanged and is given by
 (\ref{eqn:characteristiclength scale}). 
The scaling behavior within Regime II is governed by the slowest decay rate that corresponds 
to the reduced system obtained once all terms that involve a delay time are ignored. 
The appearance of a flat Regime III that is sandwiched between Regimes I and II
depends on
the maximum value of $Y_{\bm{M}i}(t\geq \tau)$. 
The length scale $(\delta x_2)_i$ associated with this regime may be estimated using
\begin{equation}\label{eqn:estimatedx2system}
(\delta x_2)_i\sim
\text{max}|\bm{M}_{\bm{Y}}(t)\cdot\bm{\mathit{f}}|_i^{\{\text{max}h_0/a_i,1\}},
\end{equation}
where $\bm{\mathit{f}}$ corresponds to the forcing direction in the chemical space. 
If $(\delta x_2)_i$ is sufficiently large, Regime III occupies all length scales larger than $\delta x_c$. 

Note that to deduce the set of Eqs. (\ref{eqn:delayconcdiff_system})
the general case for which  $\bm{c}_1,\bm{c}_1'$ have no zero entries was considered. 
Asymmetrical couplings may result in 
either $\bm{c}_1$ or $\bm{c}_1'$ having zero entries.
For these zero entries subdominant eigenfunctions 
 need to be considered
in which case the expression (\ref{eqn:generalsystemY_M}) for 
$Y_{\bm{M}}(t)$ needs to be modified in order to represent the exponential behavior of these subdominant eigenfunctions.  
A case for which either $\bm{c}_1$ or $\bm{c}_1'$ have zero entries is the delay plankton model considered in \S\ref{subsec:numerics_delayplanktonmodel}.

The theoretical analysis above has assumed that  both matrices 
$\partial\bm{\mathcal F}/\partial\bm{C}$ and $\partial\bm{\mathcal F}/\partial\bm{C}_{-\tau}$ are constant. The analysis may be extended  to a system for which the reactions are non-linear and therefore the matrices are not constant. In this case the  rates of convergence of the reaction processes  
will depend on the trajectory of the fluid parcel. For large enough trajectory times and in a flow that is uniformly chaotic, these rates are expected to be independent of the  fluid parcel trajectory. In the infinite-time limit these rates, defined as chemical Lyapunov exponents \cite{Neufeld_etal1999}, may be expected to converge to a fixed value \cite{Neufeld_etal1999}. 

\bibliographystyle{plain}

\bibliography{bib}


 \end{document}